\documentclass[12pt]{article}
\pdfoutput=1
\usepackage{amsmath,amssymb,amsthm,mathtools,mathrsfs,bm,mathdots,graphicx,float,array,multirow,multicol,rotfloat,caption,subcaption,enumerate,geometry,parskip,adjustbox,footnote,booktabs,etoolbox,tikz,tikz-cd,pdflscape,csquotes,lscape,rotating,empheq,xfrac,arydshln,afterpage,pifont,anyfontsize}
\usepackage[para]{threeparttable}
\usepackage[all]{xy}
\usepackage[shortlabels]{enumitem}
\usepackage[normalem]{ulem}
\usepackage[vcentermath]{youngtab}
\usepackage[boxsize=.8em]{ytableau}
\usepackage[numbers,sort&compress]{natbib}
\usepackage[framemethod=TikZ]{mdframed}
\usepackage[hidelinks]{hyperref}
\usepackage{cleveref}
\BeforeBeginEnvironment{ques}{\savenotes}
\AfterEndEnvironment{ques}{\spewnotes}
\numberwithin{equation}{section}
\numberwithin{figure}{section}
\graphicspath{{figs/}}
\allowdisplaybreaks
\makeatletter
\usetikzlibrary{arrows,arrows.meta,shapes,shapes.misc,patterns,decorations.pathmorphing,decorations.pathreplacing,positioning,chains,fit}
\pgfdeclarepatternformonly{coarsedots}{\pgfqpoint{-1pt}{-1pt}}{\pgfqpoint{5pt}{5pt}}{\pgfqpoint{12pt}{12pt}}{
    \pgfpathcircle{\pgfqpoint{.5pt}{.5pt}}{.5pt}
    \pgfusepath{fill}}
\tikzset{gauge/.style={rounded rectangle, draw=black!100, thick, minimum size=5mm},  gaugeD/.style={rounded rectangle, draw=black!100,double,thick,minimum size=5mm},  empty/.style={rounded rectangle, draw=white!100, thick, minimum size=5mm}, flavor/.style={rectangle, draw=black!100, thick, minimum size=5mm},flavorD/.style={rectangle, draw=black!100, double,thick, minimum size=5mm}}
\hypersetup{colorlinks=true}
\hypersetup{linkcolor=black}
\hypersetup{citecolor=black}
\hypersetup{urlcolor=black}
\theoremstyle{plain}
\newtheorem*{thm*}{Theorem}

\newcounter{ques}
\renewcommand{\theques}{\arabic{ques}}
\newenvironment{ques}[2][]{%
    \refstepcounter{ques}%
    \ifstrempty{#1}
    {\mdfsetup{%
        frametitle={%
        \tikz[baseline=(current bounding box.east),outer sep=0pt]
        \node[anchor=east,rectangle,draw=blue!40,line width=.5mm,fill=white]{\strut Question~\theques};}}
    }
    {\mdfsetup{%
        frametitle={%
        \tikz[baseline=(current bounding box.east),outer sep=0pt]
        \node[anchor=east,rectangle]{\strut Question~\theques:~#1};}}%
    }
\mdfsetup{%
innertopmargin=2pt,linecolor=blue!40,%
linewidth=2pt,topline=true,%
frametitleaboveskip=\dimexpr-\ht\strutbox\relax%
}
\begin{mdframed}[skipabove=.9\baselineskip]\relax\label{#2}}{%
\end{mdframed}}

\newtheorem*{conj:3dmirror}{Conjecture \ref{conj:3dmirror}}

\def\CN{\mathcal{N}}

\def\CD{\mathcal{D}}

\def\CF{\mathcal{F}}
\def\IZ{\mathbb{Z}}
\DeclareMathOperator{\tr}{Tr}

\DeclareMathOperator{\lcm}{lcm}

\newcommand{\Lpagenumber}{\ifdim\textwidth=\linewidth\else\bgroup
  \dimendef\margin=0 
  \ifodd\value{page}\margin=\oddsidemargin
  \else\margin=\evensidemargin
  \fi
  \raisebox{\dimexpr+2\topmargin-\headheight-\headsep-0.5\linewidth}[0pt][0pt]{%
    \rlap{\hspace{\dimexpr\margin+\textheight+3.5\footskip}%
    \llap{\rotatebox{90}{\thepage}}}}%
\egroup\fi}
\AddToHook{shipout/background}{\Lpagenumber}%
\makeatother

\begin{document}

\begin{titlepage}
\vspace*{-3cm} 
\begin{flushright}
{\tt DESY-24-125}\\
\end{flushright}
\begin{center}
\vspace{1.7cm}
{\LARGE\bfseries Generalized symmetry constraints \\[0.3em] on deformed 4d (S)CFTs}
\vspace{1cm}

{\large
Monica Jinwoo Kang,$^1$ Craig Lawrie,$^2$ Ki-Hong Lee,$^3$ and Jaewon Song$^3$}\\
\vspace{.8cm}
{$^1$ Department of Physics and Astronomy, University of Pennsylvania\\
Philadelphia, PA 19104, U.S.A.}\par
\vspace{.2cm}
{$^2$ Deutsches Elektronen-Synchrotron DESY,\\ Notkestr.~85, 22607 Hamburg, Germany}\par
\vspace{.2cm}
{$^3$ Department of Physics, Korea Advanced Institute of Science and Technology\\
Daejeon 34141, Republic of Korea}\par
\vspace{.3cm}

\scalebox{.78}{\tt monica6@sas.upenn.edu, craig.lawrie1729@gmail.com, 
khlee11812@gmail.com,
jaewon.song@kaist.ac.kr}\par
\vspace{1.2cm}

\textbf{Abstract}
\end{center}

We explore the consequence of generalized symmetries in four-dimensional $\mathcal{N}=1$ superconformal field theories. First, we classify all possible supersymmetric gauge theories with a simple gauge group that have a nontrivial one-form symmetry and flows to a superconformal field theory. Upon identifying unbroken discrete zero-form symmetries from the ABJ anomaly, we find that many of these theories have mixed zero-form/one-form 't Hooft anomalies. Then we classify the relevant deformations of these SCFTs that preserve the anomaly. From this mixed anomaly together with the anomalies of the discrete zero-form symmetries, we find obstructions for the relevant deformations of these SCFTs to flow to a trivially gapped phase. We also study non-Lagrangian SCFTs formed by gauging copies of Argyres--Douglas theories and constrain their deformations. In particular, we explore a new duality between the diagonal gauging of two $\mathcal{D}_3(SU(N))$ theories and $SU(N)$ gauge theory with two adjoints. We also repeat our analysis for a host of non-supersymmetric gauge theories having nontrivial one-form symmetry including examples that appear to flow to Bank--Zaks type CFTs. 

\vspace{1cm}
\vfill 
\end{titlepage}

\tableofcontents
\newpage

\section{Introduction}\label{sec:intro}

The infrared (IR) behavior of asymptotically-free gauge theories in four dimensions demonstrates a wealth of subtle non-perturbative physics; the most well-known examples are confinement and dynamical chiral symmetry breaking in QCD. For a given gauge theory, it is generally challenging to analyze the particular infrared behavior realized due to the dominance of strongly-coupled effects along the flow into the IR. 
Progress in understanding the IR is typically made by restricting to gauge theories with additional symmetries, especially supersymmetry. The power of the symmetry approach lies in that it provides consistency constraints that narrow down the possible IR physics. In fact, in fortunate situations, the symmetry completely fixes the infrared behavior. 

A powerful tool for constraining the infrared structure is 't Hooft anomaly matching \cite{tHooft:1979rat}. Consider a quantum field theory $\mathcal{T}_\text{UV}$, and let $S$ denote an arbitrary subgroup of the full (zero-form) global symmetry of this theory. $S$ may incorporate spacetime symmetries such as the Lorentz group or time-reversal symmetry, as well as internal symmetries. The symmetry $S$ is anomalous if there exists an $S$-transformation such that the partition function of the quantum field theory is rotated by a constant phase which cannot be canceled by the addition of a local counterterm. In \cite{tHooft:1979rat}, it was shown that if there exists a renormalization group flow to a new QFT: 
\begin{equation}
    \mathcal{T}_\text{UV} \,\,\rightarrow\,\, \mathcal{T}_\text{IR} \,,
\end{equation}
then $\mathcal{T}_\text{IR}$ must have the same anomalous transformations of the partition function as $\mathcal{T}_\text{UV}$. Therefore, if a theory has a nontrivial 't Hooft anomaly for a symmetry group $S$, then either the infrared is $S$-preserving and gapless, $S$-preserving and gapped with a nontrivial extended operator sector, or else $S$ is spontaneously broken. Crudely, there does not exist a trivially-gapped vaccum, that is, a unique gapped vacuum without topological degrees of freedom.

In recent years a more general notion of symmetry has come to the fore \cite{Gaiotto:2014kfa}. From this perspective, a global symmetry with symmetry group $G$ is associated with codimension-one topological defects, where the fusion of the defects corresponds to the multiplication law in $G$. A natural generalization is a $p$-form symmetry, with symmetry group $G$, which is instead associated with codimension-$(p+1)$ topological defects, which again fuse according to the group multiplication rule.\footnote{For a small sample of recent reviews on generalized symmetries, see \cite{Sharpe:2015mja,Schafer-Nameki:2023jdn,Cordova:2022ruw,Luo:2023ive,Brennan:2023mmt, Bhardwaj:2023kri}.} This generalized notation of symmetry is particularly powerful, as it retains the anomaly matching behavior under renormalization group flows -- anomalies involving higher-form symmetries therefore provide a new arsenal with which to constrain the infrared physics. One striking example was studied in \cite{Gaiotto:2017yup}, where a consequence of the fact that pure 4d $SU(N)$ Yang--Mills theory has a $\mathbb{Z}_N$ one-form symmetry and, at $\theta = \pi$, has a $\mathbb{Z}_2$ zero-form symmetry associated with time-reversal was observed. When $N$ is even there is a mixed 't Hooft anomaly between these symmetries, and thus the infrared theory can be described via one of the following.
\begin{enumerate}
    \item[(i)] The theory is gapless with a single vacuum. This appears unlikely, at any value of $\theta$, from both large $N$ arguments \cite{Witten:1998uka,DiVecchia:1980yfw,Witten:1980sp,Witten:1979vv,Witten:1978bc} and from lattice simulations (see, e.g., \cite{Lucini:2012gg} for a review).
    \item[(ii)] The theory is gapped with a single vacuum and the anomaly is saturated by a time-reversal preserving topological field theory (TQFT). In fact, the existence of a time-reversal preserving TQFT was ruled out in \cite{Cordova:2019bsd}, as any such TQFT violates reflection positivity.
    \item[(iii)] The time-reversal symmetry is spontaneously broken and the vacuum is twofold degenerate.
    \item[(iv)] The $\mathbb{Z}_N$ one-form symmetry is broken. However, it is believed that pure $SU(N)$ Yang--Mills confines for all $\theta$, which implies that the one-form symmetry is unbroken \cite{Gaiotto:2014kfa}.
\end{enumerate} 
Thus, we can see how the existence of the 't Hooft anomaly involving the one-form symmetry constrains the infrared behavior. In particular, the authors of \cite{Gaiotto:2017yup} combine the anomaly with the expectation that pure $SU(N \gg 1)$ Yang--Mills is confined and gapped to argue that (iii) is the most likely infrared behavior. This is further supported by recent results in lattice gauge theory \cite{Hirasawa:2024vns}. 

In this paper, we study 4d supersymmetric gauge theories where this generalized symmetry perspective provides nontrivial constraints on the infrared physics. Consider an asymptotically-free 4d $\mathcal{N}=1$ gauge theory; if there exists a well-defined $U(1)_R$ R-symmetry,\footnote{The precise notion of a ``well-defined'' $U(1)_R$ is provided in Section \ref{sec:2}.} then it is widely believed that the gauge theory flows to an SCFT in the infrared. In this paper, our concern is with the theories that have no such $U(1)_R$ symmetry. The poster-child for such a theory is pure $\mathcal{N}=1$ super-Yang--Mills, with $G = SU(N)$. The theory has a $\mathbb{Z}_N$ one-form symmetry, and the naive $U(1)$ axial symmetry is broken to $\mathbb{Z}_{2N}$ by an ABJ anomaly. There exists a mixed 't Hooft anomaly between the $\mathbb{Z}_N$ one-form symmetry and the $\mathbb{Z}_N$ subgroup of the zero-form symmetry; this anomaly then provides a constraint on the possible infrared behavior. In fact, the infrared behavior of pure $\mathcal{N}=1$ Yang--Mills can be studied explicitly; see  \cite{Witten:1982df,Veneziano:1982ah}, or else \cite{Terning:2006bq} for a textbook treatment. The $\mathbb{Z}_{2N}$ zero-form symmetry is spontaneously broken to $\mathbb{Z}_2$ via the gaugino bilinear acquiring a vacuum expectation value:
\begin{equation}
    \langle \lambda \lambda \rangle \,\,=\,\, \exp\left( \frac{2\pi i n}{N} \right) \Lambda^3 \,,
\end{equation}
where $n \in \mathbb{Z}_N$ labels the $N$ distinct vacua and $\Lambda$ is the holomorphic scale set by the divergence of the one-loop gauge coupling. We can see that this is consistent with the mixed zero-form/one-form anomaly as the anomalous subgroup of the zero-form symmetry is spontaneously broken.

For the class of theories that we are interested in, we utilize the following construction. Begin with a 4d $\mathcal{N}=1$ gauge theory, $\mathcal{T}_\text{gauge}$, with a simple gauge group, an asymptotically-free gauge coupling, and a trivial superpotential, such that there exists a well-defined $U(1)_R$ R-symmetry; this gauge theory flows to an SCFT, $\mathcal{T}_\text{SCFT}$, in the infrared. Then, consider a sequence of relevant deformations of the infrared SCFT
\begin{equation}\label{eqn:supdef}
    W = \sum_i \lambda_i \mathcal{O}_i \,,
\end{equation}
where $\mathcal{O}_i$ are relevant operators, such that all internal $U(1)$ global symmetries, which do not belong to the Cartan of any non-Abelian symmetry, are broken. The resulting deformed theory, which we refer to as $\mathcal{T}_\text{deformed}$, lacks a well-defined $U(1)_R$, and thus we expect that the theory in the infrared, $\mathcal{T}_\text{IR}$, is not an SCFT.\footnote{\label{fn:bigwarning} Of course, it is possible for $U(1)$ global symmetries to be emergent along the renormalization group flow into the infrared, in which case they can provide a well-defined $U(1)_R$ symmetry. Thus, we cannot generically rule out the existence of an IR SCFT.}  Furthermore, since supersymmetry is not explicitly broken along the flow, we expect that $\mathcal{T}_\text{IR}$ is supersymmetric.\footnote{We also consider performing a deformation by a relevant operator such that supersymmetry is broken. The consequences of the anomalies that we consider on the behavior of $\mathcal{T}_\text{IR}$ do not depend on the preservation of supersymmetry.} Pictorially, we have the following:
\begin{equation}\label{eqn:bigpicture}
    \begin{tikzcd}[row sep=5ex, column sep=14ex]
        & \mathcal{T}_\text{gauge} \arrow[d, "\text{IR}"] \\
        & \mathcal{T}_\text{SCFT} \arrow[r, "W = \sum_i \lambda_i \mathcal{O}_i", "U(1)_R \,\,\text{breaking}"'] & \mathcal{T}_\text{deformed} \arrow[d, "\text{IR}"] \\
        & & \mathcal{T}_\text{IR} &
    \end{tikzcd} 
\end{equation}
Note that the order of the superpotential deformations in equation \eqref{eqn:supdef} is important; for example some $\mathcal{O}_i$ may only become relevant after some deformations. For ease of notation only, we schematically write the sum; more generically, equation \eqref{eqn:bigpicture} involves a sequence of superpotential deformations and flows into the infrared until a $U(1)_R$ no longer exists.

How can we describe the infrared behavior of the deformed theories $\mathcal{T}_\text{deformed}$ which lack a $U(1)_R$ R-symmetry? Concretely, we seek to understand what we can learn about $\mathcal{T}_\text{IR}$ from the anomalies of $\mathcal{T}_\text{gauge}$, related via the deformation structure depicted in equation \eqref{eqn:bigpicture}. Assuming that $\mathcal{T}_\text{gauge}$ induces a mixed anomaly between the one-form symmetry and the zero-form symmetry, and a part of that mixed anomaly survives the sequence of superpotential deformations, then we can conclude that $\mathcal{T}_\text{IR}$ behaves as one of the following. 
\begin{enumerate}
	\item[(i)] A single symmetry-preserving gapless vacuum. This may be a theory of massless fermions or an interacting CFT either of which must saturate the anomaly. However, due to the unbroken supersymmetry, such a description would require the existence of an emergent $U(1)_R$ along the flow into the infrared. 
	\item[(ii)] A unique symmetry-preserving gapped vacuum where a TQFT saturates the anomaly.
	\item[(iii)] A phase where the vacuum spontaneously breaks the anomalous global symmetry. This could be a confining phase where the one-form symmetry is unbroken and the anomaly matching conditions are fulfilled by the breaking of the anomalous zero-form symmetry; to avoid overcounting when we determine the number of degenerate vacua after the symmetry breaking, it is necessary that we identify the proper global structure of the faithful symmetry group acting on the theory. When a discrete zero-form symmetry is broken, the 't Hooft anomaly can be matched by the presence of distinct local counterterms in each of the degenerate vacua. Alternatively, the one-form symmetry itself may be broken, leading to an unconfined infrared; see \cite{Hsin:2018vcg} for a discussion on the spontaneous breaking of the one-form center symmetry of 4d gauge theories. Finally, it is possible to have some combination of the above; spontaneous symmetry breaking can lead to multiple vacua, where each of the vacua may be confining, support gapless or topological degrees of freedom, etc. 
\end{enumerate}
While our focus is principally on the mixed anomaly between the discrete zero-form symmetry and the one-form symmetry, any other 't Hooft anomalies of $\mathcal{T}_\text{deformed}$ also constrain the IR behavior via 't Hooft anomaly matching; to emphasize this, we also consider the anomalies involving the discrete zero-form symmetry and other zero-form symmetries.

The structure of this paper is as follows. In Section \ref{sec:2}, we exhaustively enumerate the 4d $\mathcal{N}=1$ simple and simply-connected Lagrangian gauge theories with nontrivial one-form symmetry that flow in the infrared to an interacting SCFT. We augment this in Section \ref{sec:mixedanom} by analyzing which of these gauge theories also has a mixed anomaly between the one-form symmetry and the discrete remnant of the axial zero-form symmetry. In anticipation of clarifying the structure of the infrared phases, we study the global structure of the zero-form flavor symmetry, together with the (discrete) anomalies, in Section \ref{sec:global0}. Next, in Section \ref{sec:deformations}, we ask which of these IR SCFTs possess $\mathcal{N}=1$ or $\mathcal{N}=0$ preserving relevant deformations which maintains some of the mixed anomaly between the zero-form and one-form symmetries after deformation; the mixed anomaly provides constraints on the IR behavior of these deformed theories. In Section \ref{sec:ADtheories}, we generalize to 4d $\mathcal{N}=1$ simple and simply-connected gauge theories where the matter is instead provided by strongly-coupled Argyres--Douglas SCFTs; we focus on theories with $a=c$ and study their mixed anomalies and deformations. Finally, in Section \ref{sec:nonsusy}, we discuss a non-supersymmetric analogue of our question: we enumerate the matter content of simple and simply-connected Lagrangian 4d asymptotically-free/conformal gauge theories with nontrivial one-form symmetry, and ask about their mixed anomalies and infrared behavior. 

\section{\texorpdfstring{$\mathcal{N}=1$}{N=1} gauge theories with one-form symmetry}\label{sec:2}

Since it is our intention to study the consequences of generalized symmetry in 4d $\mathcal{N}=1$ theories, our first port of call is to construct an interesting class of theories with such generalized symmetries. We begin with Lagrangian SQFTs, and the analysis which we perform in this section is captured by the following question.

\begin{ques}{ques:first}
\renewcommand{\thempfootnote}{\arabic{mpfootnote}}
What are the 4d $\mathcal{N}=1$ Lagrangian theories with a simply-connected simple gauge group (and trivial superpotential) that flow in the infrared to an interacting SCFT with a nontrivial one-form symmetry? 
\end{ques}

Suppose that we have a simple and simply-connected gauge group $G$, and chiral multiplets that transform in the representation 
\begin{equation}\label{eqn:Rgeneral}
  \bm{R} = \bigoplus_{i=1}^s n_i \bm{R}_i \,,
\end{equation}
of $G$, where $\bm{R}_i$ are irreducible representations. Without loss of generality, we can assume that $\bm{R}_i \neq \bm{R}_j$ for $i \neq j$. We refer to these chiral multiplets as $Q_i^\alpha$, where $\alpha = 1, \cdots, n_i$. The first constraint on the association of a consistent quantum field theory with the pair $(G, \bm{R})$ is the cancellation of any gauge anomaly. This requires that $\bm{R}$ satisfies
\begin{equation}\label{eqn:CONDITION_ANOMCANC}
    \sum_{i=1}^s n_i A(\bm{R}_i) = 0 \,,
\end{equation}
where $A(\bm{R}_i)$ is the triangle anomaly coefficient of the irreducible representation $\bm{R}_i$.

\subsection{One-form symmetry}

First, we want to understand which gauge theories associated with the pair $(G, \bm{R})$ realize a nontrivial one-form symmetry. As discussed in \cite{Gaiotto:2014kfa}, a one-form symmetry arises from the presence of topological operators of codimension two, where topological means that continuous deformations of the codimension-two manifold supporting the operator do not modify any physical observables, provided the deformation does not cross a charged operator. 

When $G$ is simple and simply-connected, any one-form symmetry arises from topological Gukov--Witten operators \cite{Gukov:2006jk,Gukov:2008sn} that are labeled by the elements of the center of $G$, as explained in \cite{Heidenreich:2021xpr}, and the centers of all simple and simply-connected Lie groups are listed in Table \ref{tab:centerG}. To determine the one-form symmetry, it is necessary to determine which of these topological Gukov--Witten operators are nontrivial operators. These Gukov--Witten operators can have a nontrivial linking number with Wilson lines, which are labeled by irreducible representations of $G$. Such an operator is a nontrivial symmetry operator only if the linking number is trivial with all Wilson lines which end.
A Wilson line in a representation $\bm{r}$ can end if there exists a local charged operator also in the representation $\bm{r}$. Hence, if a Gukov--Witten operator associated with an element $k \in Z(G)$ has a nontrivial linking number (which is related to the character of the representation of $Z(G)$ induced by the representation $\bm{r}$ evaluated on $k$) with a Wilson line that can end, then the Gukov--Witten operator is trivial.

Each irreducible representation $\bm{r}$ of $G$ is associated with an irreducible representation of the center, $Z(G)$, of $G$. Characters of irreducible representations of $Z(G)$ are elements of the Pontryagin dual $\widehat{Z}(G) = \operatorname{Hom}(G, U(1))$. As the center is a cyclic group $\mathbb{Z}_M$ for $G \neq Spin(4N)$, such characters can be written as
\begin{equation}\label{eqn:nality}
    \chi_{\bm{r}}(k) = \exp\left(\frac{2 \pi i m}{M} k\right) \,,
\end{equation}
where the integer $m$ is fixed by the representation $\bm{r}$. For $G = Spin(4N)$ the center is $\mathbb{Z}_2 \times \mathbb{Z}_2$, and thus the characters are
\begin{equation}
    \chi_{\bm{r}}(k_1, k_2) = \exp\left(\pi i (m_1 k_1 + m_2 k_2) \right) \,,
\end{equation}
where now $m_1$ and $m_2$ are fixed by the choice of representation $\bm{r}$. To determine the character of the irreducible representation of $Z(G)$ for any $\bm{r}$, we need to know the integer $m$ (or the pair $(m_1, m_2)$); this can be determined from the generating representation of $G$. We specify these integers here.
\begin{itemize}
  \item For $SU(N)$ the fundamental representation $\bm{N}$ has $m = 1$; the complex conjugate representation $\overline{\bm{N}}$ has $m = -1$.
  \item For $USp(2N)$ the fundamental representation $\bm{2N}$ has $m = 1$.
  \item For $Spin(2N + 1)$ the spinor representation $\bm{S}$ has $m = 1$.
  \item For $Spin(4N + 2)$ the spinor representation $\bm{S}$ has $m = 1$.
  \item For $Spin(4N)$ the spinor representation $\bm{S}^+$ has $(m_1, m_2) = (1, 0)$ and the co-spinor representation $\bm{S}^-$ has $(m_1, m_2) = (0, 1)$.
\end{itemize}
To determine the $m$ (or $(m_1, m_2)$) for an arbitrary representation, we only need to note that they are additive under tensor products.

\begin{table}[H]   
    \centering
    \begin{tabular}{c||c|c|c|c|c|c|c}
        $G$ & $SU(N)$ & $Spin(2N+1)$ & $Spin(4N)$ & $Spin(4N+2)$ & $USp(2N)$ & $E_6$ & $E_7$ \\
        \hline
        $Z(G)$ & $\IZ_N$ & $\IZ_2 $& $\IZ_2 \times \IZ_2$ & $\IZ_4$ & $\IZ_2$ & $\IZ_3$ & $\IZ_2$ 
    \end{tabular}
    \caption{Center $Z(G)$ of the simple and simply-connected Lie group $G$. Groups not included in this table have a trivial center.}    
    \label{tab:centerG}
\end{table}

Let $\chi_i(k)$ be the characters of the irreducible representations of $Z(G)$ associated to the $\bm{R}_i$ appearing in equation \eqref{eqn:Rgeneral}. Let $\Gamma$ be the subgroup of $Z(G)$ generated in the following way:
\begin{equation}\label{eqn:Gamma}
    \Gamma = \langle \, k \in Z(G) \, | \, \chi_i(k) = 1 \text{ for all } i = 1, \cdots, s \, \rangle \, \leq \, Z(G) \,.
\end{equation}
$\Gamma$ parametrizes the elements $k \in Z(G)$ corresponding to Gukov--Witten operators that cannot be unlinked from Wilson lines ending on local operators; these Gukov--Witten operators are then nontrivial topological operators and thus generate a one-form symmetry.
Therefore, $\Gamma$ is the one-form symmetry of the $G$ gauge theory with chiral multiplets in the representation $\bm{R}$ as in equation \eqref{eqn:Rgeneral}. In order to fulfil the desired criteria listed in Question \ref{ques:first}, we require that $\bm{R}$ is such that 
\begin{equation}\label{eqn:CONDITION_1FS}
    |\Gamma| > 1 \,.
\end{equation}
We note that this calculation did not depend on the choice of superpotential, only on the matter spectrum.

Throughout this section, we have assumed that the gauge group is simply-connected. Instead, we can start by specifying the gauge algebra $\mathfrak{g}$, the representation of the chiral multiplets $\bm{R}$, and the superpotential; a priori, this data is insufficient to guarantee that the theory possesses a scalar-valued partition function on a closed spacetime manifold. Instead, it has a partition vector, which is one of the hallmarks of a relative quantum field theory (see, e.g., \cite{Witten:1996hc,Aharony:1998qu,Witten:1998wy,Moore:2004jv,Belov:2006jd,Freed:2006yc,Witten:2007ct,Henningson:2010rc,Freed:2012bs,Tachikawa:2013hya}). We briefly review the importance of relative versus absolute QFT with a view to generalizations of Question \ref{ques:first} following \cite{Aharony:2013hda,Lawrie:2023tdz,Gukov:2020btk,Argyres:2022kon} and references therein.

Generically, whenever we have a $2k$-dimensional theory which contains self-dual $(k-1)$-form gauge fields, there are $(k-1)$-dimensional dynamical objects which have charges in a charge lattice $\Lambda$. There are also $(k-1)$-dimensional heavy defects, which carry charges in a free lattice $\Lambda^* \subset \Lambda \otimes \mathbb{Q}$. On $\Lambda^*$ there is a $\mathbb{Q}$-valued bilinear pairing $b_{\Lambda^*}$, which extends the integer-valued Dirac pairing on the dynamical objects. The heavy defects whose charges are valued in $\Lambda^*$ do not typically have an integer-valued Dirac pairing amongst themselves; this leads to the theory becoming inconsistent after quantization. To resolve this difficulty, we are required to pick a maximal subset of defects, which have a mutually integer-valued Dirac pairing, as belonging to the theory. This is a choice of maximal isotropic sublattice, $\Lambda^L$, of the defect charge lattice:
\begin{equation}
    \Lambda \subseteq \Lambda^L \subseteq \Lambda^* \,.
\end{equation}
The intermediate defect group \cite{tHooft:1977nqb,tHooft:1979rtg} is defined as
\begin{equation}
    \mathbb{D} = \Lambda^*/\Lambda \,,
\end{equation}
together with a $\mathbb{Q}/\mathbb{Z}$-valued bilinear pairing $b$ inherited from the Dirac pairing $b_{\Lambda^*}$. Then, a choice of maximal isotropic sublattice is equivalent to a choice of Lagrangian subgroup of the defect group:
\begin{equation}
    L = \Lambda^L/\Lambda \subset \mathbb{D} \,.
\end{equation}
This is also known as a choice of polarization of the relative QFT.
Specifically, a polarization is a choice of $L \subset \mathbb{D}$ such that
\begin{equation}
    |L|^2 = |\mathbb{D}| \qquad \text{ and } \qquad b(\mu_1, \mu_2) = 0 \text{ mod $1$ } \quad \text{ for all } \quad \mu_1, \mu_2 \in L \,. 
\end{equation}
A relative QFT together with a polarization specifies an absolute QFT, which then has a well-defined partition function.

The one-form symmetry of the absolute theory after picking a choice of polarization $L$ of the intermediate defect group $\mathbb{D}$ is
\begin{equation}
    L^\vee = \mathbb{D}/L \,.
\end{equation}
We immediately observe the importance of the choice of polarization for the analysis in this paper: one-form symmetry is not a well-defined notion in the relative QFT! Whenever the spacetime dimension is a multiple of four, the defect group can be written as a sum 
\begin{equation}
    \mathbb{D} = \mathbb{D}^{(e)} \oplus \mathbb{D}^{(m)} \,,
\end{equation}
where $\mathbb{D}^{(e)} = \mathbb{D}^{(m)}$ and the two factors are referred to as the electric and magnetic factors, respectively.
It is straightforward to see that there are always at least two polarizations, corresponding to $L = \mathbb{D}^{(m)}$ and $L = \mathbb{D}^{(e)}$, where the first is known as the electric polarization and, when the $(k-1)$-form gauge fields are associated with a simple gauge algebra $\mathfrak{g}$, specifies the global form of the gauge group to be the simply-connected group $G$ associated with $\mathfrak{g}$.

For the 4d $\mathcal{N}=1$ gauge theories associated with the pair $(\mathfrak{g}, \bm{R})$ that we are considering, the intermediate defect group is
\begin{equation}
    \mathbb{D} = \Gamma \oplus \Gamma \,,
\end{equation}
where $\Gamma$ is the same as it was in equation \eqref{eqn:Gamma} for $G$ the simply-connected group associated with $\mathfrak{g}$. The choice of polarization changes the spectrum of line operators of the theory, such as discussed in \cite{Aharony:2013hda}. When choosing a polarization different from the electric polarization both the one-form symmetry and the structure of the mixed anomalies changes.\footnote{The simplest example demonstrating that different polarizations lead to different one-form symmetry groups occurs for $\Gamma = \mathbb{Z}_4$ where there exists a polarization such that the one-form symmetry is $L^\vee = \mathbb{Z}_2 \oplus \mathbb{Z}_2$, as opposed to the electric one-form symmetry which is $L^\vee = \mathbb{Z}_4$. See \cite{Aharony:2013hda} for more details.} We leave the analysis of the infrared fate of such non-simply-connected theories for future work.

\subsection{Conformal window}

In Question \ref{ques:first}, we are interested in gauge theories which flow in the infrared to interacting superconformal field theories. This provides a set of strong constraints on the representation $\bm{R}$ under which the chiral multiplets transform.

The first requirement is that the one-loop $\beta$-function of the gauge coupling for the gauge group $G$ is non-positive. This imposes that $\bm{R}$ must satisfy
\begin{equation}\label{eqn:oneloop}
    \sum_{i} n_i T(\bm{R}_i) \leq 3 h_G^\vee \,,
\end{equation}
where $T(\bm{R}_i)$ is the Dynkin index of the representation $\bm{R}_i$.\footnote{We use the normalization that the Dynkin index of the fundamental representation of $\mathfrak{su}(N)$ is $\sfrac{1}{2}$.} If the inequality in equation \eqref{eqn:oneloop} is saturated, then the theory is only conformal if it possesses a nontrivial conformal manifold, otherwise it is infrared-free \cite{Leigh:1995ep,Green:2010da}. For all simple and simply-connected 4d $\mathcal{N}=1$ gauge theories which saturate the equality in equation \eqref{eqn:oneloop},
the question of the existence of a conformal manifold was answered in \cite{Razamat:2020pra}.

Given an $\bm{R}$ satisfying equation \eqref{eqn:oneloop}, it is necessary for the associated theory to realize an anomaly-free superconformal R-symmetry if it is to flow to a nontrivial SCFT in the infrared. Generically, the superconformal R-symmetry is not uniquely fixed by the anomaly-free condition, and it is necessary to invoke $a$-maximization \cite{Intriligator:2003jj} to find it. Once the superconformal R-symmetry is determined, we need to make sure that all gauge-invariant operators of the putative SCFT satisfy the unitarity bound. If $\mathcal{O}$ is a chiral operator then the conformal dimension $\Delta$ of $\mathcal{O}$ is written in terms of the R-charge $R(\mathcal{O})$ as
\begin{equation}
    \Delta(\mathcal{O}) = \frac{3}{2} R(\mathcal{O}) \,.
\end{equation}
Thus, we require that 
\begin{equation}\label{eqn:unit}
    R(\mathcal{O}) > \frac{2}{3} \,,
\end{equation}
for all chiral operators. 
This unitarity bound is, in fact, sometimes violated \cite{Maruyoshi:2018nod, Agarwal:2019crm, Agarwal:2020pol, Cho:2024civ}, and in such cases we often find that the would-be unitarity-violating operator becomes free, i.e., $\Delta = 1$, and gets decoupled along the RG flow, while the remaining part of the theory becomes a unitary interacting SCFT. Such decoupling usually happens when the number of matter degrees of freedom is small, so that the quantum corrections to the classical scaling dimension are large. When such a decoupling occurs, one useful way to analyze the remaining interacting sector is to introduce a \emph{flip field} $M_\mathcal{O}$ and couple it to the operator $\mathcal{O}$ that decouples via the superpotential \cite{Barnes:2004jj, Benvenuti:2017lle, Maruyoshi:2018nod}
\begin{align}
    W = M_\mathcal{O} \mathcal{O} \,. 
\end{align}
The F-term constraint for $M_\mathcal{O}$ will remove $\mathcal{O}$ from the chiral ring, or, equivalently, the quadratic term contributes a mass term for the would-be free operator $\mathcal{O}$. Upon removing the unitary-violating operator $\mathcal{O}$,\footnote{Of course, it may not be possible to remove the unitary-violating operators by decoupling along the RG flow; this typically occurs when the number of matter degrees of freedom is sufficiently small to generate a dynamical superpotential as discussed in \cite{Affleck:1983mk} or to flow to a gapped theory as in pure SYM. In these extreme cases, we assume that the theory does not flow to an interacting SCFT in the infrared.} one should re-do the $a$-maximization procedure and re-examine the unitarity. If there is another unitarity-violating operator, we remove it via another flip field ad nauseam. Once we do not find any unitarity-violating operator, we have isolated the SCFT at the IR fixed point. This SCFT may, in fact, be a free theory, and it is necessary to verify that there remain operators which do not become free along the flow so that we have an interacting SCFT in the infrared.

In this vein, the first step is to verify that every operator belonging to the chiral ring of the theory, which can be straightforwardly enumerated from the pair $(G, \bm{R})$, satisfies the unitarity condition in equation \eqref{eqn:unit}. However, there may exist unitarity violating operators which do not belong to the chiral ring, such as were studied in \cite{Maruyoshi:2018nod, Cho:2024civ}. A more refined check which can capture such cases involves determining the superconformal index \cite{Kinney:2005ej,Romelsberger:2005eg} and verifying that there are no unitarity violating terms \cite{Beem:2012yn,Evtikhiev:2017heo}. In this work, we do not perform this additional check and instead we only focus on the chiral ring; as such, this provides a necessary, but not sufficient condition, for the existence of an interacting superconformal theory in the infrared.

Altogether, if $\bm{R}$ is such that the one-loop $\beta$-function is either asymptotically free or else conformal with at least one exactly marginal operator, and there is a well-defined superconformal R-symmetry such that, up to decoupling along the RG flow, all chiral operators satisfy the unitarity bound and at least one chiral operator does not decouple, then we expect the gauge theory associated with $(G, \bm{R})$ to flow to an interacting SCFT in the infrared. The same set of conditions leads to the conformal window of $SU(N_c)$ SQCD with $N_f$ chiral multiplets in the fundamental representation:
\begin{equation}
    \frac{3}{2} N_c < N_f < 3N_c \,.
\end{equation}
Therefore, when a gauge theory associated with the pair $(G, \bm{R})$ satisfies these conditions, we say that it lies within the conformal window. For a pair $(G, \bm{R})$ that admits a limit where the rank of the gauge group is taken to be arbitrarily large, the conformal window has been worked out for all possible $\bm{R}$ already in \cite{Agarwal:2020pol}.

\begin{table}[p]
  \centering
  \begin{threeparttable}
    \begin{tabular}{cccc}
      \toprule
      \multirow{1}{*}{$G$} & \multirow{1}{*}{$\bm{R}$} & \multirow{1}{*}{$\Gamma$} & \multirow{1}{*}{Conformal} 
      \\
       \midrule
      \multirow{1}{*}{$G$} & $n_a\, \textbf{adj}$ & \multirow{1}{*}{$Z(G)$} & $n_a = 2, 3_C$ 
      \\\midrule
      \multirow{7}{*}{$SU(2N)$} & $n_S(\bm{S^2} \oplus \bm{\overline{S^2}})$ & \multirow{7}{*}{$\mathbb{Z}_2$} & $n_S =1, 2$ 
      \\
       & $n_A(\bm{A^2} \oplus \bm{\overline{A^2}})$ & & $n_A = 2,3$ 
       \\
       & $\bm{S^2} \oplus \bm{\overline{S^2}} \oplus n_A(\bm{A^2} \oplus \bm{\overline{A^2}})$ & & $n_A =1, 2$ 
       \\
       & $\textbf{adj} \oplus \bm{S^2} \oplus \bm{\overline{S^2}}$ & & \checkmark 
       \\ 
       & $\textbf{adj} \oplus n_A(\bm{A^2} \oplus \bm{\overline{A^2}})$ & & $n_A =1, 2$ 
       \\
       & $\textbf{adj} \oplus \bm{S^2} \oplus \bm{\overline{S^2}} \oplus \bm{A^2} \oplus \bm{\overline{A^2}}$ & & $\checkmark_C$ 
       \\
       & $2\,\textbf{adj} \oplus \bm{A^2} \oplus \bm{\overline{A^2}}$ & & \checkmark 
       \\
      \midrule
       \multirow{2}{*}{$Spin(N)$} & $n_S\,\bm{S^2}$ & \multirow{2}{*}{$Z(G)$} & $n_S =1, 2$ 
       \\
       & $\bm{S^2} \oplus \textbf{adj}$ & & \checkmark 
       \\\midrule
       \multirow{6}{*}{$Spin(N)$} & $\textbf{adj} \oplus n_V \bm{V}$ & \multirow{6}{*}{$\mathbb{Z}_2$} & $ 1\leq n_V\leq 2N-4$ 
       \\
       & $2\textbf{adj} \oplus n_V \bm{V}$ & & $1 \leq n_V \leq N-2$ 
       \\
       & $\bm{S^2} \oplus n_V \bm{V}$ & & $1 \leq n_V\leq 2N-8$ 
       \\
       & $2\bm{S^2} \oplus n_V \bm{V}$ & & $1 \leq n_V\leq N-10$ 
       \\
       & $\textbf{adj} \oplus \bm{S^2} \oplus n_V \bm{V}$ & & $1 \leq n_V\leq N-6$ 
       \\
       & $n_V \bm{V}$ & & $\frac{3}{2}(N-2) < n_V < 3N-6$ 
       \\\midrule
      \multirow{3}{*}{$USp(2N)$} & $n_A \bm{A^2}$ & \multirow{3}{*}{$\mathbb{Z}_2$} & $n_A =2,  3$ 
      \\
       & $\textbf{adj} \oplus n_A \bm{A^2}$ & & $n_A =1, 2$ 
       \\
       & $2\, \textbf{adj} \oplus \bm{A^2}$ & & \checkmark 
       \\
      \bottomrule
    \end{tabular}
  \end{threeparttable}
  \caption{Simple gauge theories with one-form symmetry, $\Gamma$, with arbitrary large gauge rank. 
  $\textbf{S}^2$ and $\textbf{A}^2$ refers to rank-two symmetric and anti-symmetric representation respectively. $\mathbf{V}$ refers to the vector representation. 
  In the conformal column, we write $\checkmark_C$ to indicate the theory has vanishing one-loop $\beta$-function and that a conformal manifold exists; an unadorned $\checkmark$ indicates the gauge coupling is asymptotically free and the theory flows to an interacting SCFT. When the conformal column contains a range, a subscript $C$ on the upper bound again indicates vanishing one-loop $\beta$-function and nontrivial conformal manifold when the bound is saturated. 
  When we write $n_{\bm{r}}$ inside of $\bm{R}$, we assume that $n_{\bm{r}} \geq 1$; this prevents subtleties with enhanced one-form symmetries. In each case, the mixed axial-electric anomaly can be obtained via consultation with Table \ref{tbl:mixedanom}.
  }
  \label{tbl:genericN}
\end{table}

\begin{table}[p]
  \centering
  \begin{threeparttable}
    \begin{tabular}{cccccc}
      \toprule
      \multirow{2}{*}{$G$} & \multirow{2}{*}{$\bm{R}$} & \multirow{2}{*}{$\Gamma$} & \multirow{2}{*}{Conformal} 
      & \multicolumn{2}{c}{Mixed Anomaly} \\
       & & & 
       & Spin & Non-Spin \\\midrule
      \multirow{2}{*}{$SU(4)$} & $\bm{20'}$ & \multirow{2}{*}{$\mathbb{Z}_4$} & \checkmark 
      & \multirow{2}{*}{\checkmark} & \multirow{2}{*}{\checkmark} \\
       & $\bm{20'}\oplus \bm{15}$ & & $\checkmark_C$ 
       \\\midrule
       \multirow{6}{*}{$SU(4)$} & $\bm{20'}\oplus n_{\bm{6}}\,\bm{6}$ & \multirow{6}{*}{$\mathbb{Z}_2$} & $0\leq n_{\bm{6}} \leq 4_C$ 
       & \multirow{6}{*}{\ding{55}} & \multirow{6}{*}{\checkmark} \\
       & $\bm{10}\oplus\bm{\overline{10}}\oplus n_{\bm{6}}\,\bm{6}$ & & $0\leq n_{\bm{6}} \leq 5$ 
       \\
       & $\bm{15}\oplus\bm{10}\oplus\bm{\overline{10}}\oplus n_{\bm{6}}\,\bm{6}$ & & $0\leq n_{\bm{6}} \leq 2_C$ 
       \\
       & $n_{\bm{6}}\,\bm{6}$ & & $6< n_{\bm{6}} \leq 11$ 
       \\
       & $\bm{15}\oplus n_{\bm{6}}\,\bm{6}$ & & $1\leq n_{\bm{6}} \leq 8_C$ 
       \\
       & $2\,\bm{15}\oplus n_{\bm{6}}\,\bm{6}$ & & $0\leq n_{\bm{6}} \leq 4_C$ 
       \\\midrule
      \multirow{3}{*}{$SU(6)$} & $n_{\bm{20}} \, \bm{20}$ & \multirow{3}{*}{$\mathbb{Z}_3$} & $3\leq n_{\bm{20}} \leq 5$ 
      & \multirow{3}{*}{\checkmark} & \multirow{3}{*}{\checkmark} \\
       & $\textbf{35} \oplus n_{\bm{20}} \, \bm{20}$ &  & $1\leq n_{\bm{20}} \leq 4_C$ 
       \\
       & $2\, \textbf{35} \oplus n_{\bm{20}} \, \bm{20}$ & & $0\leq n_{\bm{20}} \leq 2_C$ 
       \\\midrule
      \multirow{3}{*}{$SU(6)$} &  $\bm{\overline{21}}\oplus 5\,\bm{15}$ & \multirow{3}{*}{$\mathbb{Z}_2$} & $\checkmark$ 
      & \multirow{3}{*}{\checkmark} & \multirow{3}{*}{\checkmark} \\
       & $4\,(\bm{15}\oplus\bm{\overline{15}})$ & & \checkmark 
       \\
       & $\bm{35}\oplus 3\,(\bm{15}\oplus\bm{\overline{15}})$ & & $\checkmark_C$ 
       \\
       \midrule
      \multirow{2}{*}{$SU(8)$} & $n_{\bm{70}} \, \bm{70}$ & \multirow{2}{*}{$\mathbb{Z}_4$} & $1\leq n_{\bm{70}} \leq 2$ 
      & \multirow{2}{*}{\checkmark} & \multirow{2}{*}{\checkmark} \\
       & $\bm{63} \oplus \bm{70}$ & & \checkmark 
       \\\midrule
      \multirow{5}{*}{$SU(8)$} & $\bm{70} \oplus \bm{36} \oplus \bm{\overline{36}}$ & \multirow{5}{*}{$\mathbb{Z}_2$} & \checkmark 
      & \multirow{5}{*}{\ding{55}} & \multirow{5}{*}{\ding{55}} \\
       & $\bm{70} \oplus n_{\bm{28}}(\bm{28} \oplus \bm{\overline{28}})$ & & $0 \leq n_{\bm{28}} \leq 2$ 
       \\
       & $\bm{63}\oplus\bm{36}\oplus 3\,\bm{\overline{28}}$ & & \checkmark 
       \\
       & $\bm{36}\oplus 3\,\bm{\overline{28}}$ & & \checkmark 
       \\
       & $\bm{36}\oplus\bm{28}\oplus 4\,\bm{\overline{28}}$ & & \checkmark 
       \\\midrule
       $SU(9)$ & $\bm{84} \oplus \bm{\overline{84}}$ & $\mathbb{Z}_{3}$ & \checkmark 
       & \ding{55} & \ding{55} \\\midrule
       \multirow{5}{*}{$SU(12)$} & $2\,\bm{\overline{78}}\oplus 4\,\bm{66}$ & \multirow{5}{*}{$\mathbb{Z}_2$} & \checkmark 
       & \multirow{5}{*}{\ding{55}} & \multirow{5}{*}{\checkmark} \\
       & $2\,\bm{\overline{78}}\oplus\bm{78}\oplus 2\,\bm{66}$ & & \checkmark 
       \\
       & $\bm{143}\oplus\bm{78}\oplus 2\,\bm{\overline{66}}$ & & \checkmark 
       \\
       & $\bm{78}\oplus 2\,\bm{\overline{66}}$ & & \checkmark 
       \\
       & $\bm{78}\oplus\bm{66}\oplus 3\,\bm{\overline{66}}$ & & \checkmark 
       \\\bottomrule
    \end{tabular}
  \end{threeparttable}
  \caption{$SU(N)$ theories with higher-rank tensors and one-form symmetry. In cases where $\overline{\bm{R}} \neq \bm{R}$, we do not write both choices. The notation is explained in the caption of Table \ref{tbl:genericN}.}
  \label{tbl:N1SU}
\end{table}

\begin{table}[p]
  \centering
  \begin{threeparttable}
    \begin{tabular}{cccccc}
      \toprule
      \multirow{2}{*}{$G$} & \multirow{2}{*}{$\bm{R}$} & \multirow{2}{*}{$\Gamma$} & \multirow{2}{*}{Conformal} 
      & \multicolumn{2}{c}{Mixed Anomaly} \\
       & & & 
       & Spin & Non-Spin \\\midrule
      \multirow{3}{*}{$Spin(7)$} & $\bm{35}$ & \multirow{3}{*}{$\mathbb{Z}_2$} & \checkmark 
      & \multirow{3}{*}{\ding{55}} & \multirow{3}{*}{\checkmark}  \\
       & $\bm{35} \oplus \bm{21}$ &  & $\checkmark_C$ 
       \\
       & $\bm{35} \oplus n_{\bm{7}}\, \bm{7}$ &  & $0\leq n_{\bm{7}} \leq 4$ 
       \\\midrule
      \multirow{6}{*}{$Spin(8)$} & $\bm{56_{v,s,c}}$ & \multirow{6}{*}{$\mathbb{Z}_2^{L,R,D}$} & \checkmark 
      & \multirow{6}{*}{\ding{55}} & \multirow{6}{*}{\checkmark} \\
       & $\bm{56_{v,s,c}} \oplus n_{v,s,c}\, \bm{8_{v,s,c}}$ & & $0\leq n_{v,s,c}  \leq 2$ 
       \\
       & $\bm{35_{v,s,c}} \oplus \bm{28}$ & & \checkmark 
       \\
       & $\bm{35_{v,s,c}} \oplus \bm{28} \oplus n_{v,s,c}\, \bm{8_{v,s,c}}$ & & $0\leq n_{v,s,c} \leq 2_C$ 
       \\
       & $\bm{28} \oplus n_{s,c}\, \bm{8_{v,s,c}}$ & & $1\leq n_{v,s,c} \leq 12_C$ 
       \\
       & $n_{v,s,c}\, \bm{8_{v,s,c}}$ & & $9<n_{s,c} \leq 17$ 
       \\
       \midrule
       \multirow{5}{*}{$Spin(12)$} & $\bm{77} \oplus \bm{66} \oplus \bm{32_{s,c}}$ & \multirow{5}{*}{$\mathbb{Z}_2^{L,R}$} & \checkmark 
       & \multirow{5}{*}{\checkmark} & \multirow{5}{*}{\checkmark} \\
        & $2\,\bm{66} \oplus n_{s,c}\,\bm{32_{s,c}}$ & & $0\leq n_{s,c} \leq 2$ 
        \\
        & $\bm{77} \oplus n_{s,c}\,\bm{32_{s,c}}$ & & $0\leq n_{s,c} \leq 3$ 
        \\
        & $\bm{66} \oplus n_{s,c}\,\bm{32_{s,c}}$ & & $1\leq n_{s,c} \leq 5_C$ 
        \\
        & $n_{s,c}\,\bm{32_{s,c}}$ & & $3\leq n_{s,c} \leq 7$ 
        \\\midrule
        \multirow{3}{*}{$Spin(16)$} & $n_{s,c} \bm{128_{s,c}}$ & \multirow{3}{*}{$\mathbb{Z}_2^{L,R}$} & $1 \leq n_{s,c} \leq 2$ 
        & \multirow{3}{*}{\ding{55}} & \multirow{3}{*}{\ding{55}} \\
         & $\bm{128_{s,c}} \oplus \bm{135}$ & & \checkmark 
         \\
         & $\bm{128_{s,c}} \oplus \bm{120}$ & & \checkmark 
         \\\midrule
      \multirow{2}{*}{$USp(4)$} & $\bm{14}$ & \multirow{2}{*}{$\mathbb{Z}_2$} & \checkmark 
      & \multirow{2}{*}{\ding{55}} & \multirow{2}{*}{\checkmark} \\
       & $\bm{14} \oplus n_{\bm{5}} \,\bm{5}$ & & $0\leq n_{\bm{5}} \leq 2_C$ 
       \\ \midrule
      \multirow{3}{*}{$USp(8)$} & $n_{\bm{42}} \, \bm{42}$ & \multirow{3}{*}{$\mathbb{Z}_2$} & $1\leq n_{\bm{42}}\leq 2$ 
      & \multirow{3}{*}{\ding{55}} & \multirow{3}{*}{\ding{55}} \\
       & $\bm{42} \oplus \bm{36}$ & & \checkmark 
       \\
       & $\bm{42} \oplus n_{\bm{27}}\,\bm{27}$ & & $0\leq n_{\bm{27}} \leq 2$ 
       \\
      \bottomrule
    \end{tabular}
  \end{threeparttable}
  \caption{Simple gauge theories with gauge group $Spin(N)$ or $USp(2N)$ and higher-rank tensors that have a one-form symmetry. In the second row of the $Spin(8)$ block, the $\bm{56}$ and the $\bm{8}$ must be both have the same $v,s,c$ subscript. In the other rows, the $v,s,c$ of different representations are uncorrelated.  The notation is explained in the caption of Table \ref{tbl:genericN}.
  }\label{tbl:N1SpinUSp}
\end{table}

To highlight this procedure, we now explore an example in detail. Consider an $SU(8)$ gauge theory with one chiral multiplet, $Q$, in the $\bm{70}$ of $SU(8)$; since $T(\bm{70}) = 10$, it is clear that the gauge coupling is asymptotically-free following equation \eqref{eqn:oneloop}. In this case, the superconformal R-symmetry is fixed by the anomaly-free condition; we find that the R-charge of the chiral multiplet is 
\begin{equation}
    R[Q] = \frac{1}{5} \,.
\end{equation}
We can see that there is a (gauge-invariant) chiral operator $Q^2$
which has R-charge
\begin{equation}
    R[Q^2] = \frac{2}{5} \leq \frac{2}{3} \,,
\end{equation}
and thus which lies below the unitarity bound. Therefore, we introduce the gauge-singlet flip field $M_{Q^2}$ and add this to our UV gauge theory, together with the superpotential coupling
\begin{equation}
    W = M_{Q^2} Q^2 \,.
\end{equation}
In the infrared, this will give rise to the same theory as our original QFT, except that the $Q^2$ operator will have been decoupled. In this process, we have introduced a new $U(1)$ global symmetry which rotates the gauge-singlet; a priori, we should consider how this $U(1)$ mixes with the superconformal R-symmetry, though in this case it is again fixed by the anomaly-free condition. We find
\begin{equation}
    R[Q] = \frac{1}{5} \,, \qquad R[M_Q^2] = \frac{8}{5} \,.
\end{equation}
Enumerating the chiral operators, we find that the operator of the minimal dimension is $Q^6$ with 
\begin{equation}
    R[Q^6] = \frac{6}{5} > \frac{2}{3} \,.
\end{equation}
Therefore, there are no more unitarity-violating operators, and there exists at least one chiral operator that does not saturate the unitarity bound. Thus, $SU(8)$ gauge theory with one chiral multiplet in the $\bm{70}$ lies within the conformal window.

In the spirit of Question \ref{ques:first}, we have listed the consistent theories associated with $(G, \bm{R})$, where $G$ is simple and simply-connected, which lie within the conformal window, and have a nontrivial one-form symmetry in Tables \ref{tbl:genericN}, \ref{tbl:N1SU}, and \ref{tbl:N1SpinUSp}.

\section{\texorpdfstring{$\mathcal{N}=1$}{N=1} theories with zero-form/one-form anomaly}\label{sec:mixedanom}

By classifying the 4d $\mathcal{N}=1$ gauge theories that have a simple and simply-connected gauge group $G$ and have a nontrivial one-form symmetry, we successfully answers Question \ref{ques:first} in Section \ref{sec:2}. This naturally brings rise a follow up question if such theories can also possess a mixed anomaly between this one-form symmetry and the axial zero-form symmetry, which refer to as a \emph{mixed axial-electric anomaly}. The focus of this section can succinctly be captured via the following question.

\begin{ques}{ques:second}
\renewcommand{\thempfootnote}{\arabic{mpfootnote}}
What are the 4d $\mathcal{N}=1$ Lagrangian theories with a simply-connected simple gauge group (and trivial superpotential) that
satisfy Question \ref{ques:first} and have a nontrivial one-form and zero-form mixed anomaly?
\end{ques}

We note that we are particularly interested in the existence of a nontrivial mixed anomaly on backgrounds with spin structure. Going beyond, we further analyze whether there are mixed anomalies on backgrounds without spin structure.

A part of the classical global symmetry of the SQFT associated with $G$ and $\bm{R}$ is simply symmetries of the Lagrangian, which rotates the chiral multiplets and the gaugino, such that
\begin{align} \label{eqn:classflav}
    \prod_i SU(n_i) \times  \prod_i U(1)_i \times U(1)_R \,.
\end{align}
Under this symmetry, the fermions inside the chiral multiplets $Q_i$ transform in the fundamental representation of $SU(n_i)$ and are charge $1$ under $U(1)_i$. Similarly, the gaugino is charge $1$ under $U(1)_R$. 

The classical flavor symmetry in equation \eqref{eqn:classflav} suffers from an Adler--Bell--Jackiw (ABJ) anomaly \cite{Adler:1969gk,Bell:1969ts}. Let $\lambda$ denote the gaugino and $\psi_i$ the chiral fermion inside of $Q_i$; let $\alpha$ and $\beta_i$ be the parameters for the $U(1)_R$ and $U(1)_i$ transformations, respectively. Then,
\begin{equation}\label{eqn:U1transforms}
    U(1)_R \, : \quad \lambda \,\, \rightarrow e^{i \alpha} \lambda \,, \qquad\qquad U(1)_i \, : \quad \psi_i \,\, \rightarrow e^{i \beta_i} \psi_i \,.
\end{equation}
In a nontrivial instanton background, the relevant measure in the path integral transforms as \cite{Fujikawa:1979ay}
\begin{align} \label{eqn:ABJ}
    [D\lambda]\prod_i [D\psi_i] \rightarrow  \operatorname{exp}\left[ \left(2h_G^\vee \alpha + \sum_i 2T(\bm{R}_i) n_i \beta_i\right) \frac{i}{16\pi^2}\int F_{\mu\nu}\widetilde{F}^{\mu\nu} \right] [D\lambda]\prod_i [D\psi_i]
\end{align}
under the $U(1)$ transformations in equation \eqref{eqn:U1transforms}. Hence, a generic global symmetry transformation specified by $(\alpha, \beta_i)$ leads to an operator-valued shift of the partition function, and thus results in an inconsistent theory. To ensure for a consistent theory, We are only allowed with the combinations of $(\alpha, \beta_i)$ satisfying
\begin{equation}\label{eqn:anomaly}
    2h_G^\vee \alpha + \sum_i 2T(\bm{R}_i) n_i \beta_i \in 2\pi \mathbb{Z} \,.
\end{equation}
In particular, we can see that this constraint breaks a linear combination of the classical $U(1)$ axial symmetries (the diagonal part of the ${U(1)_i}$) in equation \eqref{eqn:classflav} to
\begin{equation}
    \mathbb{Z}_{2L} \,, \qquad \text{ where } \qquad L = \sum_{i=1}^s n_i T(\bm{R}_i) \,.
\end{equation}
This is the discrete axial symmetry of the quantum theory. This axial symmetry acts simultaneously on the chiral fermions inside of the chiral multiplets as
\begin{equation}\label{eqn:36}
    \mathbb{Z}_{2L} \, : \quad \psi_i \,\, \rightarrow e^{\frac{\pi i \ell}{L}} \psi_i \,,
\end{equation}
for $\ell = 0, \cdots, 2L - 1$ belonging to $\mathbb{Z}_{2L}$.

One has to be a bit careful before determining this. While it is true that sometimes the action of a subgroup of $\mathbb{Z}_{2L}$ on the fermions is distinct from the action of a subgroup of the center of the gauge group, we have to carefully check such is the case here as this is not always guaranteed. In this latter case, the axial symmetry is reduced from the naive $\mathbb{Z}_{2L}$ written here; however, the analysis of the mixed anomaly in the section holds as long as the entirety of the $\mathbb{Z}_{2L}$ does not lie within the center of the gauge group. To highlight this effect, consider a $G = SU(N)$ gauge group and a chiral multiplet in an irreducible representation $\bm{R}$ of $G$. The global and gauge symmetries cause the fermion to be rotated, respectively, as
\begin{equation}
    \mathbb{Z}_{2T(\bm{R})} \,:\, \psi \rightarrow \exp\left(\frac{\pi i \ell}{T(\bm{R})}\right) \psi \,, \qquad \qquad \mathbb{Z}_N \, : \, \psi \rightarrow \exp\left(\frac{2\pi i m n}{N}\right) \psi \,,
\end{equation}
where $\ell$ is an integer between $0$ and $2T(\bm{R}) - 1$ parametrizing the global transformation, $n$ an integer between $0$ and $N - 1$ parametrizing the gauge transformation, and $m$ is the n-ality of the representation $\bm{R}$ as described around equation \eqref{eqn:nality}. If there exists a pair of $n$ and $\ell$ such that
\begin{equation}\label{eqn:careful}
    \exp\left(\frac{\pi i \ell}{T(\bm{R})}\right) = \exp\left(\frac{2\pi i m n}{N}\right) \,,
\end{equation}
then the subgroup of $\mathbb{Z}_{2L}$ generated by the element $\ell$ is, in fact, summed over in the path integral, and thus does not form a global symmetry of the theory. The generalization to multiple fermions in different representations is straightforward.

Having determined the axial symmetry, the mixed axial-electric anomaly can be captured as follows \cite{Gaiotto:2014kfa}. Suppose that we turn on a background two-form field, $B$, for the one-form symmetry; this shifts the instanton number by a fractional quantity proportional to the integral of the Pontryagin square of $B$: $\mathcal{P}(B)$.\footnote{See the appendix of \cite{Kapustin:2013qsa} for a review of the definition/properties of the Pontryagin square operation; for our purposes it is important to note that $\int \mathcal{P}(B)$ is an integer on a generic manifold, and an even integer on a spin manifold. Considering the $\mathbb{Z}_M$-valued background field $B$ as an element of $H^2(X_4, \IZ_M)$, then the Pontryagin square operation is a map $\mathcal{P}: H^2(X_4, \IZ_M) \to H^2(X_4, \IZ_{M \gcd(2, M)})$. When $M$ is odd then $\mathcal{P}(B)$ is simply $B \cup B$, whereas when $M$ is even $\mathcal{P}(B)$ reduces to $B \cup B$ modulo $N$. See the discussion in \cite{Aharony:2013hda} for more details on the Pontryagin square.} As we have seen in equation \eqref{eqn:ABJ}, an axial symmetry transformation modifies the partition function by a phase proportional to $\int F_{\mu\nu}\widetilde{F}^{\mu\nu}$; under the shift incorporating $\mathcal{P}(B)$, this phase may not be trivial under a $\mathbb{Z}_{2L}$ transformation. This is not an operator-valued shift as $B$ is a background field, however, if we were to attempt to gauge the one-form symmetry by path-integrating over all possible $B$, there would arise an inconsistency. As an anomaly is nothing other than an obstruction to gauging; this transformation of the partition function captures a mixed anomaly between the axial symmetry and the one-form symmetry.

We first consider the case where $G = SU(N)$ and the chiral multiplets in the representation $\bm{R}$ lead to a discrete axial symmetry $\mathbb{Z}_{2L}$. We further assume there exists a one-form symmetry which is $\mathbb{Z}_p$. Turning on a background two-form field $B$ for the one-form symmetry, the axial transformation, parametrized by $\ell \in \mathbb{Z}_{2L}$, transforms the partition function as
\begin{align}
    Z[B] \,\to\, \exp \left( i \alpha (2L) \nu_{\textrm{inst}} \right) Z[B] \ , 
\end{align}
where the transformation is parametrized by $\alpha = \frac{2\pi \ell}{2L}$, for $\ell \in \mathbb{Z}$, and 
\begin{equation}
    \nu_\text{inst} = \frac{1}{8\pi^2} \int F \wedge F \,,
\end{equation}
is the instanton number. It is well-known that when $G$ is a non-simply-connected Lie group, then a $G$-bundle does not necessarily have an integer instanton number \cite{Witten:2000nv,Aharony:2013hda}; similarly, in the presence of background two-form fields for the one-form symmetry, the instanton number can be non-zero modulo one. As it is important for the analysis in this paper, we briefly review the derivation of the fractional part of the instanton number, following the recent discussions in \cite{Gaiotto:2017yup, Argurio:2023lwl} and references therein.

The discrete 2-form $\IZ_p$ gauge field can be realized by a pair of 2-form and one-form $U(1)$ fields $(B,C)$ whose gauge transformation is given by
\begin{align}
    \begin{split}
    B&\rightarrow B+d\lambda\,,\\
    C&\rightarrow C+df+p\lambda\,,
    \end{split}
\end{align}
and satisfying the relation $pB=dC$ in order to ensure that $e^{i\oint_\Sigma B}$ is a $p$th root of unity on the two-cycle $\Sigma$. Then we extend the $SU(N)$ gauge field strength $F$ to a $U(N)$ gauge field strength $F'$ by
\begin{align}
    F'=F+B\,\mathbb{I}_N\,.
\end{align}
Here $\oint_\Sigma B\in\frac{2\pi}{p}\IZ$ so that the diagonal part of $F'$ has a proper algebraic value of $\IZ_p$. Now, to ensure the original $F$ being traceless, we add a Lagrange multiplier term 
\begin{align}
    \frac{1}{2\pi}u\wedge\left(\tr F'-\frac{N}{p}dC\right)\,,
\end{align}
so that $\tr F=\tr F'-NB=0$, using $pB=dC$. When the background $B$-field is turned on, the instanton number is
\begin{align}
\begin{split}
    \nu=&\frac{1}{8\pi^2}\int\tr F\wedge F=\frac{1}{8\pi^2}\int\tr\left(F'-B\,\mathbb{I}_N\right)\wedge\left(F'-B\,\mathbb{I}_N\right)\\
    =&\frac{1}{8\pi^2}\int \tr F'\wedge F'-\frac{N}{8\pi^2}\int B\wedge B\,,
    \end{split}
\end{align}
where we used the constraint $\tr F'=NB$. The first term can be further decomposed with the second Chern class $c_2=\frac{1}{8\pi^2}(\tr F'\wedge F'-\tr F'\wedge\tr F')$ as
\begin{align}
    \nu=\int c_2+\frac{1}{8\pi^2}\int \tr F'\wedge \tr F'-\frac{N}{8\pi^2}\int B\wedge B\in \IZ+\frac{N(N-1)}{8\pi^2}\int B\wedge B \,.
\end{align}
We can rescale $B=\frac{2\pi}{p}\hat{B}$ so that $\hat{B}$ is  $\oint_\Sigma\hat{B}\in \IZ$. Then the instanton number can be rewritten as
\begin{align}
    \nu\in\IZ+\frac{N(N-1)}{2p^2}\int\hat{B}\wedge\hat{B}=\IZ+\frac{q'}{p'}\int \hat{B}\wedge\hat{B}\,,
\end{align}
where $\int \hat{B}\wedge\hat{B}\in\IZ$ and the two integers $p'$ and $q'$, given by
\begin{align}\label{eqn:reduced}
    p'=2p/\gcd(2p,(N-1)N/p)\,,\quad q'=(N-1)N/p/\gcd(2p,(N-1)N/p) \,,
\end{align}
are the coprime numerator/denominator of the coefficient.
Here we note that even though the background field $\hat{B}$ is $\IZ_{p}$-valued, the instanton number, which is the measure of the anomalous phase in our case, is $\IZ_{p'}$-valued because of the numerator of the coefficient. In the extreme case when $p=N$, the anomalous phase is $\IZ_{2p}$ valued for even $N$ while it is still $\IZ_p$ valued for odd $N$. Therefore, the integrand can be safely written as $\hat{B}\cup \hat{B}$ when $N$ is odd, while the integrand should be lifted to an element of $H^4(M,\IZ_{2p})$ by the Pontryagin square when $N$ is even. Since $p$ divides $N$, we can express the instanton number similarly by embedding the $\IZ_p\subset \IZ_N$.

Therefore, in the presence of a nontrivial background $\mathbb{Z}_p$-valued two-form field $B$, the transformation of the partition function under an axial rotation can be written as
\begin{equation}\label{eqn:mixmix}
     Z[B] \,\to\, \exp\left( 2 \pi i \ell \,  \frac{N(N-1)}{2p^2} \int_X \mathcal{P}(B) \right) Z[B] \,,
\end{equation}
where $X$ is the spacetime four-manifold and $\ell$ is an arbitrary integer. From a modern perspective, a powerful technique to capture such anomalies is to rephrase them in terms of a classical gauge-invariant Lagrangian in $(d+1)$-dimensions. Let $\omega(A, B)$ be this Lagrangian, with $A$ the gauge field for the axial zero-form symmetry and $B$ the gauge field for the one-form symmetry. We let $X$ be the boundary of the $(d+1)$-dimensional spacetime, $Y$, and assume that all background fields on $X$ can be extended into gauge fields in the bulk. Such a $(d+1)$-dimensional Lagrangian captures the anomalous phase in equation \eqref{eqn:mixmix} if, under a gauge transformation with parameter $\alpha$, $A \,\, \rightarrow \,\, A^\alpha$, we have that 
\begin{equation}
    \exp\left( 2\pi i \int_Y (\omega(A^\alpha, B) - \omega(A, B))  \right) = \exp\left( \int_X \beta(\alpha, B) \right) \,,
\end{equation}
where $\beta(\alpha, B)$ is the phase appearing in equation \eqref{eqn:mixmix}. We write the axial transformation locally as $A \,\,\rightarrow\,\, A + d\lambda$ where $\lambda$ on the boundary $X$ is $\ell$, with $\ell = 0, \cdots, 2L-1$ parametrizing the $\mathbb{Z}_{2L}$ transformation. Then, it is straightforward to see that the $\omega(A, B)$ that encodes the mixed axial-electric anomaly is:
\begin{equation}\label{eqn:anomTFT}
    \omega(A, B) = \frac{N(N-1)}{2p^2} A \cup \mathcal{P}(B) \,.
\end{equation}
We note that the TQFT encoding the anomalous phase is not unique; in particular, a nontrivial anomaly (on a generic non-spin manifold) exists between the $\mathbb{Z}_{p'}$ subgroup of the $\mathbb{Z}_p$ one-form symmetry, and the subgroup $\mathbb{Z}_{L'}$ of $\mathbb{Z}_{2L}$, generated by elements $\ell$, for which the denominator of 
\begin{equation}
    \frac{N(N-1)\ell}{2p^2} \,,
\end{equation}
as a reduced fraction is not one. Alternatively, the anomaly TFT can be written in a reduced form in terms of $\mathbb{Z}_{p'}$ and $\mathbb{Z}_{L'}$ valued background fields. The denominator, $p'$, of the fractional instanton number was written in equation \eqref{eqn:reduced} as 
\begin{equation}
    \frac{1}{p'} = \frac{\gcd(2p, (N-1)\frac{N}{p})}{2p} \,,
\end{equation}
however $p'$ itself can be rational, and thus the fractional instantons (or equivalently the coefficient of the anomaly TFT) depend on the numerator of $p'$ as a reduced fraction. We can simplify the expression as follows
\begin{equation}
    p' = \begin{cases}
        \displaystyle\frac{\gcd(2,p)p}{\gcd(2p, N/p)}&\text{ if $p$ even and $N/p$ even,}\\[1em]
        \displaystyle\frac{\gcd(2,p)p}{\gcd(p, N/p)}&\text{ otherwise.}
    \end{cases}
\end{equation}
In this way, we can clearly see precisely which subgroup of the unscreened $\mathbb{Z}_p \subseteq \mathbb{Z}_N$ one-form symmetry captures the anomalous behavior; and thus write the anomaly TFT in terms of the minimally-normalized background fields.

As we can see, the phase in equation \eqref{eqn:mixmix} can be nontrivial, and such a nontrivial phase constitutes a mixed anomaly between the axial zero-form symmetry and the $\mathbb{Z}_p$ one-form symmetry. Thus, the absence of a mixed anomaly requires that 
\begin{equation}
    \frac{N(N-1)}{2p^2} \int_X \mathcal{P}(B) = 0 \,\, \operatorname{mod} \,\, 1 \,,
\end{equation}
for $X$ a generic four-manifold. On a generic four-manifold $\int_X \mathcal{P}(B)$ is an integer, and thus there is no mixed axial-electric anomaly when
\begin{equation}\label{eqn:nscond}
    \frac{N(N-1)}{2p^2} \in \mathbb{Z} \,.
\end{equation}
If $X$ is a four-manifold that admits a spin structure, then $\int_X \mathcal{P}(B)$ is, in fact, an even integer; therefore, if we consider our quantum field theory as only defined on backgrounds with spin structure, there is no mixed axial-electric anomaly whenever 
\begin{equation}\label{eqn:scond}
    \frac{N(N-1)}{p^2} \in \mathbb{Z} \,.
\end{equation}
Since Question \ref{ques:first} requires that we study theories with a nontrivial mixed anomaly on  generic four-manifolds (either with or without spin-structure), we require that the either equation \eqref{eqn:scond} or equation \eqref{eqn:nscond}, respectively, is satisfied.

A priori, there may be a local counterterm that we can add to the action which would cancel the putative anomaly in equation \eqref{eqn:mixmix}. The most general relevant local counterterm modifies the partition function as \cite{Kapustin:2014gua}
\begin{equation}
    Z[B] \quad\rightarrow\quad Z[B] \exp\left( \frac{2\pi i q}{2N} \int \mathcal{P}(B) \right) \,,
\end{equation}
where $q$ is some integer between $0$ and $2N - 1$. The axial transformation does not modify the counterterm, and thus the anomaly could only be canceled by the addition of the local counterterm if there exists a $q$ such that
\begin{equation}
    \frac{q}{2N} = \frac{q}{2N} + \frac{N(N-1)}{2p^2} \quad \mod  1 \,.
    \label{eq:axialelectricanomalycondition}
\end{equation}
Thus, any mixed axial-electric anomaly cannot be cancelled by a local counterterm.

We now turn to the $G = USp(2N)$ case.\footnote{Fractional instanton numbers of gauge theories in simple non-simply-connected Lie groups that are not $SU(N)$ can be determined following \cite{Witten:2000nv}. The basic idea is to decompose the $Spin$ and $USp$ groups into smaller subgroups including $SU(N)$ for which we know how to compute the fractionalized instanton numbers. We do not repeat the derivation from \cite{Witten:2000nv} here.} If there is a nontrivial one-form symmetry, then it is necessarily $\Gamma = \mathbb{Z}_2$; turning on a background two-form field, $B$, for this one-form symmetry and performing a $\mathbb{Z}_{2L}$ axial transformation, parametrized by $\ell$, leads to the following transformation of the partition function:
\begin{equation}
    Z[B] \to \exp\left( 2 \pi i \ell \frac{N}{4} \int \mathcal{P}(B) \right) \, Z[B] \,.
\end{equation}
There is a nontrivial axial-electric anomaly on a spin-manifold if $N$ is not even and on a non-spin-manifold if $N$ is not a multiple of $4$.
When $G = Spin(N)$ with $N$ odd, the one-form symmetry is again necessarily $\mathbb{Z}_2$. Turning on the background two-form field and performing the axial transformation, the partition function changes by the following phase:
\begin{equation}
    Z[B] \to \exp\left( 2 \pi i \ell \frac{1}{2} \int \mathcal{P}(B) \right) \, Z[B] \,.
\end{equation}
We can see that there is never a mixed anomaly on spin-manifolds, but there is always a mixed anomaly on non-spin-manifolds.

\begin{table}[t]
    \centering
    \begin{threeparttable}   
        \begin{tabular}{cccc}
            \toprule
            \multirow{2}{*}{$G$} & \multirow{2}{*}{$\Gamma$} & \multicolumn{2}{c}{Mixed Anomaly} \\ & & Spin & Non-Spin \\\midrule
            $SU(pM)$ & $\mathbb{Z}_p$ & $\begin{gathered}\checkmark \text{ if } p \nmid M \\ \text{\ding{55}} \text{ if } p \mid M \end{gathered}$ & $\begin{aligned}&\text{\ding{55}} \text{ if } \left(M^2 - \frac{M}{p}\right) \in 2\mathbb{Z}  \\  &\checkmark \text{ otherwise } \end{aligned}$\\\midrule
            $USp(2N)$ & $\mathbb{Z}_2$ & $\begin{gathered}\checkmark \text{ if } 2 \nmid N \\ \text{\ding{55}} \text{ if } 2 \mid N \end{gathered}$ & $\begin{gathered}\checkmark \text{ if } 4 \nmid N \\ \text{\ding{55}} \text{ if } 4 \mid N \end{gathered}$\\\midrule
            $Spin(2M + 1)$ & $\mathbb{Z}_2$ & \ding{55} & \checkmark \\\midrule
            \multirow{2}{*}{$Spin(4M + 2)$} & $\mathbb{Z}_2$ & \ding{55} & \checkmark \\
             & $\mathbb{Z}_4$ & \checkmark & \checkmark \\\midrule
            \multirow{4}{*}{$Spin(4M)$} & $\mathbb{Z}_2 \times \mathbb{Z}_2$ & \checkmark & \checkmark \\
             & $\mathbb{Z}_2^{L,R}$ & $\begin{gathered}\checkmark \text{ if } 2 \nmid M \\ \text{\ding{55}} \text{ if } 2 \mid M \end{gathered}$ & $\begin{gathered}\checkmark \text{ if } 4 \nmid M \\ \text{\ding{55}} \text{ if } 4 \mid M \end{gathered}$ \\
             & $\mathbb{Z}_2^D$ & \ding{55} & \checkmark \\
            \bottomrule 
        \end{tabular}
    \end{threeparttable}
    \caption{For a $G$ gauge theory with matter in a representation $\bm{R}$ such that there exists a one-form symmetry $\Gamma$, a mixed axial-electric anomaly exists under the conditions listed here.}
    \label{tbl:mixedanom}
\end{table}

Next, we consider $G = Spin(N)$ where $N = 4M + 2$. We denote the one-form symmetry as $\mathbb{Z}_p$, where $p = 2$ or $p = 4$. The transformation of the partition function under the $\mathbb{Z}_{2L}$ in the presence of a background two-form field for the one-form symmetry is
\begin{equation}\label{eqn:spin4M2mixanom}
    Z[B] \to \exp\left( 2 \pi i \ell \frac{N}{p^2} \int \mathcal{P}(B) \right) \, Z[B] \,.
\end{equation}
Therefore, there is an axial-electric anomaly whenever
\begin{equation}
    \frac{2N}{p^2} \not\in \mathbb{Z} \qquad \text{ or } \qquad \frac{N}{p^2} \not\in \mathbb{Z} \,,
\end{equation}
for spin-manifolds and non-spin-manifolds, respectively.
Finally, we turn to the more intricate case where $G = Spin(N)$ with $N = 4M$. When the full $\mathbb{Z}_2^L \times \mathbb{Z}_2^R$ one-form symmetry is realized, we can turn on background fields $B_L$ and $B_R$ for each factor; then, under an axial transformation, the partition function is rotated by the following phase:
\begin{equation}\label{eqn:spin4Nshift}
    Z[B_L, B_R] \to \exp\left( 2 \pi i \ell \frac{1}{4} \int (M+1)\mathcal{P}(B_L + B_R) - \mathcal{P}(B_L) - \mathcal{P}(B_R) \right) Z[B_L, B_R] \,.
\end{equation}
This anomaly is always nontrivial, regardless of the spin structure. We can consider, without loss of generality, two separate $\mathbb{Z}_2$ subgroups of the center: either of the two factors, or the diagonal. First, we consider one-form symmetry $\mathbb{Z}_2^L$; the change in the partition function becomes
\begin{equation}
    Z[B_L] \to \exp\left( 2 \pi i \ell \frac{1}{4} \int M\mathcal{P}(B_L) \right) Z[B_L] \,.
\end{equation}
That is, there is a nontrivial mixed axial-electric anomaly on a spin manifold if $M$ is not $\in 2\mathbb{Z}$, and on a non-spin manifold if $M$ is not $\in 4\mathbb{Z}$. Now we consider the case where the one-form symmetry is the diagonal $\mathbb{Z}_2^D$, with background field $B_L = B_R = B$; the axial symmetry transforms the partition function as
\begin{equation}
    Z[B] \to \exp\left( 2 \pi i \ell \frac{1}{2} \int \mathcal{P}(B) \right) Z[B] \,.
\end{equation}
On spin manifolds there is never a mixed anomaly, whereas on non-spin manifolds there is always a mixed anomaly.

The presence of a nontrivial mixed axial-electric anomaly does not depend on the value of $L$ for the $\mathbb{Z}_{2L}$ axial symmetry, only that $L \geq 1$ which is necessary for the zero-form symmetry to exist. Since the mixed anomaly does not depend on the specifics of the matter spectrum $\bm{R}$, but only on the pair $(G, \Gamma)$, we have summarized whether there exists a mixed anomaly for each such pair in Table \ref{tbl:mixedanom}. It is straightforward to use Table \ref{tbl:mixedanom} to read off whether or not each entry in Tables \ref{tbl:genericN}, \ref{tbl:N1SU}, and \ref{tbl:N1SpinUSp} has a nontrivial mixed axial-electric anomaly. 

\section{Global zero-form symmetry and anomalies}\label{sec:global0}

We enumerated in Section \ref{sec:2} the superconformal theories with nontrivial one-form symmetry, where each is realized as a fixed point of a simple gauge theory. These theories possess a variety of global symmetry factors, as seen around equations \eqref{eqn:classflav} and \eqref{eqn:36}. However, the transformations of each of these factors are not always independent; there can be transformations of two different factors that act on the degrees of freedom in an identical manner. It is important to identify these redundancies in the global symmetry group, as it is crucial for determining nontrivial anomalies arising from the axial one-form symmetry. In fact, the only discrete subgroup of the flavor group that has a nontrivial anomaly with the one-form symmetry is the axial symmetry. Furthermore, when we consider $U(1)_R$-breaking deformations in Section \ref{sec:deformations}, the global structure of the zero-form symmetry group affects the number of vacua in an spontaneous symmetry breaking phase.  In this section, we will describe a methodology to find the global form of zero-form symmetries for an arbitrary supersymmetric Lagrangian simple gauge theory and describe the corresponding anomalies involving the discrete factors.

\subsection{The global zero-form symmetry} \label{subsec:ABJ}

Looking into the theories listed in Section \ref{sec:2}, we can write the zero-form symmetry of an arbitrary $G$ gauge theory as follows. For a gauge theory with simple gauge group $G$ coupled to chiral multiplets in the representation 
\begin{align}
    n_1\mathbf{R}_1\oplus n_2\mathbf{R}_2\oplus\cdots\oplus n_m\mathbf{R}_m \,,\quad n_i,m\in\mathbb{N} \,,
\end{align}
its zero-form symmetry can be written as
\begin{align}
    U(1)_R\times\prod_{i=1}^{m}U(n_i)=U(1)_R \times\prod_{i=1}^m U(1)_i\times SU(n_i)/\IZ_{n_i} \,,
\end{align}
where each $U(1)_i$ rotates the $i$th chiral multiplet as 
\begin{align}
   U(1)_i: (\phi_i,\psi_i)\mapsto e^{i\theta_i}(\phi_i,\psi_i) \,.
\end{align} 
We then find $(m-1)$ combinations of ABJ anomaly-free $U(1)'_i$ symmetries, whose generators are defined as
\begin{align}
    Q_{i}'=Q_i-\frac{n_iC_2(\mathbf{R}_i)}{n_{i+1}C_2(\mathbf{R}_{i+1})}\, Q_{i+1}\,,\quad i=1\,\cdots\,m-1\,,
\end{align}
where $Q_i$ is the generator of $U(1)_i$ symmetry and $C_2(\mathbf{R}_i)$ is the Dynkin index of $\mathbf{R}_i$ representation. Then the $U(1)'_i$ acts on $i$th and $(i+1)$th chiral fermion fields by
\begin{align}
U(1)'_i: (\psi_i,\psi_{i+1})\mapsto\left( e^{i\theta_i}\,\psi_i\,,\ e^{-\frac{n_iC_2(\mathbf R_i)}{n_{i+1}C_2(\mathbf R_{i+1})}i\theta_i}\,\psi_{i+1}\right)\,,
\end{align}
and the same for the bosonic fields $(\phi_i,\phi_{i+1})$.

Due to the existence of the ABJ anomaly, we know that a combination of $U(1)$ symmetries is broken to a discrete symmetry which acts on the fields of the chiral multiplets as
\begin{align}
    \IZ^A=\IZ_{2\sum_i{n_iC_2(\mathbf{R}_i)}}: (\phi_i,\psi_i)\mapsto e^{\frac{2\pi ik_A}{2\sum_i n_iC_2(\mathbf{R}_i)}} (\phi_i,\psi_i)\,,
\end{align}
which satisfies equation \eqref{eqn:anomaly}. Then, we can define an anomaly-free R-symmetry as
\begin{align}\label{eqn:Raction}
\begin{split}
    R: \begin{cases}
    \displaystyle
    \,\lambda \mapsto e^{i\theta_R}\lambda \,,\\[.5em]
    \displaystyle
    \phi_i \mapsto e^{\left(1-\frac{h_G^\vee}{\sum_in_iC_2(\mathbf{R}_i)}\right)i\theta_R}\, \phi_i\,, \\[.5em]
    \displaystyle
    \psi_i \mapsto e^{\left(-\frac{h_G^\vee}{\sum_in_iC_2(\mathbf{R}_i)}\right)i\theta_R}\, \psi_i\,,
    \end{cases}
    \end{split}
\end{align}
where $\lambda$ is the gaugino. It is important to note that the above R-symmetry is NOT the superconformal R-symmetry at the IR fixed point. If a gauge theory flows to an IR SCFT, the above R-symmetry is dressed with $U(1)_i'$ symmetries to form the superconformal R-symmetry, whose coefficients can be found via $a$-maximization \cite{Intriligator:2003jj}.

Note that the Abelian zero-form symmetries we computed above are not independent of each other. Then it is necessary to eliminate overlapping discrete symmetries to remove any redundancy. Thus we will determine the shared discrete symmetries by reducing to total five separate cases:
\begin{enumerate}[(1)]
    \item $\mathbf{U(1)'_i}$ -- $\mathbf{U(1)'_j}$,
    \item $\mathbf{U(1)'_i}$ -- $\mathbf{U(1)_R}$,
    \item $\mathbf{U(1)_i'}$ -- $\mathbf{\IZ^A}$,
    \item $\mathbf{U(1)_R}$ -- $\mathbf{\IZ^A}$,
    \item $\mathbf{(U(1)_i'\times \IZ^A)-Z(G)}$.
\end{enumerate}
We will visit these cases individually below, and then combine these results to finally generate the global structure of the zero-form symmetry group. In case (5), we are considering the overlap between the Abelian zero-form symmetries and the center of the \emph{gauge} group -- this is important as these transformations must then be integrated over in the path integral; they do not thus form part of the global symmetry.

\paragraph{$\mathbf{U(1)'_i}$ -- $\mathbf{U(1)'_j}$:}
Since each $U(1)'_i$ only on the $i$th and $(i+1)$th chiral multiplets, then $U(1)'_i$ and $U(1)'_j$ can only overlap when $i=j$ or $i=j\pm 1$. Without loss of generality, we assume $j=i+1$. Each $U(1)'_i$ symmetry transforms $(\psi_i,\psi_{i+1},\psi_{i+2})$ as
\begin{align}
\begin{split}
    U(1)'_i: (\psi_i,\psi_{i+1},\psi_{i+2})\mapsto &\left( e^{i\theta_i}\,\psi_i\,,\ e^{-\frac{n_iC_2(\mathbf{R}_i)}{n_{i+1}C_2(\mathbf{R}_{i+1})}i\theta_i}\,\psi_{i+1}\,,\ \psi_{i+2}\right)\,,\\
    U(1)'_{i+1}: (\psi_i,\psi_{i+1},\psi_{i+2})\mapsto &\left(\psi_i\,,\ e^{i\theta_{i+1}}\,\psi_{i+1}\,,\ e^{-\frac{n_{i+1}C_2(\mathbf{R}_{i+1})}{n_{i+2}C_2(\mathbf{R}_{i+2})}i\theta_{i+1}}\,\psi_{i+2}\right)\,.
    \end{split}
\end{align}
Thus, the overlapping symmetry should satisfy the condition
\begin{align}
    \frac{\theta_i}{2\pi}=k_i\,,\quad \frac{\theta_{i+1}}{2\pi}=-\frac{n_iC_2(\mathbf{R}_i)}{n_{i+1}C_2(\mathbf{R}_{i+1})}\frac{\theta_i}{2\pi}+k_{i+1}\,,\quad -\frac{n_{i+1}C_2(\mathbf{R}_{i+1})}{n_{i+2}C_2(\mathbf{R}_{i+2})}\frac{\theta_{i+1}}{2\pi}=k_{i+2} \,.
    \label{eq:symcond1}
\end{align}
Inserting the third equation to the second one, we obtain
\begin{align}
    \frac{\theta_{i+1}}{2\pi}=-\frac{n_iC_2(\mathbf{R}_i)k_i}{n_{i+1}C_2(\mathbf{R}_{i+1})}+k_{i+1}=-\frac{n_{i+2}C_2(\mathbf{R}_{i+2})k_{i+2}}{n_{i+1}C_2(\mathbf{R}_{i+1})}.
    \label{eq:symcond2}
\end{align}
The first equality in equation \eqref{eq:symcond1} tells us that $\theta_{i+1}$ is in units of 
\begin{align}
    2\pi\, \frac{\gcd\left(n_iC_2(\mathbf{R}_i),n_{i+1}C_2(\mathbf{R}_{i+1})\right)}{n_{i+1}C_2(\mathbf{R}_{i+1})} \,,
\end{align}
while the equality in equation \eqref{eq:symcond2} indicates that $\theta_{i+1}$ is also in the unit of 
\begin{align}
    2\pi\, \frac{\gcd\left(n_{i+2}C_2(\mathbf{R}_{i+2}),n_{i+1}C_2(\mathbf{R}_{i+1})\right)}{n_{i+1}C_2(\mathbf{R}_{i+1})} \,.
\end{align}
Hence, we can determine that
\begin{align}
    \frac{\theta_{i+1}}{2\pi}\in \frac{\gcd(\tilde n_{i+1},\lcm (\tilde n_i, \tilde n_{i+2}))}{\tilde{n}_{i+1}}\IZ\,,
\end{align}
where $\tilde n_i\equiv n_iC_2(\mathbf{R}_i)$ and we used the identity $\lcm\left(\gcd (a,c),\gcd (b,c)\right)=\gcd (c,\lcm (a,b))$. Therefore, we conclude that there is an overlapping 
\begin{align}
    \IZ_{\tilde n_{i+1}/\gcd (\tilde n_{i+1},\,\lcm (\tilde n_i, \tilde n_{i+2}))}
\end{align}
among $U(1)'_i$ and $U(1)'_{i+1}$ that acts only on $(\phi_{i+1},\psi_{i+1})$.

\paragraph{$\mathbf{U(1)'_i}$ -- $\mathbf{U(1)_R}$:} While the individual $U(1)'_i$ symmetry can never overlap with the R-symmetry, since each rotates only a single $i$th or $(i+1)$th chiral matters respectively, there is a overlap between a linear combination of $U(1)'_i$ symmetries and R-symmetry. First of all, the intersection of the two symmetries should act trivially on the gaugino $\lambda$. That requires
\begin{align}
    \theta_R=2\pi k_R \,,
\end{align}
in the R-symmetry transformation given in equation \eqref{eqn:Raction}. This ensures the $\phi_i$ and $\psi_i$ in the same chiral multiplet get the same phase which can be part of flavor symmetry. Then, the $i$th fermion $\psi_i$ gains the phase 
\begin{align}
    \psi_i\mapsto \exp\left[-\frac{2\pi ih_G^\vee k_R}{\sum_i\tilde n_i}\right]\psi_i=\exp\left[i\left(\theta_i-\frac{\tilde n_{i-1}}{\tilde n_i}\theta_{i-1}\right)\right]\psi_i\,.
    \label{eqn:FRsym}
\end{align}
We can solve this system recursively for the $\theta$s to determine the overlap. We find
\begin{align}
\begin{split}
    \frac{\theta_1}{2\pi}&=-\frac{h_G^\vee k_R}{\sum_i \tilde n_i}+k_1\,,\\
    \frac{\theta_2}{2\pi}&=-\frac{h_G^\vee k_R}{\sum_i \tilde n_i}+\frac{\tilde n_1}{\tilde n_2}\frac{\theta_1}{2\pi}+k_2=-\frac{h_G^\vee k_R}{\sum_i \tilde n_i}\frac{\sum_{j=1}^2 \tilde n_j}{\tilde n_2}+\frac{\sum_{j=1}^2 \tilde n_jk_j}{\tilde n_2}\,,\\
    &\ \vdots\\
    \frac{\theta_p}{2\pi}&=-\frac{h_G^\vee k_R}{\sum_i \tilde n_i}\frac{\sum_{j=1}^p\tilde n_j}{\tilde n_p}+\frac{\sum_{j=1}^p\tilde n_jk_j}{\tilde n_p}\,,\\
    &\ \vdots\\
    \frac{\theta_{m-1}}{2\pi}&=-\frac{h_G^\vee k_R}{\sum_i \tilde n_i}\frac{\sum_{j=1}^{m-1}\tilde n_j}{\tilde n_{m-1}}+\frac{\sum_{j=1}^{m-1}\tilde n_jk_j}{\tilde n_{m-1}}\,,
\end{split}
\end{align}
where $k_i\in\mathbb{Z}$. 
Note that this transformation rotates the last ($m$th) chiral fermion as
\begin{align}
    \psi_m\mapsto \exp\left[-\frac{\tilde n_{m-1}}{\tilde n_m}i\theta_{m-1}\right]\psi_m=\exp\left[-\frac{2\pi ih_G^\vee k_R}{\sum_i\tilde n_i}\right]\psi_m\,,
\end{align}
which restricts the possible values of $k_R$ by the relation
\begin{align}
    -\frac{\tilde n_{m-1}}{\tilde n_m}\left(-\frac{h_G^\vee k_R}{\sum_i \tilde n_i}\frac{\sum_{j=1}^{m-1}\tilde n_j}{\tilde n_{m-1}}+\frac{\sum_{j=1}^{m-1}\tilde n_jk_j}{\tilde n_{m-1}}\right)&=-\frac{h_G^\vee k_R}{\sum_i\tilde n_i}+k_m \,,\\
    \Longrightarrow\ h_G^\vee& k_R=\sum_i \tilde n_ik_i\,.
\end{align}
As $k_i$ are arbitrary integers, we determine that
\begin{align}
    h_G^\vee k_R\in \gcd (\tilde n_1,\cdots ,\tilde n_m, h_G^\vee)\IZ \,. 
\end{align}
Therefore, the $U(1)_R$ symmetry and the $U(1)_i'$ symmetries share the discrete symmetry
\begin{align}\label{eqn:FR}
    \IZ_{\sum_i\tilde n_i/\gcd (\tilde n_1,\cdots,\tilde n_m,h_G^\vee)}\,,
\end{align}
that acts as above, and which should appear as a quotient in the global form of the flavor symmetry.

\paragraph{$\mathbf{U(1)_i'}$ -- $\mathbf{\IZ^A}$:}
The common symmetry between flavor $U(1)'_i$ and discrete axial $\IZ^A$ is almost the same as the second case above (i.e., between $U(1)'_i$ and $U(1)_R$), since $\IZ^A$ action is almost the same as in equation \eqref{eqn:FRsym}, but with $h_G^\vee k_R$ replaced by $-k_A/2$. Hence, with the analogous derivation, we find that 
\begin{align}
    k_A\in 2\gcd (\tilde n_1,\cdots,\tilde n_m)\IZ \,,
\end{align}
and thus the overlapping symmetry between $U(1)_i'$ and $\IZ^A$ is 
\begin{align}\label{eqn:FA}
    \IZ_{\sum_i\tilde n_i/\gcd (\tilde n_1,\cdots,\tilde n_m)}\,.
\end{align}
We note that the overlap between the $U(1)'_i$ and $U(1)_R$ that was determined in equation \eqref{eqn:FR} acts as a subgroup of overlap between the $U(1)_i$ and the $\mathbb{Z}^A$ -- therefore, we only need to quotient the global symmetry group by the action of equation \eqref{eqn:FA}.

\paragraph{$\mathbf{U(1)_R}$ -- $\mathbf{\IZ^A}$:}
Finally, the intersection between the R-symmetry and discrete axial symmetry should be generated by $\theta_R=2\pi k_R$, again because $\IZ^A$ does not rotate the gaugino. Such a subgroup of $U(1)_R$ is a strict subset of $\IZ^A$ that corresponds to
\begin{align}
    k_A=2h_G^\vee k_R\,.
\end{align}
Hence, the common discrete symmetry action between $U(1)_R$ and $\IZ^A$ is
\begin{align}\label{eqn:RA}
    \IZ_{\sum_i\tilde n_i/\gcd (\sum_i\tilde n_i,h_G^\vee)}\,.
\end{align}

\paragraph{$\mathbf{(U(1)_i'\times \IZ^A)-Z(G)}$:} Finally, we consider the global Abelian symmetry transformations that are absorbed by the gauge symmetry. In particular, we determine just the global symmetry transformations that are contained within the center, $Z(G)$, of the gauge group.
Let the charge of $i$th chiral multiplet under $Z(G)=\IZ_{|Z(G)|}$ be $z_i$.\footnote{For notational convenience, we do not consider the cases where $G=Spin(4n)$.} Then the center of the gauge symmetry rotates the chiral multiplets via
\begin{align}
    Z(G):\psi_i\mapsto e^{\frac{2\pi i z_ik_G}{|Z(G)|}}\,\psi_i\,,
\end{align}
for the $Z(G)$ element $k_G=0,\cdots,|Z(G)|-1$, while the vector multiplet is left invariant. Then we consider a transformation by $U(1)_i'\times \IZ^A$ characterized by the elements $(\theta_i,k_A)$, which rotates each chiral field by
\begin{align}
\psi_i\mapsto e^{i\Theta_i(\theta_i,k_A)}\,\psi_i\,,\quad i\Theta_i(\theta_i,k_A)\equiv i\theta_i-\frac{\widetilde{n}_{i-1}}{\widetilde{n}_i}i\theta_{i-1}+\frac{2\pi ik_A}{2\sum_i\widetilde{n}_i}\,,
\end{align}
where $k_A$ is $\IZ_{2\sum_i\widetilde{n}_i}$-valued. We look for $(\theta_i,k_A)$ satisfying the condition
\begin{align}
    \frac{2\pi z_ik_G}{|Z(G)|}=\Theta_i(\theta_{i-1},\theta_i,k_A)+2\pi k_i\,.
\end{align}
To extract the explicit condition on $k_A$, we first take the sum
\begin{align}
\sum_i\widetilde{n}_i\frac{2\pi z_ik_G}{|Z(G)|}=\sum_i\widetilde{n}_i\left(\Theta_i(\theta_{i-1},\theta_i,k_A)+2\pi k_i\right)
=\pi k_A+2\pi\sum_i\widetilde{n}_ik_i\,.
\end{align}
Then we can directly get the condition for $k_G$ from 
\begin{align}
k_A=\frac{2\sum_i\widetilde{n}_iz_i}{|Z(G)|}k_G-2\sum_i\widetilde{n}_ik_i\,.
\end{align}
Hence, the allowed values of $k_G$ requires to be quantized in the unit of
\begin{align}\label{eqn:global-gauge}
    k_G\in \frac{|Z(G)|}{\gcd(|Z(G)|,2\sum_i\widetilde{n}_iz_i)}\IZ\mod\,|Z(G)|\,.
\end{align}
Now, with the allowed values of $k_G$, we can find arbitrary $\IZ_{2\sum_i\widetilde{n}_i}$--valued $k_A$. The allowed $k_G$ is an element of
\begin{align}
    \IZ_{\mathcal{G}}=\IZ_{\gcd(|Z(G)|,2\sum_i\widetilde{n}_iz_i)} \,,
\end{align}
where we have defined the positive integer $\mathcal{G}$.
Note that the true action of this center symmetry on the chiral multiplets is given by
\begin{align}
    \psi_i\mapsto e^{\frac{2\pi iz_i k_G'}{\mathcal{G}}}\,\psi_i\,,\quad k_G'=\frac{|Z(G)|}{\mathcal{G}}k_G\,.
\end{align}
If every center charge $z_i$ of chiral multiplets is in the unit of $z_c=\gcd(z_1,\cdots,z_m)$, then the actual phase angle is in the unit of 
\begin{align}
    \frac{2\pi}{\mathcal{G}/\gcd(\mathcal{G},z_c)},
\end{align}
which is an element of 
\begin{align}
    \IZ_{\mathcal{G}/\gcd(\mathcal{G},z_c)}\,.
\end{align}
Now we extract the condition for $\theta_i$ through recursion which must be such that the following system of equations is satisfied:
\begin{align}
\begin{split}
    \frac{\theta_1}{2\pi}=&\,\frac{z_1k_G}{|Z(G)|}-\frac{ k_A}{2\sum_i\widetilde{n}_i}-k_1\,,\\
    \frac{\theta_2}{2\pi}=&\,\frac{\widetilde{n}_1}{\widetilde{n}_2}\frac{\theta_1}{2\pi}+\frac{z_2k_G}{|Z(G)|}-\frac{ k_A}{2\sum_i\widetilde{n}_i}-k_2\\
    =&\,\frac{k_G}{|Z(G)|}\frac{\sum_{i=1}^2\widetilde{n}_iz_i}{\widetilde{n}_2}-\frac{k_A}{2\sum_i\widetilde{n}_i}\frac{\sum_{i=1}^2\widetilde{n}_i}{\widetilde{n}_2}-\frac{\sum_{i=1}^2\widetilde{n}_ik_i}{\widetilde{n}_2}\,,\\
    \vdots\,&\\
    \frac{\theta_{m-1}}{2\pi}=&\,\frac{k_G}{|Z(G)|}\frac{\sum_{i=1}^{m-1}\widetilde{n}_iz_i}{\widetilde{n}_{m-1}}-\frac{k_A}{2\sum_i\widetilde{n}_i}\frac{\sum_{i=1}^{m-1}\widetilde{n}_i}{\widetilde{n}_{m-1}}-\frac{\sum_{i=1}^{m-1}\widetilde{n}_ik_{m-1}}{\widetilde{n}_{m-1}}\,.
    \end{split}
\end{align}
The phase angle that $\psi_m$ acquires provides a consistency condition
\begin{align}
    \frac{z_mk_G}{|Z(G)|}=-\frac{\widetilde{n}_{m-1}}{\widetilde{n}_m}\frac{\theta_{m-1}}{2\pi}+\frac{k_A}{2\sum_i\widetilde{n}_i}+k_m\,,
\end{align}
which is same as in equation \eqref{eqn:global-gauge}. Therefore, we have verified that
\begin{align}
    \IZ_{\mathcal{G}/\gcd(\mathcal{G},z_c)}
\end{align}
is consistently the subgroup of global symmetry group that is gauged, and where the action is explicitly specified above.

\paragraph{The global structure:}
Now that we have determined the overlap between all the Abelian factors (both continuous and discrete) in the global symmetry, after taking into account the ABJ anomaly, we determined which subgroup of the global symmetry sits within the center of the gauge group, we are ready to present the full global structure of the gauge theory. Taking everything into account, we find that the Abelian part of the zero-form symmetry of a $G$-gauge theory with chiral multiplets in the $n_1\mathbf{R}_1\oplus\cdots\oplus n_m\mathbf{R}_m$ representation is given by
\begin{align}
    \frac{
    U(1)^{m-1}\times U(1)_R\times Z_{2\sum_i\tilde n_i}}{\displaystyle \left(\prod_{j=1}^{m-2}\IZ_{\tilde n_{j+1}/\gcd (\tilde n_{j+1},\lcm (\tilde n_j,\tilde n_{j+2}))}\right)\times \IZ_{\sum_i\tilde n_i/\gcd (\tilde n_1,\cdots,\tilde n_m)}\times \IZ_{\sum_i\tilde n_i/\gcd (\sum_i\tilde n_i,h_G^\vee)}\times\IZ_{\mathcal{G}/\gcd(\mathcal{G},z_c)}}\,.
\end{align}
The various quantities appearing in this equation are as defined throughout this section, and similarly the explicit action with which the quotient acts is specified above.

\subsection{Anomalies}

Similar to the case of a continuous symmetry, a discrete symmetry can have a 't Hooft anomaly or a mixed anomaly \cite{Ibanez:1991hv, Csaki:1997aw, Hsieh:2018ifc}, and they remain invariant under renormalization group flows. A simple way of computing this is via embedding the discrete symmetry into $U(1)$ symmetry, and then computing the triangle anomaly, as in the case of a continuous symmetry. The difference lies in the matching of 't Hooft anomalies: the 't Hooft anomaly for the discrete symmetries is not required to match exactly. They are only required to match up to the contributions from the massive Majorana fermions. For example, when the discrete symmetry is $\IZ_N$ with $N$ even, a massive Majorana fermion with charge $N/2$ can contribute to the anomaly in the UV, but vanishes in the IR.

By considering free fermions with the symmetry group $\textrm{Spin}(4) \times \IZ_N$, the 't Hooft anomaly ($\IZ_N^3$) and the mixed-gravitational anomaly ($\IZ_N-(\textrm{grav})^2)$ are computed in \cite{Hsieh:2018ifc} respectively as
\begin{align}\label{eq:disanom}
\begin{aligned}
    \IZ_N^3 :\quad& (N+1)(N+2)\Delta s^3 &&\mod\ 6N\,,\\ 
    \IZ_N-(\textrm{grav})^2 :\quad& 2\Delta s &&\mod\ N\,,
\end{aligned}
\end{align}
where $\Delta s^3$ and $\Delta s$ are the difference of cubic and linear charge sums of left- and right-chirality fermions, respectively:
\begin{align} \label{eq:disanomNaive}
    \Delta s^3=\sum q_L^3-\sum q_R^3 \ , \quad
    \Delta s=\sum q_L-\sum q_R \,.
\end{align} 
Note that they are identical to the usual `t Hooft and mixed-gravitational anomalies computed by the triangle diagram via embedding $\IZ_N$ into $U(1)$. The anomalies, given in equation \eqref{eq:disanom}, have to match along renormalization group flows. Notice that the matching condition, given by equation \eqref{eq:disanom}, is more stringent than the naive matching of the ``anomaly coefficients'' that can be computed from equation \eqref{eq:disanomNaive}.  We compute discrete anomalies for the axial symmetry for all the gauge theories we classified in Section \ref{sec:2}. The result is presented in Tables \ref{tbl:discreteAnomaly}, \ref{tbl:discretegaugingII}, and \ref{tbl:discretegaugingIII}.\footnote{It is important to note that these anomalies are that of the full infrared theory, which may involve decoupled sectors; to determine the anomaly of the interacting IR SCFT, it is necessary to remove the contributions from these decoupled degrees of freedom.} 

We note that when there is a one-form symmetry, it is possible that the discrete zero-form anomaly is modified by certain one-form backgrounds \cite{Delmastro:2022pfo, Brennan:2022tyl}. When there is a mixed anomaly between zero-form and one-form symmetries, zero-form 't Hooft anomalies can be shifted by changing the one-form symmetry background, or the ``fractionalization class''. While this phenomena does happen in many of the theories we consider, it does not affect our conclusions regarding the IR phase of the deformed gauge theories. Hence, we do not compute their effects in this paper.

\afterpage{
\begin{landscape}
\pagestyle{empty}
\begin{table}[H]
  \centering
  \resizebox{!}{.27\paperheight}{%
  \begin{threeparttable}
    \begin{tabular}{ccc}
      \toprule
      $G$ & $\bm{R}$ & Discrete anomaly \\\midrule
      \multirow{1}{*}{$G$} & $n_a\, \textbf{adj}$ & \begin{tabular}{c}
          $2(N-1)(N+1)n_a(n_aN+1)(2n_aN+1)\mod12n_aN$\\
          $-2n_a\mod2n_aN$
      \end{tabular} \\\midrule
      \multirow{13}{*}{$SU(2N)$} & $n_S(\bm{S^2} \oplus \bm{\overline{S^2}})$ & \begin{tabular}{c}
      $8n_S^3N(N+1)(2N^2+1)+4n_SN(2N+1) \mod 24(N+1)n_S$\\
      $4n_SN(2N+1)\mod 4n_S(N+1)$ 
      \end{tabular}\\
      & $n_A(\bm{A^2} \oplus \bm{\overline{A^2}})$ & \begin{tabular}{c}
       $8n_A^3N(N-1)(2N^2+1)+4n_AN(2N-1)\mod 24(N-1)n_A$\\
       $4n_AN(2N-1)\quad\mod 4n_A(N-1)$
       \end{tabular}  \\
       & $\bm{S^2} \oplus \bm{\overline{S^2}} \oplus n_A(\bm{A^2} \oplus \bm{\overline{A^2}})$ & \begin{tabular}{c}
       $4N(2Nn_A+2N-2n_A+3)(2Nn_A+2N-n_A+1)(4Nn_A+4N-4n_A+5)\mod 24(Nn_A+N-n_A+1)$\\
       $4N(2Nn_A+2N-2n_A+3)\mod 4(n_AN+N-4n_A+4)$
       \end{tabular} \\
       & $\textbf{adj} \oplus \bm{S^2} \oplus \bm{\overline{S^2}}$ & \begin{tabular}{c}$(8N+5)(8N+6)(8N^2+2N-1)\mod 24(2N+1)$\\
       $4N+2\mod 8N+4$
       \end{tabular} \\ 
       & $\textbf{adj} \oplus n_A(\bm{A^2} \oplus \bm{\overline{A^2}})$ & \begin{tabular}{c}
       $(4n_AN+4N-4n_A+1)(4n_AN+4N-4n_A+2)(4n_AN^2+4N^2-2n_AN-1)\mod24(n_AN+N-n_A)$\\
       $2(2n_AN-1)\mod4(n_AN+N-n_A)$
       \end{tabular} \\
       & $\textbf{adj} \oplus \bm{S^2} \oplus \bm{\overline{S^2}} \oplus \bm{A^2} \oplus \bm{\overline{A^2}}$ & \begin{tabular}{c}
       $16N^2+36N-2\mod 72N$\\
       $8N^2-1\mod 12N$
       \end{tabular} \\
       & $2\,\textbf{adj} \oplus \bm{A^2} \oplus \bm{\overline{A^2}}$ & \begin{tabular}{c}
       $2(12N-3)(12N-2)(6N^2-N-1)\mod 24(3N-1)$\\
       $4(6N^2-N-1)\mod4(3N-1)$
       \end{tabular}
       \\\midrule
       \multirow{3}{*}{$Spin(N)$} & $n_S\,\bm{S^2}$ & \begin{tabular}{c}
       $n_SN(N+1)(n_S(2N+4)+1)(n_S(N+2)+1)\mod12n_S(N+2)$\\
       $n_SN(N+1)\mod2N+4$
       \end{tabular} \\
       & $\bm{S^2} \oplus \textbf{adj}$ & \begin{tabular}{c}
       $N^2(4N+1)(4N+2)\mod24N$\\
       $2N^2\mod 4N$
       \end{tabular} \\\midrule
       \multirow{11}{*}{$Spin(N)$} & $\textbf{adj} \oplus n_V \bm{V}$ & \begin{tabular}{c}
       $N(N+n_V-1)(N+2n_V-1)(2N+2n_V-3)\mod12(N+n_V-2)$\\
       $N(N+n_V-1)\mod2(N+n_V-2)$
       \end{tabular} \\
       & $2\textbf{adj} \oplus n_V \bm{V}$ & \begin{tabular}{c}
       $2N(N+n_V-1)(2N+n_V-3)(4N+2n_V-7)\mod12(2N+n_V-4)$\\
       $2N(N+n_V-1)\mod2(2N+n_V-4)$
       \end{tabular} \\
       & $\bm{S^2} \oplus n_V \bm{V}$ & \begin{tabular}{c}
       $(N+n_V+3)(2N+2n_V+5)(N^2+2n_VN+N-2)\mod12(N+n_V+2)$\\
       $N^2+2n_VN+N-2\mod2(N+n_V+2)$
       \end{tabular} \\
       & $2\bm{S^2} \oplus n_V \bm{V}$ & \begin{tabular}{c}
       $2(2N+n_V+5)(4N+2n_V+9)(N^2+n_VN+N-2)\mod12(2N+n_V+4)$\\
       $N^2+n_VN+N-2\mod2(2N+n_V+4)$
       \end{tabular} \\
       & $\textbf{adj} \oplus \bm{S^2} \oplus n_V \bm{V}$ & \begin{tabular}{c}
       $2(2N+n_V+1)(4N+2n_V+1)(N^2+n_VN-1)\mod12(2N+n_V)$\\
       $2(N^2+n_VN-1)\mod2(2N+n_V)$
       \end{tabular} \\
       & $n_V \bm{V}$ & \begin{tabular}{c}
       $2n_VN(n_V+1)(2n_V+1)\mod12n_V$\\
       $0\mod2n_V$
       \end{tabular} \\\midrule
      \multirow{5}{*}{$USp(2N)$} & $n_A \bm{A^2}$ & \begin{tabular}{c}
      $2(n_AN-n_A+1)(2n_AN-2n_A+1)(2n_AN^2-n_AN-1)\mod12n_A(N-1)$\\
      $2(2n_AN^2-n_AN-1)\mod2n_A(N-1)$
      \end{tabular} \\
       & $\textbf{adj} \oplus n_A \bm{A^2}$ & \begin{tabular}{c}
       $2(2N-1)(n_AN+N+1)(n_AN+N-n_A+2)(2n_AN+2N-2n_A+3)\mod12(n_AN+N-n_A+1)$\\
       $2(2N-1)(n_AN+N+1)\mod2(n_AN+N-n_A+1)$
       \end{tabular} \\
       & $2\, \textbf{adj} \oplus \bm{A^2}$ & \begin{tabular}{c}
       $2(n_AN+2N-n_A+3)(2n_AN+4N-2n_A+5)(2n_AN^2+4N^2-n_aN+2N-1)\mod12(n_AN+2N-n_A+2)$\\
       $2(2n_AN^2+4N^2-n_aN+2N-1)\mod2(n_AN+2N-n_A+2)$
       \end{tabular} \\
      \bottomrule
    \end{tabular}
  \end{threeparttable}}
  \caption{Discrete anomalies for the axial symmetry $\IZ_{2L}$. For each theory, the first and second row denotes 't Hooft and mixed-gravitational anomalies for the axial symmetry.}\label{tbl:discreteAnomaly}
\end{table}
\end{landscape}

\begin{table}[H]
  \centering
  \resizebox{!}{.335\paperheight}{%
  \begin{threeparttable}
    \begin{tabular}{cccc}
      \toprule
      $G$ & $\bm{R}$ & Discrete anomaly & Discrete anomaly preserving operator\\\midrule
      \multirow{3}{*}{$SU(4)$} & $\bm{20'}$ & \begin{tabular}{c}
           $72\mod96$   \\
             $8\mod16$
       \end{tabular} & none  \\
       & $\bm{20'}\oplus \bm{15}$ & \begin{tabular}{c}
       $142\mod144$\\
       $22\mod24$ \end{tabular} & $\tr Q_1^2\,,\,\tr Q_2^2$ \\\midrule
       \multirow{11}{*}{$SU(4)$} & $\bm{20'}\oplus n_{\bm{6}}\,\bm{6}$ & \begin{tabular}{c}
       $8(n_{\mathbf{6}}^2+n_{\mathbf{6}}-63)\mod12(n_{\mathbf{6}}+8)$\\
       $-56\mod2(n_{\bm6}+8)$ \end{tabular} & --  \\
       & $\bm{10}\oplus\bm{\overline{10}}\oplus n_{\bm{6}}\,\bm{6}$ & \begin{tabular}{c}
       $8n_{\bm6}^2-320\mod12(n_{\bm6}+6)$\\
       $-32\mod 2(n_{\bm6}+6)$
       \end{tabular} & -- \\
       & $\bm{15}\oplus\bm{10}\oplus\bm{\overline{10}}\oplus n_{\bm{6}}\,\bm{6}$ & \begin{tabular}{c}
       $8n_{\bm{6}}^2+10n_{\bm{6}}-750\mod12(n_{\bm6}+10)$\\
       $-50\mod 2(n_{\bm6}+10)$
       \end{tabular} & --  \\
       & $n_{\bm{6}}\,\bm{6}$ & \begin{tabular}{c}
       $0\mod 12n_{\bm6}$\\
       $0\mod 2n_{\bm6}$
       \end{tabular} & --  \\
       & $\bm{15}\oplus n_{\bm{6}}\,\bm{6}$ & \begin{tabular}{c}
       $6n_{\bm6}+6\mod12(n_{\bm6}+4)$\\
       $-18\mod 2(n_{\bm6}+4)$
       \end{tabular} & -- \\
       & $2\,\bm{15}\oplus n_{\bm{6}}\,\bm{6}$ & \begin{tabular}{c}
       $-36\mod12(n_{\bm6}+8)$\\
       $-36\mod2(n_{\bm6}+8)$
       \end{tabular} & --  \\\midrule
      \multirow{5}{*}{$SU(6)$} & $n_{\bm{20}} \, \bm{20}$ & \begin{tabular}{c}
      $4n_{\bm{20}}\mod36n_{\bm{20}}$\\
      $4n_{\bm{20}}\mod6n_{\bm{20}}$
      \end{tabular} & --  \\
       & $\textbf{35} \oplus n_{\bm{20}} \, \bm{20}$ & \begin{tabular}{c}
       $22n_{\bm{20}}+34\mod36(n_{\bm{20}}+2)$\\
       $4n_{\bm{20}}-2\mod6(n_{\bm{20}}+2)$
       \end{tabular} & --  \\
       & $2\, \textbf{35} \oplus n_{\bm{20}} \, \bm{20}$ & \begin{tabular}{c}
       $4n_{\bm{20}}+68\mod36(n_{\bm{20}}+4)$\\
       $4n_{\bm{20}}+68\mod6(n_{\bm{20}}+4)$
       \end{tabular} & -- \\\midrule
      \multirow{5}{*}{$SU(6)$} &  $\bm{\overline{21}}\oplus 5\,\bm{15}$ & \begin{tabular}{c}
      $24\mod144$\\
      $24\mod28$
      \end{tabular} & $ \epsilon Q_2^3$ \\
       & $4\,(\bm{15}\oplus\bm{\overline{15}})$ & \begin{tabular}{c}
       $48\mod192$\\
       $16\mod32$
       \end{tabular} & $\epsilon\,Q_1^3\,,\epsilon\,Q_2^3\,,\,Q_1Q_2$ \\
       & $\bm{35}\oplus 3\,(\bm{15}\oplus\bm{\overline{15}})$ & \begin{tabular}{c}
       $142\mod216$\\
       $34\mod36$
       \end{tabular} & $\tr Q_1^2\,,\,Q_2Q_3$ \\
       \midrule
      \multirow{3}{*}{$SU(8)$} & $n_{\bm{70}} \, \bm{70}$ & \begin{tabular}{c}
      $40n_{\bm{70}}^3+20n_{\bm{70}}\mod120n_{\bm{70}}$\\
      $0\mod20n_{\bm{70}}$
      \end{tabular} & -- \\
       & $\bm{63} \oplus \bm{70}$ & \begin{tabular}{c}
      $158\mod216$\\
      $14\mod36$
      \end{tabular} & $\tr Q_1^2\,,\,\epsilon Q_2^2$ \\\midrule
      \multirow{9}{*}{$SU(8)$} & $\bm{70} \oplus \bm{36} \oplus \bm{\overline{36}}$ & \begin{tabular}{c}
      $204\mod 240$\\
      $4\mod40$
      \end{tabular} & $\tr Q_1^2\,,\,\tr Q_1^3\,,\,Q_2Q_3\,,\,Q_1Q_2Q_3$ \\
       & $\bm{70} \oplus n_{\bm{28}}(\bm{28} \oplus \bm{\overline{28}})$ & \begin{tabular}{c}
           $48n_{\bm{28}}^2-180\mod24(3n_{\bm{28}}+5)$\\
           $4n_{\bm{28}}-40\mod4(3n_{\bm{28}}+5)$
       \end{tabular} & -- \\
       & $\bm{63}\oplus\bm{36}\oplus 3\,\bm{\overline{28}}$ & \begin{tabular}{c}
       $234\mod264$\\
       $14\mod44$
       \end{tabular} & $\tr Q_1^2\,,\,\tr Q_1^3\,,\,Q_1Q_2Q_3$ \\
       & $\bm{36}\oplus 3\,\bm{\overline{28}}$ & \begin{tabular}{c}
       $72\mod168$\\
       $16\mod28$
       \end{tabular} & $Q_1^2Q_2^2\,,\,\epsilon Q_2^4$  \\
       & $\bm{36}\oplus\bm{28}\oplus 4\,\bm{\overline{28}}$ & \begin{tabular}{c}
       $192\mod240$\\
       $32\mod40$
       \end{tabular} & $Q_2Q_3$  \\\midrule
       $SU(9)$ & $\bm{84} \oplus \bm{\overline{84}}$ & \begin{tabular}{c}
       $84\mod252$\\
       $0\mod42$
       \end{tabular} & $\epsilon Q_1^3\,,\,\epsilon Q_2^3$ \\\midrule
       \multirow{9}{*}{$SU(12)$} & $2\,\bm{\overline{78}}\oplus 4\,\bm{66}$ & \begin{tabular}{c}
       $24\mod408$\\
       $24\mod68$
       \end{tabular} & none  \\
       & $2\,\bm{\overline{78}}\oplus\bm{78}\oplus 2\,\bm{66}$ & \begin{tabular}{c}
       $360\mod372$\\
       $50\mod62$
       \end{tabular} & $Q_1Q_2$  \\
       & $\bm{143}\oplus\bm{78}\oplus 2\,\bm{\overline{66}}$ & \begin{tabular}{c}
       $300\mod348$\\
       $10\mod58$
       \end{tabular} & $\tr Q_1^2\,,\,\tr Q_1^3\,,\,Q_1Q_2Q_3$  \\
       & $\bm{78}\oplus 2\,\bm{\overline{66}}$ & \begin{tabular}{c}
       $12\mod204$\\
       $12\mod34$
       \end{tabular} & none  \\
       & $\bm{78}\oplus\bm{66}\oplus 3\,\bm{\overline{66}}$ & \begin{tabular}{c}
       $36\mod324$\\
       $36\mod54$
       \end{tabular} & $\tr Q_1^2\,,\,\tr Q_1^3\,,\,Q_2Q_3\,,\,Q_1Q_2Q_3$  \\\bottomrule
    \end{tabular}
  \end{threeparttable}}
  \caption{Discrete anomalies for the axial symmetries of the outlier unitary theories from Table \ref{tbl:N1SU}. See the caption of Table \ref{tbl:discreteAnomaly} for a description. For rows where the matter spectrum does not depend on an integer parameter, we have schematically enumerated the relevant operators for which the deformation preserves some of the discrete anomaly. Here the $Q_i$ are the chiral multiplets in the same order as listed in the $\bm{R}$ column.}
  \label{tbl:discretegaugingII}
\end{table}

\begin{table}[H]
  \centering
  \resizebox{!}{.36\paperheight}{%
  \begin{threeparttable}
    \begin{tabular}{cccc}
      \toprule
      \multirow{2}{*}{$G$} & \multirow{2}{*}{$\bm{R}$} & \multirow{2}{*}{Discrete anomaly} & \multirow{1}{*}{Discrete anomaly} \\
        & & & preserving operator \\\midrule
      \multirow{5}{*}{$Spin(7)$} & $\bm{35}$ & \begin{tabular}{c}
           $90\mod120$   \\
             $10\mod20$
       \end{tabular} & none  \\
       & $\bm{35}\oplus \bm{21}$ & \begin{tabular}{c}
       $112\mod180$\\
       $22\mod30$ \end{tabular} & $\tr Q_1^2\,,\,\tr Q_2^2$ \\
       & $\bm{35}\oplus n_{\bm7}\bm{7}$ & \begin{tabular}{c}
       $4n_{\bm7}^3+10n_{\bm7}^2+4n_{\bm7}+2970\mod12(n_{\bm7}+10)$\\
       $-70\mod2(n_{\bm7}+10)$ \end{tabular} & -- \\\midrule
       \multirow{11}{*}{$Spin(8)$} & $\bm{56_{v,s,c}}$ & \begin{tabular}{c}
       $112\mod180$\\
       $22\mod30$ \end{tabular} & none  \\
       & $\bm{56_{v,s,c}}\oplus n_{v,s,c}\,\bm{8_{v,s,c}}$ & \begin{tabular}{c}
       $8n_{v,s,c}^3+8n_{v,s,c}^2+4n_{v,s,c}+25132\mod12(n_{v,s,c}+15)$\\
       $-128\mod 2(n_{v,s,c}+15)$
       \end{tabular} & -- \\
       & $\bm{35_{v,s,c}}\oplus\bm{28}$ & \begin{tabular}{c}
       $30\mod192$\\
       $30\mod32$
       \end{tabular} & $Q_1^2\,,\,Q_1^3\,,\,Q_2^2\,,\,Q_1Q_2^2$  \\
       & $\bm{35_{v,s,c}}\oplus\bm{28}+n_{v,s,c}\,\bm{8_{v,s,c}}$ & \begin{tabular}{c}
       $8n_{v,s,c}^3+4n_{v,s,c}^2+6n_{v,s,c}+31710\mod 12(n_{v,s,c}+16)$\\
       $-130\mod 2(n_{v,s,c}+16)$
       \end{tabular} & --  \\
       & $\bm{28}+n_{v,s,c}\,\bm{8_{v,s,c}}$ & \begin{tabular}{c}
       $8n_{v,s,c}^3+4n_{v,s,c}^2+4n_{v,s,c}+1568\mod12(n_{v,s,c}+6)$\\
       $-40\mod 2(n_{v,s,c}+6)$
       \end{tabular} & -- \\
       & $n_{v,s,c}\,\bm{8_{v,s,c}}$ & \begin{tabular}{c}
       $8n_{v,s,c}^3+4n_{v,s,c}\mod12n_{v,s,c}$\\
       $0\mod2n_{v,s,c}$
       \end{tabular} & --  \\\midrule
      \multirow{9}{*}{$Spin(12)$} & $\bm{77}\oplus\bm{66}\oplus\bm{32_{s,c}}$ & \begin{tabular}{c}
      $294\mod336$\\
      $14\mod56$
      \end{tabular} & $Q_1^2\,,\,Q_1^3\,,\,Q_2^2\,,\,Q_2Q_3^2$  \\
       & $2\,\bm{66}\oplus n_{s,c}\,\bm{32_{s,c}}$ & \begin{tabular}{c}
       $32n_{s,c}^3+32n_{s,c}^2+3144\mod48(n_{s,c}+5)$\\
       $-56\mod8(n_{s,c}+5)$
       \end{tabular} & --  \\
       & $\bm{77}\oplus n_{s,c}\,\bm{32_{s,c}}$ & \begin{tabular}{c}
       $32n_{s,c}^3+16n_{s,c}^2+8n_{s,c}+1134\mod24(2n_{s,c}+7)$\\
       $-70\mod4(n_{s,c}+7)$
       \end{tabular} & -- \\ 
       & $\bm{66}\oplus n_{s,c}\,\bm{32_{s,c}}$ & \begin{tabular}{c}
       $32n_{s,c}^3+16n_{s,c}^2+372\mod24(2n_{s,c}+5)$\\
       $-28\mod4(n_{s,c}+5)$
       \end{tabular} & -- \\ 
       & $n_{s,c}\,\bm{32_{s,c}}$ & \begin{tabular}{c}
       $16n_{s,c}(2n_{s,c}^2+1)\mod48n_{s,c}$\\
       $0\mod8n_{s,c}$
       \end{tabular} & -- \\\midrule
      \multirow{5}{*}{$Spin(16)$} &  $n_{s,c}\,\bm{128_{s,c}}$ & \begin{tabular}{c}
      $64n_{s,c}(2n_{s,c}^2+1)\mod192n_{s,c}$\\
      $0\mod32n_{s,c}$
      \end{tabular} & -- \\
       & $\bm{128_{s,c}}\oplus\bm{135}$ & \begin{tabular}{c}
       $186\mod408$\\
       $50\mod68$
       \end{tabular} & $Q_1^2$ \\
       & $\bm{128_{s,c}}\oplus\bm{120}$ & \begin{tabular}{c}
       $136\mod360$\\
       $16\mod60$
       \end{tabular} & $Q_1^2$ \\
       \midrule
      \multirow{3}{*}{$USp(4)$} & $\bm{14}$ & \begin{tabular}{c}
      $0\mod84$\\
      $0\mod14$
      \end{tabular} & -- \\
       & $\bm{14} \oplus n_{\bm5}\,\bm{5}$ & \begin{tabular}{c}
      $8n_{\bm5}^3+6n_{\bm5}^2+4n_{\bm5}+2436\mod12(n_{\bm5}+7)$\\
      $-42\mod2(n_{\bm5}+7)$
      \end{tabular} & -- \\\midrule
      \multirow{5}{*}{$USp(8)$} & $n_{\bm{42}}\,\bm{42}$ & \begin{tabular}{c}
      $0\mod84n_{\bm{42}}$\\
      $0\mod14n_{\bm{42}}$
      \end{tabular} & none \\
       & $\bm{42}\oplus\bm{36}$ & \begin{tabular}{c}
           $12\mod144$\\
           $12\mod24$
       \end{tabular} & $Q_1^2$ \\
       & $\bm{42}\oplus n_{\bm{27}}\,\bm{27}$ & \begin{tabular}{c}
       $6(3n_{\bm{42}}^2+4n_{\bm{42}}-14)\mod12(3n_{\bm{42}}+7)$\\
       $6n_{\bm{42}}-28\mod2(3n_{\bm{42}}+7)$
       \end{tabular} & -- \\\bottomrule
    \end{tabular}
  \end{threeparttable}}
  \caption{Discrete anomalies for the axial symmetries of the outlier orthosymplectic theories from Table \ref{tbl:N1SpinUSp}. See the caption of Tables \ref{tbl:discreteAnomaly} and \ref{tbl:discretegaugingII} for a description.}
  \label{tbl:discretegaugingIII}
\end{table}
}

\section{Relevant deformations}\label{sec:deformations}

Thus far, in Questions \ref{ques:first} and \ref{ques:second}, we have focused on gauge theories which flow in the infrared to superconformal field theories. In this sense, the existence of the mixed zero-form/one-form anomaly constrains the infrared behavior only insofar as the infrared SCFT must saturate the anomaly. However, many of these SCFTs possess relevant operators, and performing a superpotential deformation with respect to these operators triggers a flow to a new infrared fixed point. After a sequence of such deformations, the infrared behavior may no longer be superconformal,\footnote{In particular, the deformations + ABJ anomaly may break all classical Abelian symmetries, and thus there is no candidate $U(1)$ R-symmetry with which to form the 4d $\mathcal{N}=1$ superconformal algebra. It is still possible for an unexpected $U(1)$ to emerge along the flow into the infrared, and thus to restore superconformal symmetry in the infrared.} as depicted in equation \eqref{eqn:bigpicture}, and we would like to understand the possible consistent infrared behaviors. As usual, anomalies provide constraints on the infrared physics -- as the infrared degrees of freedom must saturate the anomalies. While it is generally believed that any set of anomalies can arise from a symmetry-preserving conformal field theory, when the infrared behavior is not of this form, the anomalies provide a powerful constraint on the alternative, anomaly-saturating, infrared physics. 

We can also consider non-supersymmetric deformations by a relevant operator, for example performing the deformation with just the super-primary of the multiplet associated with the relevant operator, instead of the whole multiplet. Interestingly, whether the deformation is supersymmetric or not does not have a significant effect on the mixed anomaly of the residual zero-form/one-form symmetries in the infrared; therefore similar conclusions can be drawn about the behavior in the infrared. For this reason, we do not overly emphasize the non-supersymmetric deformations in this section.

The purpose of this section is succinctly captured via the following question.\footnote{In this section, we focus only on mixed anomalies on spin manifolds. The extension to theories defined on non-spin manifolds is straightforward.}
\begin{ques}{ques:fourth}
\renewcommand{\thempfootnote}{\arabic{mpfootnote}}
What are the theories that solve Question \ref{ques:second} and preserve one-form and zero-form mixed anomaly upon (supersymmetric or non-supersymmetric) relevant deformations?
\end{ques}

There are a multitude of simple gauge theories with a mixed zero-form/one-form anomaly listed in the tables in Section \ref{sec:2}. It would be tedious to write here the exhaustive analysis of each of these (families of) theories; instead, we have selected several representative theories which evince an interesting landscape of deformations interplaying with the mixed anomaly.

\subsection{\texorpdfstring{$SU(8)$}{SU(8)} with 4-index anti-symmetric matter}

In this subsection we consider an $SU(8)$ gauge theory with $n_{\bm{70}}$ chiral multiplets in the 70-dimensional $4$-index anti-symmetric representation of the gauge group. We expect the theory to flow in the infrared to an interacting conformal field theory when $n_{\bm{70}} = 1, 2$. 
 
The chiral multiplet(s) screen only part of the $\mathbb{Z}_8$ center, and we find a preserved $\mathbb{Z}_4$ one-form symmetry. ABJ anomaly cancellation gives rise to a discrete zero-form symmetry 
\begin{equation}
    Z^{A}=\mathbb{Z}_{20n_{\bm{70}}} \,.
\end{equation}
The presence of the mixed anomaly between this zero-form symmetry and the one-form symmetry is immediately evident. In the appropriate fractional instanton background, we find that the partition function transforms under a discrete zero-form transformation as
\begin{equation}
    Z[B] \to \exp\left( i \alpha (2L) \nu \right) Z[B] = \exp\left(2 \pi i \ell \frac{7}{2} \int \frac{\mathcal{P}(B)}{2} \right) Z[B] \,,
\end{equation}
where $\ell = 0, \cdots, 20n_{\bm{70}} - 1$. When $\ell$ is not even, this is a nontrivial phase.

The superconformal R-symmetry is fixed by demanding the absence of the ABJ anomaly for $R$:
\begin{align}
    0 = \tr R GG = 8 + n_{\bm{70}} T(\mathbf{70}) (R-1) = 8 + 10 n_{\bm{70}} (R-1).
\end{align}
Hence, for $n_{\bm{70}}=1$ and $n_{\bm{70}}=2$, the chiral multiplet $Q$ has $R$-charge $R = 1/5$ and $R=3/5$, respectively. When $n_{\bm{70}}=2$, there is no relevant operator except for the mass term; thus, we focus on $n_{\bm{70}}=1$ case.

When $n_{\bm{70}}=1$, the gauge-invariant operator of the form $Q^2$ has its $R$-charge below the unitarity bound. It follows that this operator gets decoupled along the renormalization group flow. Then, to isolate the interacting part of the IR fixed point theory, we introduce a flip field $X_{Q^2}$ and add a superpotential coupling in the ultraviolet: $W=X_{Q^2} Q^2$.

The faithful zero-form symmetry of this theory is
\begin{align}
    \frac{U(1)_R \times \IZ_{20}}{\IZ_5} \,, 
\end{align}
where the $\IZ_5$ quotient comes from
\begin{align}
    \begin{split}
    U(1)_R:&\quad \lambda \to e^{i \alpha} \lambda, \quad Q \to e^{-i\alpha/5}, \quad \psi \to e^{-4i\alpha/5}\psi \\
    Z^{A}=\IZ_{20} \subset U(1)_A:&\quad \lambda \to \lambda, \,\qquad Q \to e^{i\beta} Q, \quad\;\; \psi \to e^{i\beta} \psi \ , 
    \end{split}
\end{align}
with $\alpha = 2\pi n$, $\beta = 2\pi n/5$ with $n\in \IZ_5$. 
If we consider the faithful symmetry that acts on the 4-index tensor matter field (including the gauge-group action), we have
\begin{align} \label{eq:su8w70faithful}
    \frac{SU(8) \times U(1)_R \times \IZ_{20}}{\IZ_8 \times \IZ_5} \ ,
\end{align}
from $SU(8)/\mathbb{Z}_4 \times U(1)_R \times \IZ_{20}$, where the quotient is enlarged from $\IZ_4$ to $\IZ_8$. This comes from the generator $e^{\frac{2\pi i \ell}{8}} \bm{1} \subset SU(8)$ that acts as
\begin{align}
    (Q, \psi) \to e^{-\pi i \ell} (Q, \psi) = - (Q, \psi) \,. 
\end{align}
This demonstrates that when a Wilson line in the rank-4 tensor representation is screened by the matter field $Q$ in the rank-4 tensor representation, the ``$\IZ_2$'' global symmetry part from $SU(8)/\mathbb{Z}_4\times \IZ_{20}$ is not screened.  
Hence, if we turn on the background gauge field for this ``$\IZ_2$'' symmetry, this will affect Wilson lines that carry the one-form charges $\IZ_4$. 
Combining it with the fact that the following short exact sequence
\begin{align} \label{eq:SESsu8w70}
    1 \to \IZ_4 \to \IZ_8 \to \IZ_2 \to 1
\end{align}
does not split,\footnote{Note that $\IZ_4$ comes from the one-form symmetry, and $\IZ_2$ comes from the zero-form symmetry part of $\IZ_{20}$ that the matter field carries.} we find that this theory has a nontrivial ``2-group'' symmetry. It is important to remember that this is actually a \emph{central extension}, not a ``true'' 2-group.\footnote{See \cite{Kang:2023uvm} for details on central extensions and 2-groups.} This sequence in equation \eqref{eq:SESsu8w70} corresponds to the unique nontrivial element of $H^2(\mathbb{Z}_2,\mathbb{Z}_4)$, which is the group whose element classifies equivalence classes of central extensions of $\mathbb{Z}_2$ by $\mathbb{Z}_4$.

Given the base $X$, we are trying to lift a principle $B\mathbb{Z}_2$ bundle to $B\mathbb{Z}_8$ bundle. This corresponds to factoring a continuous map $X\to B^2 \mathbb{Z}_2$ through $B^2 \mathbb{Z}_8 \to B^2 \mathbb{Z}_2$. Its cofiber is $B^3 \mathbb{Z}_4$, so the obstruction class for this extension condition is given by a homotopy class of the induced map 
\begin{align}
    X\to\left( B^2 \mathbb{Z}_2\to\right) B^3 \mathbb{Z}_4 \,. 
\end{align}
Note that the map $B^2 \mathbb{Z}_2\to B^3 \mathbb{Z}_4$, which corresponds to a class $H^2(\mathbb{Z}_2,\mathbb{Z}_4)$, is given by the extension class of the given nonsplit central extension 
\begin{align}
    1\to B\mathbb{Z}_4\to B\mathbb{Z}_8\to B\mathbb{Z}_2\to 1 \,.
\end{align}
For any finite groups $G$ and $H$, any map $BG\to BH$ is homotopic to the map induced by some  group homomorphism $G\to H$. Hence we see that the given central extension of $B\mathbb{Z}_n$s corresponds to the nonsplit central extension given in equation \eqref{eq:SESsu8w70}.

Here, we are considering $B\mathbb{Z}_n$s in order to make sense of the obstruction classes $H^2(X,\mathbb{Z}_2)$ and $H^2(X,\mathbb{Z}_4)$. Otherwise the obstruction classes lie on $H^1$. Although delooping everything makes no difference here (as everything is Abelian), it would be interesting to ask why those delooped finite groups appear naturally in a physical sense.

While this construction includes this paragraph redundantly entirely, but let us present in a recent physics literature friendly fashion on its implications. There exists a Bockstein homomorphism $\beta: H^2(X, \IZ_2) \to H^3(X, \IZ_4)$ associated to the central extension as taken in equation \eqref{eq:SESsu8w70}, which provide relevant cochains. Two of them $E,w_2 \in H^2 (X, \IZ_4)$ corresponds respectively to the background gauge field for the $\IZ_4$ symmetry and the obstruction to lift $SU(8)/\IZ_4$ bundle to $SU(8)$ bundle, and another two cochains $v_2,a_2 \in H^2(X, \IZ_2)$ responds to obstructions to lift $SU(8)/\IZ_8$ bundle to $SU(8)/\IZ_4$ bundle and $\IZ_{20}/\IZ_2 $ to $\IZ_{20}$ bundle, where $E=w_2$ and $v_2=a_2$ from equation \eqref{eq:su8w70faithful}. Then we are left with only a pair $(a_2=v_2, E=w_2)$ that controls the obstruction of lifting the $SU(8)/\IZ_8$ bundle to $SU(8)$ bundle, and the $w_2$ is not necessarily closed:
\begin{align}\label{eqn:2group}
    \delta E =\delta w_2 = \beta(v_2)= \beta(a_2) \,.
\end{align}
Thus, in the presence of a background field for the $\IZ_2$ symmetry ($a_2$), background field ($E$) for the one-form symmetry fails to be flat, analogous to the Green--Schwarz mechanism. 

We can now consider superpotential deformations of these infrared SCFTs via relevant operators. Let us continue with the case of $n_{\bm{70}} = 1$. As we discussed in the beginning of this subsection, it is necessary to add a flip field, $X_{Q^2}$, which flips the $Q^2$ operator that decouples along the flow into the infrared. We have the relevant operators
\begin{align}
\setlength{\arraycolsep}{7pt}
\renewcommand{\arraystretch}{1.8}
    \begin{array}{c|cccc}
         & Q^4 & Q^6 & Q^8 & X_{Q^2} \\\hline
        R\text{-charge} & \dfrac{4}{5} & \dfrac{6}{5} & \dfrac{8}{5} & \dfrac{8}{5} \\
        \mathbb{Z}_{20}\text{-charge} & 4 & 6 & 8 & 18
    \end{array}
\end{align}
with their $R$-charges and also their $\mathbb{Z}_{20}$ charges. A deformation by one of these operators will break the $\IZ_{20}$ zero-form symmetry. Let us denote the generator of the zero-form symmetry $\omega$, so that
\begin{align}
    \IZ_{20} = \langle \omega \rangle.
\end{align}
The $Q^6$ deformation breaks the zero-form symmetry $\mathbb{Z}_{20}$ to a $\mathbb{Z}_2$ generated by the element $\omega^{10}$. Since nontrivial phases of the partition function only arise for odd elements of $\mathbb{Z}_{20}$, it follows that there is no mixed anomaly between this $\mathbb{Z}_2$ zero-form symmetry and the $\mathbb{Z}_4$ one-form symmetry. 

The situation is different for the cases of the $Q^4$ and $Q^8$ deformations. For these cases, the zero-form symmetry $\mathbb{Z}_{20}$ is broken to the $\mathbb{Z}_4$ subgroup generated by the element $\omega^5$. Since this element is not even, there exists a nontrivial mixed anomaly between the $\mathbb{Z}_4$ zero-form symmetry, and the $\mathbb{Z}_4$ one-form symmetry. Thus, the SCFT deformed by $Q^4$ or $Q^8$ cannot flow to a trivially gapped theory; furthermore, since there is no $U(1)_R$ in the infrared (modulo emergent symmetries), we believe that the anomaly is not saturated by an infrared SCFT. This leaves the option for the anomaly to be saturated by topological degrees of freedom (ruled out by \cite{Cordova:2019bsd}), or else for spontaneous symmetry breaking to occur. We note that the equality relation in equation \eqref{eqn:2group} holds under these RG flows via superpotential deformations.

\subsection{\texorpdfstring{$SU(2N)$}{SU(2N)} with symmetric matter}

We now explore a non-chiral theory. Consider $SU(2N)$ gauge theory with a pair of chiral multiplets, $Q$ in the two-index symmetric representation and its conjugate $\widetilde{Q}$.\footnote{We assume $N > 1$ to differentiate the symmetric matter from the adjoint matter studied in the next subsection.}
This theory has a $\mathbb{Z}_2$ one-form symmetry. There is a 
\begin{equation}
    U(1)_B \times U(1)_A \times U(1)_R \,,
\end{equation}
classical continuous global symmetry, and under the combined gauge and global symmetries the chiral multiplets transform as follows:
\begin{align}
\setlength{\arraycolsep}{6pt}
\renewcommand{\arraystretch}{1.8}
    \begin{array}{c|cccc}
         & SU(2N) & U(1)_B & U(1)_A & U(1)_R \\
         \hline 
         Q & \ydiagram{2} & 1 & 1 & \dfrac{1}{N+1} \\
         \widetilde{Q} & \overline{\ydiagram{2}} & -1 & 1 & \dfrac{1}{N+1} 
    \end{array} \,,
\end{align}
where the R-charges are fixed via the anomaly-free condition. The $U(1)_A$ axial symmetry is broken via the ABJ anomaly such that
\begin{equation}
    U(1)_A \,\,\rightarrow\,\, \mathbb{Z}_{4N + 4} \,,
\end{equation}
as explained around equation \eqref{eqn:ABJ}. However, these discrete axial transformations are not independent of the rest of the global symmetries, as we discussed in Section \ref{sec:global0}. In particular, we can see that the $\mathbb{Z}_2 \subset \mathbb{Z}_{4N+4}$ overlaps with $\mathbb{Z}_2 \subset U(1)_B$. Similarly, $\mathbb{Z}_{N+1} \subset \mathbb{Z}_{4N+4}$ overlaps with $\mathbb{Z}_{N+1} \subset U(1)_R$. Altogether, the global structure of the zero-form flavor is
\begin{equation}\label{eqn:globalSU2N}
    \frac{U(1)_B \times U(1)_R \times \mathbb{Z}_{4N+4}}{\mathbb{Z}_2 \times \mathbb{Z}_{N+1}} \,,
\end{equation}
where the quotiented action is as written above.\footnote{For simplicity, we have ignored the overlap between the $U(1)_B$ and the $U(1)_R$. See Section \ref{sec:global0} for the inclusion of this additional quotient.} 

Furthermore, there can be elements of the zero-form symmetry in equation \eqref{eqn:globalSU2N} which overlap with gauge transformations; in particular, we want to explore gauge transformations in the center of $SU(2N)$ and their overlap with the discrete axial symmetry. The subgroup of the axial symmetry
\begin{equation}\label{eqn:pineapple}
    \mathbb{Z}_{\gcd(N,4)} \subseteq \mathbb{Z}_{4N + 4} \,,
\end{equation}
generated by the element
\begin{equation}\label{eqn:tomato}
    \ell = \frac{4N + 4}{\gcd(N,4)}
\end{equation}
lies inside of the center of the gauge group. Conveniently, when $N$ is odd, which will be the main subject of this section, there is no such overlap as the element in equation \eqref{eqn:tomato} is the identity inside of $\mathbb{Z}_{4N + 4}$.

With the knowledge of the symmetry structure in hand, we analyze the anomalies. Under a $\mathbb{Z}_{4N+4}$ axial transformation parameterized by $\ell$, with a background field for the $\mathbb{Z}_2$ one-form symmetry, the partition function picks up the following phase:
\begin{equation}\label{eqn:macadamia}
    Z[B] \,\,\rightarrow\,\, \exp\left( 2\pi i \frac{N(2N-1)\ell}{2} \int \frac{\mathcal{P}(B)}{2} \right) Z[B] \,.
\end{equation}
When $N$ is even, it is clear that this phase is always trivial, and thus there is no mixed anomaly in the infrared SCFT. However, when $N$ is odd, the $\mathbb{Z}_{4N+4}$ transformations triggered by odd integers $\ell$ have a nontrivial phase. Therefore, the infrared SCFT obtained from $SU(2N)$ gauge theory with $\bm{S}^2 \oplus \overline{\bm{S}^2}$ when $N$ is odd has a mixed anomaly. We can also study the pure and mixed-gravitational anomalies of the $\mathbb{Z}_{4N+4}$ zero-form symmetry, as explained in Section \ref{sec:global0}. Following equation \eqref{eq:disanom}, these anomalies are, respectively, captured by the following integers:
\begin{equation}\label{eqn:symmatdisc}
  \begin{aligned}
    (4N+5)(4N+6)\times2N(2N+1)\,\,&\equiv\,\, 4\left(2N^2+N\right) \,& &\mod 24(N+1)\,,\\
    2\times2N(2N+1)\,\,&\equiv\,\, 4\,& &\mod 4(N+1)\,.
  \end{aligned}
\end{equation}
As we can see, regardless of the value of $N$, these quantities are never trivial. Indeed, as we note shortly, several operators decouple along the flow, and the anomalies in equation \eqref{eqn:symmatdisc} also include the contribution from those (gapless) decoupled sectors. By taking the flip fields $X_{\tr(Q\widetilde{Q})^i}$ whose $\IZ_{4N+4}$ charges are $-2i$ for $1\leq i\leq s\equiv\left\lfloor\frac{N+1}{3}\right\rfloor$ into account, we can also determine the $\mathbb{Z}_{4N+4}$ anomalies for the interacting sector in the infrared; we find
\begin{equation}\label{eqn:symmatdiscINTER}
\begin{aligned}
    &(4N+5)(4N+6)\times\left(2\times N(2N+1)+\sum_{i=1}^s(-2i)^3\right) \\ &\qquad\qquad \equiv4\left( 2N^2+N+8(N+1)^2s(s+1)-s^2(s+1)^2\right)\mod\,24(N+1)\,,\\
    &2\times\left(2\times N(2N+1)+\sum_{i=1}^s(-2i)\right)\equiv 2\left(2-s(s+1)\right)\mod\,4(N+1)\,.
\end{aligned}
\end{equation}
Thus, we see that the anomaly coefficients for the interacting SCFT in the infrared are nontrivial for any $N$.

We now consider an enumeration of the relevant (single-trace) operators of the infrared SCFT obtained from this gauge theory. The first class of operators are referred to as the mesonic operators, which take the form
\begin{equation}
    M_q \,=\, \tr (Q\widetilde{Q})^q \,,\quad q=1,\cdots, N.
\end{equation}
Under the $U(1)_B \times U(1)_R \times \mathbb{Z}_{4N + 4}$, these operators have the following charges:
\begin{equation}\label{eqn:Mqcharges}
    B[M_q] = 0 \,, \qquad R[M_q] = \frac{2q}{N+1} \,, \qquad A[M_q] = 2q \mod 4(N+1) \,.
\end{equation}
We notice that when 
\begin{equation}
    q \leq s ,
\end{equation}
the R-charge of $M_q$ either reaches or falls below the unitary bound and the associated operator thus decouples along the flow into the infrared. To incorporate this, we introduce gauge-neutral chiral multiplets (often called flip fields) $X_{M_q}$ that have opposite charges for each such $q$, and add a term in the UV superpotential
\begin{align}
    W = X_{M_q} M_q
\end{align}
 for each. This gives us the interacting SCFT in the infrared. The second class of relevant operators that we consider are the baryonic operators, formed out of the chiral multiplets as
\begin{equation}\label{eqn:baryons}
    B \,=\, \operatorname{det}Q^{2N} \,, \qquad \widetilde{B} \,=\, \operatorname{det}\widetilde{Q}^{2N} \,.
\end{equation}
Under the global symmetries the charges are
\begin{equation}
  \begin{aligned}
    B[B] &= 2N \,, &\qquad R[B] &= \frac{2N}{N+1} \,, &\qquad A[B] &= 2N \mod 4(N+1) \,, \\
    B[\widetilde{B}] &= -2N \,, &\qquad R[\widetilde{B}] &= \frac{2N}{N+1} \,, &\qquad A[\widetilde{B}] &= 2N \mod 4(N+1) \,.
  \end{aligned}
\end{equation}

Now that we have enumerated the relevant operators of interest, we turn to an analysis of the superpotential deformations of the SCFT. We first consider a superpotential of the form 
\begin{equation}\label{eqn:mesondef}
    W = M_q \,,
\end{equation}
where $M_q$ is a mesonic operator belonging to the interacting SCFT. Due to the charge of the meson under the discrete axial symmetry, given in equation \eqref{eqn:Mqcharges}, the superpotential breaks the axial symmetry:
\begin{equation}
    \mathbb{Z}_{4N + 4} \,\, \rightarrow \,\, \mathbb{Z}_{\gcd(2q, 4N+4)} \,.
\end{equation}
We are especially interested in deformations which preserve the mixed anomaly between the (unbroken subgroup of the) axial zero-form symmetry and the $\mathbb{Z}_2$ one-form symmetry. From equation \eqref{eqn:macadamia}, we can see that in the $\mathbb{Z}_2$ fractional instanton background, the $\mathbb{Z}_{\gcd(2q, 4N+4)}$ axial transformation with parameter $\ell$ transforms the partition function by the phase
\begin{equation}
    \exp\left( 2\pi i \frac{N(2N-1)(4N+4)\ell}{2\gcd(2q, 4N+4)} \int \frac{\mathcal{P}(B)}{2} \right)\,.
\end{equation}
where we recall that, in this section, we are interested in anomalies that exist when the QFT is defined on a spin background. Therefore, a mixed anomaly exists after the mesonic deformation by operator $M_q$ whenever $N$ is odd and
\begin{equation}
  N\text{ is odd }\quad\&\quad\frac{N+1}{\gcd(q, 2N + 2)} \not\in\mathbb{Z} \,.
\end{equation}
In fact, there are various theories where these conditions are satisfied. One of the simplest explicit examples where the mixed anomaly is nontrivial is the $\tr (Q\widetilde{Q})^4$ deformation of the infrared SCFT associated with the $SU(10)$ gauge theory with symmetric matter. We note that these mesonic deformations do not break all of the Abelian symmetries since the $U(1)_B$ survives, however, it is known that baryonic symmetries do not mix to form the superconformal R-symmetry \cite{Intriligator:2003jj}, and therefore the infrared of such deformed SCFTs is not believed to be superconformal.

Even when the mixed axial-electric anomaly is trivial after deformation, this is only one check of whether or not it is consistent for the infrared theory to be trivially gapped. We can also check the remnants of the discrete anomalies in equation \eqref{eqn:symmatdiscINTER} after deformation -- if any are nontrivial then we learn that the deformed theory is not trivially gapped. The remaining $\IZ_{\gcd(2q, 4N+4)}=\IZ_{2\rho}$ symmetry after the superpotential deformation has 't Hooft anomalies characterized by the two integers 
\begin{equation}
\begin{aligned}
    &(2\rho+1)(2\rho+2)\times\left(2\times N(2N+1)+\sum_{i=1}^s(-2i)^3\right) \\ &\qquad\qquad \equiv-8\left((\rho-1)(\rho+1)+2\rho^2s(s+1)-s^2(s+1)^2\right)\mod\,12\rho\,,\\
    &2\times\left(2\times N(2N+1)+\sum_{i=1}^s(-2i)\right)\equiv 2\left(2-s(s+1)\right)\mod\,2\rho \,.
\end{aligned}
\end{equation}
Here we remind the reader that $s\equiv\left\lfloor\frac{N+1}{3}\right\rfloor$ is the power of the (apparently) heaviest decoupling operator $\tr (Q\widetilde{Q})^s$. The $s$-dependent terms in the anomaly coefficients remove the contribution by the free (massless) chiral multiplets to the anomaly coefficients. A simple numerical check finds the anomaly is nontrivial for arbitrary $N$ and $\rho$. Therefore, the 't Hooft anomaly of the zero-form symmetry enforces that the deformed theory is nontrivially gapped.

In fact, the remnant of the discrete axial symmetry for which we have just discussed the anomaly is not the only discrete symmetry after deformation. The deformation also breaks the $U(1)_R$ to a discrete subgroup. When we turn on equation \eqref{eqn:mesondef}, at the IR fixed point, the R-charge for the $Q, \widetilde{Q}$ becomes $R[Q]=R[\widetilde{Q}]=1/q$. Therefore, the $U(1)_R$ symmetry is broken due to the ABJ anomaly
\begin{align}
    r \equiv 2 \tr RGG = 2 \left( 2N + 2 \left(\frac{1}{q}-1 \right)\frac{2N+2}{2} \right) = \frac{4(N+1-q)}{q} \,. 
\end{align}
When $r$ is an integer, the $U(1)_R$ symmetry breaks to $\IZ_r$. When $r$ is not an integer, but a rational number given as $r=s/t$ with coprime $s, t$, we find that the anomaly phase factor in \eqref{eqn:ABJ} trivialize for the angle $\alpha = 2\pi m / (s/t) = 2\pi m t/s $. Since $(s, t)=1$, we have $m=0, 1, \ldots, s-1$ so that the unbroken symmetry becomes $\IZ_s = \IZ_{rt}$. When the $1/\nu$-fractional instantons are present, we repeat the same analysis to find the angles $\alpha = 2\pi m / (r \nu)$. 

Let us consider some examples. First, consider the deformation by the operator $\tr Q\widetilde{Q}$: this is simply a mass deformation and the infrared behavior is known to be the same as pure $SU(2N)$ $\mathcal{N}=1$ super-Yang--Mills.\footnote{At the $W=0$ fixed point, the quadratic operator gets decoupled and should be replaced by its flip field. Here we add this the coupling/mass at higher energy scale than the the dynamical scale at which $Q\tilde{Q}$ becomes free and gets decoupled. This alters the trajectory of RG flow.} The $\mathbb{Z}_{4N+4}$ axial symmetry is broken to $\mathbb{Z}_2$, and the $U(1)_R$ is broken to $\mathbb{Z}_{4N}$. We have to consider the global structure as written in equation \eqref{eqn:globalSU2N}. If $N$ is odd, we have
\begin{equation}
    \frac{\mathbb{Z}_{4N} \times \mathbb{Z}_2}{\mathbb{Z}_2} \, = \, \mathbb{Z}_{4N} \,,
\end{equation}
whereas if $N$ is even then we naively find that the global symmetry is 
\begin{equation}
    \mathbb{Z}_{4N} \times \mathbb{Z}_2 \,,
\end{equation}
however, taking into account equation \eqref{eqn:pineapple}, we notice that the $\mathbb{Z}_2$ factor is contained inside of the center of the gauge group. Thus, in either case, the global symmetry is $\mathbb{Z}_{4N}$, 
as expected from the super-Yang--Mills theory. We can see that the $\mathbb{Z}_2$ one-form symmetry has a mixed anomaly with this $\mathbb{Z}_{4N}$. In fact, in this case, the $\mathbb{Z}_2$ one-form symmetry enhances to $\mathbb{Z}_{2N}$ which gives rise to the standard $\mathbb{Z}_{4N}$--$\mathbb{Z}_{2N}$ mixed zero-form/one-form anomaly of pure $\mathcal{N}=1$ SYM; the anomaly is saturated by $2N$ degenerate vacua in the infrared.

For $q=2$, $U(1)_R$ is broken to $\IZ_{2N-2}$ which has a non-vanishing anomaly with the $\IZ_2$ one-form symmetry. Therefore, the IR theory for $q=2$ cannot be trivially gapped, for all $N$. As we discussed before, it has an additional 't Hooft anomaly for even $N$. 

For $q=4$, $U(1)_R$ is broken to $\IZ_{N-3}$. Now, this zero-form symmetry can have a mixed anomaly with the $\IZ_2$ one-form symmetry if $N$ is odd since upon gauging $\IZ_2$ one-form symmetry, we can activate fractional instantons of half-integer instanton numbers so that the zero-form symmetry is broken to $\IZ_{\frac{N-3}{2}}$. When $N$ is even, the instantons are still integral and there is no mixed anomaly for this case. 

Finally, we can consider superpotential deformations via the baryonic operators in equation \eqref{eqn:baryons}, instead of the mesonic operators. Such deformations break the baryonic symmetry $U(1)_B$. Consider turning on the superpotential 
\begin{equation}\label{eqn:barry}
    W = B \,,
\end{equation}
in the infrared SCFT. Before deformation the classical global symmetry in the UV can be written as 
\begin{equation}
    U(1)_S \times U(1)_{\widetilde{S}} \times U(1)_R \,,
\end{equation}
where the $U(1)_S$, $U(1)_{\widetilde{S}}$ act only on the $Q$, $\widetilde{Q}$, respectively; for the baryonic deformation this is a more useful presentation than that given in equation \eqref{eqn:globalSU2N}. After deformation, the superpotential in equation \eqref{eqn:barry} and the ABJ anomaly breaks the global symmetry to
\begin{equation}
    U(1)_R \times \mathbb{Z}_{2N + 2} \,,
\end{equation}
where the $\mathbb{Z}^A = \mathbb{Z}_{2N + 2}$ acts only on $\widetilde{Q}$ and not $Q$. To determine the existence of a mixed anomaly, we note that $\tr \widetilde{S}GG$ is nontrivial, as determined from the UV, and thus there exists a mixed anomaly with the $\mathbb{Z}_2$ one-form symmetry in the usual way by studying the fractional instanton background. 
 
In fact, this deformation generally leads to a superconformal theory. There is an unbroken $U(1)_R$ R-symmetry under which the charges of the operators can be determined from
\begin{equation}
    R[Q] = \frac{1}{N} \qquad \text{ and } \qquad R[\widetilde{Q}] = \frac{N-1}{N(N+1)} \,.
\end{equation}
Certain mesonic operators become unitarity-violating under this R-symmetry, and must be decoupled via the introduction of flip fields; after this process the resulting IR is superconformal. This infrared SCFT also has relevant operators, coming from either the mesons or the other baryon; we can further deform the theory via one of these operators to obtain a theory without a $U(1)_R$ R-symmetry, and track the mixed anomaly after that subsequent deformation. The $N=2$ case is special since 
\begin{equation}
    R[\widetilde{B}] = \frac{2}{3} \,,
\end{equation}
and thus the other baryon decouples after the deformation in equation \eqref{eqn:barry}. We leave the enumeration of further deformations and their anomalies for the reader.

\subsection{\texorpdfstring{$SU(N)$}{SU(N)} with adjoint matter}

As our next example, we consider $SU(N)$ gauge theory coupled to adjoint-valued chiral multiplets. This theory flows to an SCFT in the infrared if there are two or three adjoint chiral multiplets; the latter, with the appropriate cubic superpotential, is $\CN=4$ super-Yang--Mills. Hence, we consider in this subsection the former case: $SU(N)$ gauge theory coupled to two adjoint-valued chiral multiplets.

The adjoint matters $Q$ preserve the center one-form $\IZ_N$ symmetry, while it breaks the axial zero-form $U(1)_A$ into $\IZ_{4N}$. The R-charge of $Q$ is fixed as $1/2$ by the anomaly-free condition. Consequently, the relevant operators at the IR fixed point of the gauge theory are given by the Casimir operators
\begin{align}
    \tr Q_IQ_J\,,\quad \tr Q_IQ_JQ_K \,,
\end{align}
where $I,J,K=1,2$ are the flavor indices. 

We first consider the mass deformation $\tr Q_IQ_J$. If we turn on the mass term of $Q_I$ only,
\begin{align}
    W=\tr Q_I^2\,,
\end{align}
the zero-form symmetry is broken into 
\begin{align}
    \frac{SU(2)\times \IZ_{4N}\times U(1)_R}{\IZ_2 \times \mathbb{Z}_2}\quad\rightarrow\quad \IZ_{2N}\times \IZ_2\times U(1)_R' \,,
\end{align} 
where the new R-symmetry $U(1)_R'$ charges the $Q_{J\neq I}$ by 1 whereas $Q_I$ remains neutral. The $\IZ_{2N}$ part rotates the $Q_{J\neq I}$, while the $\IZ_2$ part of the zero-form symmetry rotates $Q_I$ by the angle $\theta=0,\,\pi$ and is generated by
\begin{align}
    \left(\left(\begin{matrix}
    i & 0\\
    0 & -i
    \end{matrix}\right),\, \omega^N\right)
    \in (SU(2) \times \IZ_{4N})/\IZ_2
\end{align}
with respect to the $4N$-th root of unity $\omega$. This is consistent with the fact that the $\IZ_2$ does not act faithfully after $Q_I$ is integrated out. Hence, the $\IZ_2$ does not have mixed anomaly with the one-form $\IZ_2$ symmetry.
It follows that the remaining symmetries are those of the $\CN=2$ super-Yang--Mills theory (without the corresponding superpotential).

Now we consider giving mass to both of chiral matters by
\begin{align}
    W=\tr Q_1Q_2\,.
\end{align}
The preserved zero-form symmetry is
\begin{align}
    \frac{SU(2)\times \IZ_{4N}\times U(1)_R}{\IZ_2 \times \mathbb{Z}_2}\quad\rightarrow\quad SU(2)/\IZ_2\times\IZ_2\times U(1)_R'\,,
\end{align}
where the new R-symmetry $U(1)_R'$ does not rotate $Q_{1,2}$. Similarly, the $\IZ_2$ zero-form symmetry does not have mixed anomaly with the one-form $\IZ_2$ symmetry, which is consistent with the fact that such $\IZ_2$ does not act faithfully in the IR as well as the $SU(2)$. There is another way of giving masses to $Q_1$ or $Q_2$ by turning on 
\begin{align}
    W=\tr Q_1^2+\tr Q_2^2\,.
\end{align}
In this case, $SU(2)$ is further broken to $\IZ_2$, and the analysis remains the same.

We now consider cubic deformations of the form $\tr Q_IQ_JQ_K$. Then we can consider deforming the theory by the cubic superpotential 
\begin{align}
W=\tr Q_1^3\quad \text{or}\quad \tr Q_1^2Q_2\,.
\end{align}
The former breaks the zero-form symmetry by
\begin{align}
    \frac{SU(2)\times \IZ_{4N}\times U(1)_R}{\IZ_2 \times \mathbb{Z}_2}\quad\rightarrow\quad \IZ_{2N}\times\IZ_\text{gcd(3,N)}\times U(1)_R'\,.
\end{align}
Note that the $\IZ_{2N}$ part rotates $Q_2$ whereas instead of getting the $\IZ_3$, rotating $Q_1\mapsto \exp{\left[\frac{2\pi i}{3}\right]}Q_1$, we get $\IZ_{\gcd (3,N)}$ due to the ABJ anomaly. The new R-symmetry $U(1)'_R$ charges the chiral matters $Q_1$ and $Q_2$ by $1/3$ and $2/3$, respectively. 
If $\gcd (3,N)=3$, the symmetry transformation by the elements $(k_1,k_2)\in\IZ_{2N}\times \IZ_3$ generates additional phase
\begin{align}
    \exp\left[i\left(\frac{2\pi k_1}{2N}+\frac{2\pi k_2}{3}\right)(2N)\int F\wedge F\right]\,.
    \label{eq:phaseZ2NxZ3}
\end{align}
Hence, both $\IZ_{2N}$ and $\IZ_3$ (if $3\mid N$) zero-form symmetries have mixed anomalies with one-form $\IZ_N$ symmetry.
The additional phase in equation \eqref{eq:phaseZ2NxZ3} is fractional under the fractional instanton background once the $\IZ_N$ one-form symmetry is gauged. 

We now consider the other cubic superpotential deformation $\tr Q_1^2Q_2$, which preserves a different zero-form symmetry 
\begin{align}
     \frac{SU(2)\times \IZ_{4N}\times U(1)_R}{\IZ_2 \times \mathbb{Z}_2}\quad\rightarrow\quad \IZ_{2N}\times\IZ_{\gcd (3,N)}\times U(1)_R'\,.
\end{align}
The $\IZ_{2N}$ now acts on both chiral matter fields instead of just $Q_2$; the action is
\begin{align}
    (Q_1,Q_2)\mapsto (\omega^kQ_1,\omega^{-2k}Q_2) \,,
\end{align}
generated by the $2N$th root of unity $\omega$. There also exists the remaining axial $\IZ_3$ symmetry that rotates $Q_i\mapsto \exp\left[\frac{2\pi ik}{3}\right]Q_i$ when $\gcd(3,N)=3,$ while it is broken by the ABJ anomaly when $\gcd(3,N)=1$. The preserved R-symmetry $U(1)_R'$ charges $Q_1$ by 1, whereas $Q_2$ is not charged. Correspondingly, the operators 
\begin{align}
    \tr Q_2^2\,,\quad\tr Q_2^3\,,\quad\cdots\,,\quad\tr Q_2^N \,,
\end{align} 
decouple from the rest of the theory. There exist mixed anomalies between the $\IZ_N$ one-form symmetry and both discrete zero-form symmetries $\IZ_{2N}$ and $\IZ_{\gcd (3,N)}$.

\subsection{\texorpdfstring{$Spin(2N)$}{Spin(2N)} with symmetric matter}

Now that we have explored several examples with special unitary gauge groups, we turn to an example where the gauge group is orthosymplectic. We consider $G = Spin(2N)$ gauge theory with a single chiral multiplet in the rank-two symmetric traceless representation.\footnote{We can consider the same matter spectrum for $Spin(2N+1)$ gauge theory. However, following Table \ref{tbl:mixedanom}, this theory does not have any mixed anomaly between the one-form symmetry and the axial symmetry. As such, we do not study this case here.} As the chiral multiplet does not screen any of the center of $G$, the one-form symmetry of the theory is $Z(G)$. The classical zero-form symmetries are 
\begin{equation}
    U(1)_R \times U(1)_A \,,
\end{equation}
and the axial symmetry is broken as $U(1)_A \rightarrow \mathbb{Z}_{4N+4}$. The infrared R-charge of the chiral multiplet, $S$, is fixed by gauge-anomaly cancellation to be
\begin{equation}
    R[S] \,=\, \frac{2}{N+1} \,.
\end{equation}
We define the following operators
\begin{equation}
    S_q = \tr S^q \,.
\end{equation}
We can see that when 
\begin{equation}
    q \leq \left\lfloor \frac{N+1}{3} \right\rfloor \qquad \Rightarrow \qquad R[S_q] \leq \frac{2}{3} \,,
\end{equation}
and thus these operators decouple from the theory along the flow into the infrared. To automatically decouple these operators, we introduce a flip field for each such operator, and then the infrared theory will be the interacting SCFT of interest; however, we will suppress these flip fields in the discussion that follows. The $S_q$ with 
\begin{equation}\label{eqn:blah}
    \left\lfloor\frac{N+1}{3}\right\rfloor < q < N+1 \,,
\end{equation}
are relevant operators belong to the infrared SCFT. These SCFTs have been explored in \cite{Agarwal:2020pol}.

Before exploring the deformations of such SCFTs, we first briefly review the anomalies involving the discrete axial symmetry. As described in Section \ref{sec:mixedanom}, the structure of the mixed anomaly between the zero-form and one-form symmetries depends on the parity of $N$. For $N = 2M + 1$, we see from equation \eqref{eqn:spin4M2mixanom} that, on manifolds admitting spin structure, only transformations belonging to the 
\begin{equation}
    \mathbb{Z}_{N+1} \subset \mathbb{Z}_{4N+4} \,,
\end{equation}
subgroup of the axial symmetry leave the partition function invariant. For $N = 2M$, it is the subgroup
\begin{equation}
    \mathbb{Z}_{2N+2} \subset \mathbb{Z}_{4N+4} \,,
\end{equation}
that leaves the partition function invariant, as observed from equation \eqref{eqn:spin4Nshift}. In both situations, this demonstrates the presence of a mixed axial-electric anomaly. We can also study the $\mathbb{Z}_{4N+4}^3$ and mixed $\mathbb{Z}_{4N+4}$-gravitational anomalies using equation \eqref{eq:disanom}. These two anomalies are captured by the following integers, respectively:
\begin{equation}\label{eqn:spinNdiscanom}
    \begin{aligned}
        (4N+5)(4N+6)\times N(2N+1) \,\,&\mod 24(N+1) \,,  \\
          2\times N(2N+1) \,\,&\mod 4(N+1) \,.
    \end{aligned}
\end{equation}
It is easy to see that these 't Hooft anomalies are not trivial, since none of the above numbers is multiplier of $N+1$. This agrees with the expectation that the theory will flow to an IR SCFT plus a collection of free chiral multiplets \cite{Agarwal:2020pol}; note that the anomalies in equation \eqref{eqn:spinNdiscanom} include the contributions from the decoupled $S_q$ operators. We can also capture the anomaly coefficients of the IR SCFT only by including the contributions from the flip fields as well. They end up being
\begin{equation}
    \begin{aligned}
           2N(8N+7)-(4N+5)(4N+6)\left(\frac{1}{4}s^2(s+1)^2-1\right)&\,\,\mod\,24(N+1) \,,\\
         2N(2N+1)-s(s+1)+2&\,\,\mod\,4(N+1)\,,
    \end{aligned}
\end{equation}
where $s\equiv\left\lfloor\frac{N+1}{3}\right\rfloor$. By a straightforward numerical check, we find that the anomaly coefficients are nontrivial, obstructing the theory to be trivially gapped in the IR. 

Now, we are ready to examine the deformations of these SCFTs via the $S_q$ operators. Consider the deformation 
\begin{equation}
    W = S_q \,,
\end{equation}
where $q$ is in the range given in equation \eqref{eqn:blah}. This deformation explicitly breaks the axial zero-form symmetry as
\begin{equation}\label{eqn:bowmakers}
    \mathbb{Z}_{4N + 4} \,\,\rightarrow\,\, \mathbb{Z}_{\gcd(4N + 4, q)} \,,
\end{equation}
as well as breaking the $U(1)_R$ R-symmetry. To determine the existence of a mixed axial-electric anomaly after the deformation, we again separate the discussion depending on the parity of $N$. If $N = 2M + 1$ then a mixed anomaly exists only if 
\begin{equation}
    4 \,\nmid\, \frac{4N + 4}{\gcd(4N + 4, q)} \,,
\end{equation}
that is, if the generator of the $\mathbb{Z}_{\gcd(4N + 4, q)}$ subgroup in equation \eqref{eqn:bowmakers} is not a multiple of four. A low rank example of such a theory with a mixed anomaly is $Spin(10)$ gauge theory deformed in the infrared by the $S_4$ operator. A similar analysis holds when $N = 2M$: we find that a mixed anomaly exists after deformation by $S_q$ when $q$ is such that
\begin{equation}
    2 \,\nmid\, \frac{4N + 4}{\gcd(4N + 4, q)} \,.
\end{equation}
In this case, a simple example is a $Spin(12)$ gauge theory deformed by the $S_4$ operator; we discuss this example explicitly anon.

Similarly, we can constrain the infrared behavior by studying the anomalies of the remnant $\mathbb{Z}_{\gcd(4N + 4, q)}$ zero-form symmetry. We define
\begin{align}
    \ell = \gcd(4N + 4, q)
\end{align}
for simplicity. The 't Hooft anomalies for this zero-form symmetry are
\begin{equation}\label{eqn:shrugemoji}
    \begin{aligned}
        (\ell+1)(\ell+2) \left(N(2N+1)-\frac{1}{4}s^2(s+1)^2+1\right)\,\,&\mod 6\ell\,, \\
         2\times \left(N(2N+1)-\frac{1}{2}s(s+1)+1\right)\,\,&\mod\ell\,.
    \end{aligned}
\end{equation}
The anomaly coefficients are nontrivial for generic $N$ and $q$, providing additional constraints on the theory being trivially gapped in the IR.

\subsubsection*{An example: \texorpdfstring{$Spin(12)$}{Spin(12)} with symmetric matter}

To highlight the analysis in the subsection, we now explore an explicit example. Let us consider a $G = Spin(12)$ gauge theory together with a single chiral multiplet $Q$ in the $\bm{77}$, i.e., in the $2$-symmetric representation. The infrared R-charge of the chiral multiplet is fixed by gauge anomaly cancellation to be
\begin{equation}
    R[Q] = \frac{2}{7} \,,
\end{equation}
and the classical axial $U(1)$ flavor symmetry is broken by the ABJ anomaly to $\mathbb{Z}_{28}$
where the charge of the chiral multiplet under this residual discrete symmetry is
\begin{equation}
    A[Q] = 1 \,.
\end{equation}

We can see that the operator $Q^2$ has conformal dimension less than one, and so it must be flipped. We introduce a new chiral multiplet $X_{Q^2}$ which has 
\begin{equation}
    R[X_{Q^2}] = \frac{10}{7} \,, \qquad F[X_{Q^2}] = - 2 \,,
\end{equation}
and add a superpotential term $W = X_{Q^2}Q^2$ in the ultraviolet. Then, the infrared theory is the same as the infrared of our original theory up to decoupled operators, and the following relevant operators exist in the infrared:\footnote{Note: these are independent operators, so $Q^6 \neq Q^2 Q^2 Q^2 \neq Q^3 Q^3$, etc. In the notation of the earlier part of this section, these operators, excluding the flip field, are called, respectively, $S_3$, $S_4$, $S_5$, $S_3^2$, $S_6$.}
\begin{equation}\label{eqn:spin12rel}
    Q^3\,,\,\, Q^4\,,\,\, Q^5\,,\,\, (Q^3)^2\,,\,\, Q^6\,,\,\, X_{Q^2} \,.
\end{equation}
Since the 2-symmetric matter is uncharged under the $\mathbb{Z}_2 \times \mathbb{Z}_2$ center of $Spin(12)$, the chiral multiplet does not screen any of the one-form symmetry. There is an anomaly between the $\mathbb{Z}_{28}$ discrete zero-form symmetry and the $\mathbb{Z}_2 \times \mathbb{Z}_2$ one-form symmetry. Under a $\mathbb{Z}_{28}$ one-form symmetry transformation, the partition function transforms as in equation \eqref{eqn:spin4Nshift}. Therefore, on a manifold with spin structure, a $\mathbb{Z}_{28}$ transformation parametrized by an odd integer $\ell$ leads to a nontrivial phase of the partition function.

Turning on a superpotential deformation by any of the relevant operators in equation \eqref{eqn:spin12rel} triggers a renormalization group flow that leads to an infrared theory where the $U(1)_R$ is anomalous, and thus conformal symmetry appears to be broken. Under the superpotential deformation by a relevant operator $\mathcal{O}$, the discrete flavor symmetry is broken via:
\begin{equation}
    \mathbb{Z}_{28} \,\,\rightarrow\,\, \mathbb{Z}_{\gcd(A[\mathcal{O}], 28)} \,.
\end{equation}
Therefore, the deformations triggered by $X_{Q^2}$, $Q^4$, and $Q^6$ preserve a discrete zero-form symmetry which is $\IZ_2$, $\IZ_4$, and $\IZ_2$, respectively. Some of these deformations preserve a part of the mixed zero-form/one-form anomaly.\footnote{Deformations by the $Q^3$ and $Q^5$ operators break all of the discrete zero-form symmetry, and thus there can be no mixed anomaly with the one-form symmetry.} In particular, the $X_{Q^2}$ and $Q^6$ deformations break the $\mathbb{Z}_{28}$ to the subgroup generated by the $\ell = 14$ element which is even and thus there is no remaining mixed anomaly; however the $Q^4$ deformation breaks the $\mathbb{Z}_{28}$ to the $\mathbb{Z}_4$ subgroup generated by the \emph{odd} $\ell = 7$ element.

In conclusion, the $Q^4$ deformation of infrared SCFT coming from $Spin(12)$ gauge theory with one chiral multiplet in the $\bm{77}$ representation, has no $U(1)_R$ symmetry visible from the ultraviolet, and has a mixed anomaly between a $\mathbb{Z}_4$ zero-form symmetry and the $\mathbb{Z}_2 \times \mathbb{Z}_2$ one-form symmetry. The absence of the $U(1)_R$ rules out a symmetry preserving gapless phase, and the existence of the mixed anomaly obstructs a trivially gapped phase. The result of \cite{Cordova:2019bsd} together with the $Spin(12)$ generalization of the anomaly TFT in equation \eqref{eqn:anomTFT}, indicates that there does not exist a symmetry-preserving gapped phase consistent with this mixed anomaly. Therefore, if the theory confines in the infrared, the $\mathbb{Z}_4$ zero-form symmetry must be spontaneously broken; the minimal breaking pattern consistent with the anomaly is $\mathbb{Z}_4 \rightarrow\mathbb{Z}_2$,
with two degenerate vacua. We leave a more detailed analysis of the vacuum structure of this theory for future work.

\subsection{\texorpdfstring{$Spin(N)$}{Spin(N)} with symmetric and adjoint matter}

To conclude, we consider an example where there are several chiral multiplets which are not in the same (or conjugate) representations of the gauge group. Let $G = Spin(N)$ and let $A$ denote a chiral multiplet in the adjoint representation of $G$, and $S$ a chiral multiplet in the two-index traceless symmetric representation. The matter spectrum does not screen any of the center of the gauge group, so the one-form symmetry is $Z(G)$. The ABJ anomaly breaks one linear combination of the classical Abelian symmetries to $\mathbb{Z}_{4N}$.

There exists a mixed anomaly between the $\mathbb{Z}_{4N}$ zero-form symmetry and the one-form symmetry when $N$ is even; this can be seen directly via consultation with Table \ref{tbl:mixedanom}, as well as in the previous subsection. We can also study the 't Hooft anomalies of the zero-form symmetries involving the $\mathbb{Z}_{4N}$; we find that they are captured by the integers
\begin{equation}
  \begin{aligned}
    (4N+1)(4N+2)N^2\,\,&\mod 24N\,,\\
    2N^2\,\,&\mod 4N \,.
  \end{aligned}
\end{equation}
The mixed $\mathbb{Z}_{4N}$-gravitational anomaly is trivial when $N$ is even. For the cubic anomaly, we can rewrite via
\begin{equation}
    \begin{aligned}
        8N^2(N+1)(2N+1)-6N^2(2N+1)&\equiv -6N^2(2N+1) \,,\\
    \equiv -12N^2(N+1)+6N^2&\equiv 6N^2\,\,\mod 24N \,.
    \end{aligned}
\end{equation}
We can see that this anomaly is only trivial when $N$ is a multiple of four. 

The SCFT that arises in the infrared of this $Spin(N)$ gauge theory has several relevant operators. Due to the existence of chiral multiplets in different representations, the superconformal R-symmetry is not fixed via anomaly-cancellation, but instead it must be determined from $a$-maximization; this leads to the operators having irrational R-charges. For $G = Spin(7)$, we find
\begin{equation}
    (R[S], R[A]) \sim (0.640501,0.647098) \,,
\end{equation}
and which decrease monotonically with $N$, and in the large $N$ limit we find that they asymptote as follows:
\begin{equation}
    (R[S], R[A]) \,\,\xrightarrow{\,\,N \rightarrow \infty\,\,}\,\, \left(\sfrac{1}{2},\sfrac{1}{2}\right) \,. 
\end{equation}
We note that these are charges are sufficiently large that no operators decouple along the flow into the infrared, so we do not need to introduce any flip fields. The relevant operators, for any $N$, are thus of the form
\begin{equation}
    \mathcal{O}_{q_S, q_A} = S^{q_S} A^{q_A} \qquad \text{ where } \qquad q_S + q_A = 2, 3 \,.
\end{equation}

The superpotential deformation of the infrared SCFT by the relevant operator $\mathcal{O}_{q_S, q_A}$ breaks the zero-form symmetry
\begin{equation}
    \mathbb{Z}_{4N} \,\, \rightarrow \,\,\mathbb{Z}_{\ell} \qquad \text{ where } \qquad \ell = \gcd(q_S + q_S, 4N) \,.
\end{equation}
The conditions for the existence of a mixed anomaly after deformation depend on whether $N = 4M + 2$ or $N = 4M$ due to the presence of quarter or half instantons, respectively. There exists a mixed anomaly whenever 
\begin{equation}
  \begin{aligned}
    4 \,\nmid\, \frac{4N}{\ell} \qquad &\text{ if }\,\, N = 4M + 2 \,, \\
    2 \,\nmid\, \frac{4N}{\ell} \qquad &\text{ if }\,\, N = 4M \,, \\
  \end{aligned}
\end{equation}
and, since there was no mixed anomaly in the undeformed theory, there is no mixed anomaly when $N$ is odd. We can also compute the cubic and mixed-gravitational anomalies for the $\mathbb{Z}_\ell$; we find that they are
\begin{equation}
\begin{aligned}
     (\ell+1)(\ell+2)N^2\,\,&\mod 6\ell\,,  \\
     2N^2\,\,&\mod\ell \,.
\end{aligned}
\end{equation}
Whatever the nature of the infrared after this superpotential deformation, the degrees of freedom must saturate these three anomalies. While a single deformation of the form $\mathcal{O}_{q_S, q_A}$ does not lead to an infrared theory without a $U(1)$ R-symmetry, a sequence of relevant deformations may continue to preserve the anomalies, and thus lead to an infrared theory that is not a unique gapless vacuum. We leave the detailed analysis of the landscape of deformations and their IR fates for future work. 

\section{Gauged Argyres--Douglas theories and dualities}\label{sec:ADtheories}

Throughout this paper thus far, we have considered Lagrangian $\mathcal{N}=1$ simple gauge theories and their superpotential deformations. However, a priori, it is an unnecessary restriction to focus only on Lagrangian field theories. Instead, we may consider a simple gauge group coupled to nontrivial conformal fields theories (sometimes referred to as strongly-coupled matter). Such theories can also have mixed zero-form/one-form anomalies, and we can analyze whether the infrared fate of such theories after certain superpotential deformations are constrained by the existence of these mixed anomalies.

In particular, in this section, we focus on superpotential deformations of SCFTs obtained via the diagonal gauging of the common flavor symmetry of a collection of 4d $\mathcal{N}=2$ Argyres--Douglas theories. We consider the Argyres--Douglas theories known as $\mathcal{D}_p(G)$, where $G$ is a simple and simply-connected Lie group and $p \geq 2$ is an integer \cite{Xie:2012hs, Cecotti:2012jx, Cecotti:2013lda, Wang:2015mra}. For generic values of $p$, such theories have a $G$ flavor symmetry, though this can be enhanced for particular values of $p$. Diagonal $\mathcal{N}=1$ gauging of the $G$ flavor symmetry for a collection of SCFTs, 
\begin{equation}
    \left\{\mathcal{D}_{p_i}(G) \, | \, i = 1, \cdots, M\right\} \,,
\end{equation}
was studied in \cite{Kang:2021ccs}, in particular to classify when the resulting 4d $\mathcal{N}=1$ theory flows in the infrared to an SCFT. Particular focus was placed on the subset of such theories that have identical central charges, which is correlated with a variety of interesting physical features as recently explored in \cite{Buican:2020moo,Kang:2021lic,Kang:2021ccs,Kang:2022vab,Kang:2023dsa,LANDSCAPE}.

Exploring the superpotential deformations of such 4d $\mathcal{N}=1$ SCFTs has led to novel dualities between deformations of non-Lagrangian theories and Lagrangian QFTs \cite{Kang:2023dsa}; see also \cite{Bajeot:2023gyl,Maruyoshi:2023mnv,Benvenuti:2024glr,Hwang:2024hhy}. In particular, in \cite{Kang:2023dsa}, a variety of evidence was given that the diagonal gauging of the $SU(2N+1)$ flavor symmetry of three copies of $\mathcal{D}_2(SU(2N+1))$ flows in the infrared to a point on the ($\mathcal{N}=1$)-preserving conformal manifold of $\mathcal{N}=4$ super-Yang--Mills with gauge group $SU(2N+1)$. This evidence previously provided was on the level of the gauge algebra, as opposed to the global form of the gauge group; alternatively said, we demonstrated a duality between relative quantum field theories, before the choice of polarization was made. In this section, we study the mixed axial-electric anomaly on both sides of the duality, and thus provide evidence for the duality that is sensitive to the global form of the gauge group. Similarly, we explore the duality between the diagonal gauging of two copies of the $\mathcal{D}_3(G)$ theory and the two-adjoint gauge theory \cite{LANDSCAPE} from this anomaly perspective. We do not attempt the answer the question of the constraints on the IR fate of $U(1)_R$-breaking deformations in this landscape of non-Lagrangian $a=c$ theories. The landscape of such theories is itself expansive \cite{LANDSCAPE}, and we leave the exploration of the mixed zero-form/one-form and discrete anomalies across that landscape for future work. 

\subsection{The mixed axial-electric anomaly}

We are interested in studying theories $\mathcal{D}_p(G)$ that satisfy $\gcd(p, h_G^\vee) = 1$, where $h_G^\vee$ is the dual Coxeter number of the Lie algebra associated with $G$. The $\gcd$ condition fixes that the flavor symmetry is precisely $G$ -- it is not enhanced -- and the flavor central charge is:\footnote{For the standard properties of the $\mathcal{D}_p(G)$ theories, with our normalizations and conventions, we refer the reader either to the previous work of the current authors \cite{Kang:2021ccs}, or to \cite{Couzens:2023kyf}.}
\begin{equation}
    k_G = \frac{2(p-1)}{p} h_G^\vee \,.
\end{equation}
It is believed that the global form of the flavor group, when $\gcd(p, h_G^\vee) = 1$ is the simply-connected Lie group $G$ with ADE Lie algebra $\mathfrak{g}$. Some evidence for this comes from the analysis of protected sectors of the theory; for example, the refined Schur indices of such $\mathcal{D}_p(G)$ theories involve only representations of $G$ with trivial n-ality \cite{Song:2017oew}. In other words, when $G$ is gauged, the strongly-coupled matter does not screen any of the topological Gukov--Witten operators in the center of $G$.

To highlight the structure of the anomalies under gauging of the flavor symmetry, in this subsection, we assume that we are ($\mathcal{N}=1$)-gauging the flavor symmetry of a single such $\mathcal{D}_p(G)$. In fact, such a gauging does not flow in the infrared to an interacting SCFT \cite{Kang:2021ccs} as chiral operators in the gauged theory acquire negative R-charges which generate a dynamical runaway superpotential similar to the Affleck--Dine--Seiberg superpotential \cite{Affleck:1983mk}. However, this assumption is sufficient to highlight the relevant features, before we move on to study specific examples with multiple copies of $\mathcal{D}_p(G)$ in Sections \ref{sec:fox} and \ref{sec:hound}.

As an $\mathcal{N}=2$ SCFT, $\mathcal{D}_p(G)$ has an R-symmetry which is $SU(2)_I \times U(1)_r$, 
and after gauging $G$ via an $\mathcal{N}=1$ vector multiplet, this is broken to $U(1)_{R_0} \times U(1)_F$, where the generators of these two Abelian factors are
\begin{equation} \label{eq:R0F}
    R_0 = \frac{1}{3} r + \frac{4}{3} I_3 \ , \quad F = -r + 2I_3 \,,
\end{equation}
respectively. Here, we have used $r$ to denote the generator of $U(1)_r$ and $I_3$ for the generator of the Cartan of the $SU(2)_I$. We have chosen the normalization of the $U(1)_r$ charge such that the Coulomb branch operators have conformal dimension $\Delta = \sfrac{r}{2}$. There is a mixed anomaly between the $U(1)_F$ global symmetry and the newly-introduced $G$ gauge symmetry:
\begin{equation}
    \tr F GG = \frac{p-1}{p}h^\vee_G \,.
\end{equation}
This reflects the fact that the $U(1)_F$ suffers from an ABJ anomaly and is broken to a discrete subgroup. Under a $U(1)_F$ transformation, parametrized by an angle $\alpha$, the partition function transforms as
\begin{equation}
    Z[B] \,\, \rightarrow \,\, \exp\left( i \alpha (2 \tr F GG) \nu_\text{inst} \right) Z[B] \,\,=\,\, \exp\left( i \alpha \frac{2(p-1)}{p} h_G^\vee \nu_\text{inst} \right) Z[B] \,,
\end{equation}
where $\nu_\text{inst}$ as usual is the instanton number and we have written the dependence of the partition function on the background two-form gauge field $B$ capturing the one-form symmetry. As we can see, when no background $B$ is turned on and $\nu_\text{inst}$ is an integer, only angles $\alpha$ of the form
\begin{equation}
    \alpha = \frac{2\pi p m}{2(p-1)h_G^\vee} \,,
\end{equation}
where $m$ is an integer, leave the partition function invariant. Therefore the $U(1)_F$ is broken to a discrete subgroup:
\begin{equation}
    U(1)_F \,\,\rightarrow\,\, \mathbb{Z}_M \,,\qquad\qquad M = \begin{cases}
        2(p-1)h_G^\vee &\text{ if $p$ odd, } \\
        (p-1)h_G^\vee &\text{ if $p$ even. }
    \end{cases}
\end{equation}
In deriving this result, we recall that we are assuming $\gcd(p, h_G^\vee) = 1$. 

When turning on a nontrivial center-valued background field $B$ for the one-form symmetry, the instanton number fractionalizes, and this induces a mixed anomaly between the $\mathbb{Z}_M$ zero-form symmetry and the $Z_G$ one-form symmetry. If we consider now $G = SU(N)$, the anomalous phase of the partition function is
\begin{equation}\label{eqn:wcoyote}
    \exp\left( 2 \pi i m \,  \frac{(N-1)}{2N} \int_X \mathcal{P}(B) \right) \,,
\end{equation}
where $m$ is an arbitrary integer. It is clear that there always exists a mixed anomaly since $\gcd(N-1, N) = 1$. We may also be interested in knowing if there exists a collection of axial rotations which lead to a trivial phase. When $N$ is odd, the partition function admits a trivial phase whenever
\begin{equation}
    m = N m' \,,
\end{equation}
for any integer $m'$. When $N$ is even there is a trivial phase for 
\begin{equation}
    m = 2N m' \,,
\end{equation}
with $m'$ an arbitrary integer; since $p$ must be odd to satisfy $\gcd(p, N) = 1$ we can see that $2N
\mid M$. Therefore, altogether, there exists a non-anomalous subgroup
\begin{equation}
    \mathbb{Z}_{M'} \subset \mathbb{Z}_M \,, \qquad\qquad M' = \begin{cases}
        2(p-1) &\text{ if $p$ odd and $N$ odd, } \\
        (p-1) &\text{ if $p$ even and $N$ odd, } \\ (p-1) &\text{ if $p$ odd and $N$ even. }
    \end{cases}
\end{equation}
If we assume that the QFT is only defined on manifolds that admit a spin structure, then this subgroup of the discrete axial symmetry that does not have a mixed anomaly with the one-form symmetry enhances.

\subsection{The non-Lagrangian dual of \texorpdfstring{$\mathcal{N}=4$}{N=4} SYM}\label{sec:fox}

We begin by exploring the 4d $\mathcal{N}=1$ asymptotically-free gauge theory obtained via the diagonal $(\mathcal{N}=1$)-gauging of the $SU(2n+1)$ flavor symmetries of three copies of $\mathcal{D}_2(SU(2n+1))$. This construction is depicted in Figure \ref{fig:N4SYMRGflow}. In \cite{Kang:2023dsa}, it was observed that this gauge theory is infrared-dual to $\mathcal{N}=4$ super-Yang--Mills; more specifically it flows to a point in the ($\mathcal{N}=1$)-preserving conformal manifold of $\mathcal{N}=4$ super-Yang--Mills with gauge group $SU(2n+1)$. It this section, we study the matching of the anomalies involving the discrete zero-form symmetry across this duality.

We label each of the three copies of $\mathcal{D}_p(SU(2n+1))$ by $i = 1, 2, 3$. The $\mathcal{N}=1$ gauging breaks the $\mathcal{N}=2$ R-symmetry to $U(1)_{R_0} \times U(1)_{F_i}$. For ease of notation, we let $F_i$ denote the generator of $U(1)_{F_i}$, and then we can see that the $U(1)$ generated by
\begin{equation}
    A = F_1 + F_2 + F_3 \,,
\end{equation}
suffers from an ABJ anomaly, and thus is broken to some discrete subgroup in the quantum theory. Specifically, we find that 
\begin{equation}
    U(1)_{F_1} \times U(1)_{F_2} \times U(1)_{F_3} \,\,\,\rightarrow\,\,  U(1)_{\mathcal{F}_1} \times U(1)_{\mathcal{F}_2} \times \mathbb{Z}_{3(2n+1)} \,,
\end{equation}
where the two surviving $U(1)$ factors are generated by the linear combinations
\begin{equation}
    \mathcal{F}_1 = F_1 - F_2 \,, \qquad\qquad \mathcal{F}_2 = F_2 - F_3 \,.
\end{equation}
In fact, as in Section \ref{sec:global0}, it is important to clarify the global structure of the zero-form flavor symmetry. It can be seen directly that the global structure is
\begin{equation}\label{eqn:globenc}
    \frac{U(1)_{\mathcal{F}_1} \times U(1)_{\mathcal{F}_2} \times \mathbb{Z}_{3(2n+1)}}{\mathbb{Z}_3} \,,
\end{equation}
where the diagonal $\mathbb{Z}_3$ inside of the $U(1)_{\mathcal{F}_1} \times U(1)_{\mathcal{F}_2}$ associated with the pair of elements
\begin{equation}
    (\omega, \omega^{-1}) \,,
\end{equation}
where $\omega$ is a third root of unity, is identified with the obvious $\mathbb{Z}_3$ inside of $\mathbb{Z}_{3(2n+1)}$. Finally, we recall that $U(1)_{R_0}$ is not the R-symmetry of the infrared SCFT; the superconformal R-symmetry is determined by symmetry and gauge anomaly cancellation to be the linear combination
\begin{equation}
    R = R_0 + \sum_j \varepsilon_j \mathcal{F}_j \,, \qquad\qquad \varepsilon_1 = \varepsilon_2 = -\frac{1}{3} \,.
\end{equation}

\afterpage{
\begin{figure}[t]
    \centering
    \scalebox{1}{
    \begin{tikzpicture}[declare function={rr=2*(1+0.2*sin(3*\t-10)+0.1*rnd);},xscale=.7,yscale=.35]
    \draw[scale=1.9,rotate=30,fill=yellow!30] plot[smooth cycle,variable=\t,samples at={0,45,...,315},xshift=-8mm,yshift=-36mm] (\t:rr);
    \node[anchor=south west,yshift=4mm](At) at (0,3.6) {$\mathcal{D}_{2}(SU(2n+1))$};
    \node[anchor=south west,yshift=3mm](A1) at (-3.5,-2.2) {$\mathcal{D}_{2}(SU(2n+1))$};
    \node[anchor=south west, draw,rounded rectangle, inner sep=3pt, minimum size=5mm, text height=3mm,yshift=2mm](A2) at (1,1) {$SU(2n+1)$};
    \node[anchor=south west,yshift=3mm](A3) at (3.5,-2.2) {$\mathcal{D}_{2}(SU(2n+1))$};
    \draw (A1)--(A2)--(A3);
    \draw (At)--(A2);
    \draw[->,thick] (2,-.6)--(2,-4.5) node[midway]{RG Flow$\qquad\qquad\quad$};
    \node[circle, scale=.6, fill=black] at (2,-4.9) {}; \draw[thick,xshift=50mm,yshift=-69mm,rotate=220,scale=1.3] plot [smooth, tension=.7] coordinates { (.2,-.2) (0.3,-1.4) (2,-2) (2.8,-3.2)};
    \draw[->,thick,blue] (1.8,-5.1)--(.8,-6.4) node[midway] {};
    \node[anchor=south west,xshift=23mm,yshift=-20mm] at (-.3,0) {$\mathcal{M}_{\mathcal{N}=4}$};
    \node[anchor=south west,xshift=20mm,yshift=-35mm] at (-.3,0) {$\mathcal{M}_{\mathcal{N}=1}$};
    \end{tikzpicture}
    }
    \caption{The 4d $\CN=1$ gauge theory formed out of gauging Argyres--Douglas theories which flows in the infrared to a point in the ($\CN=1$)-preserving conformal manifold ($\mathcal{M}_{\mathcal{N}=1}$) of the $\CN=4$ SYM theory with $G = SU(2n + 1)$.}
    \label{fig:N4SYMRGflow}
\end{figure}
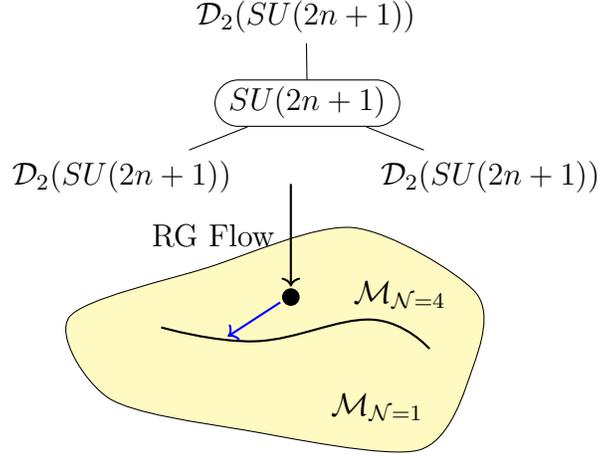
}

It is straightforward to see from equation \eqref{eqn:wcoyote} that there is a mixed anomaly between the axial $\mathbb{Z}_{3(2n+1)}$ zero-form symmetry and the $\mathbb{Z}_{2n+1}$ one-form symmetry. Furthermore, we can see that when we perform an axial transformation of the form 
\begin{equation}
    m = (2n + 1)m' \,,
\end{equation}
for $m'$ an integer, the anomalous phase trivializes. Therefore, there exists a subgroup 
\begin{equation}
    \mathbb{Z}_3 \subset \mathbb{Z}_{3(2n+1)} \,,
\end{equation}
for which the mixed zero-form/one-form anomaly does not exist. In addition to the mixed anomaly involving the one-form symmetry, there are two other anomalies involving the $\mathbb{Z}_{3(2n+1)}$ zero-form symmetry that are natural to study. These are the pure discrete and mixed discrete-gravitational anomalies discussed in Section \ref{sec:global0}. These are characterized by two integers which, in this case, are
\begin{equation}
    \begin{aligned}
         (6n+4)(6n+5)\times\tr A'^3&\mod\,18(2n+1) \,, \\
         2\times\tr A'&\mod\,3(2n+1)\,, 
    \end{aligned}
\end{equation}
where $A'$ is the generator of the $\IZ_{3(2n+1)}$. Recall that the $\IZ_{3(2n+1)}$ subgroup originates from the axial symmetry generator $A=F_1+F_2+F_3$. Thus, the coefficients $(\tr A'^3,\tr A')$ are proportional to $(\tr A^3,\tr A)$ up to normalization. They are
\begin{align}
\begin{split}
   & \tr A^3=\sum_{i=1}^3\tr (-r+2I_3)^3_i=\sum_{i=1}^3\tr(-r^3-12r I_3^2)_i\,,\\
   & \tr A=\sum_{i=1}^3\tr(-r+2I_3)_i=-\sum_{i=1}^3\tr r_i\,.
\end{split}
\end{align}
Using the relations,
\begin{align}
\begin{split}
    \tr r=\tr r^3=48(a-c)\,,&\qquad\tr r I_3^2=4a-2c\,,\\
    a=\frac{(4p-1)(p-1)}{48p}\operatorname{dim}(G)\,,&\qquad c=\frac{p-1}{12}\operatorname{dim}(G)\,,
\end{split}
\end{align}
for $\CD_p(G)$ theories with $\gcd(p,h^\vee_G)=1$, the anomaly coefficients of the axial symmetry generated by $A$ are 
\begin{align}
    \tr A^3=-3\times(4n^2+4n)\,,\qquad\tr A=\frac{3}{2}(4n^2+4n)\,.
\end{align}
To fix the normalization, we consider how $A$ and $A'$ are related. The axial symmetry generated by $A$ is broken by the ABJ anomaly which is proportional to
\begin{align}
    \tr AGG=\sum_{i=1}^3(-r+2I_3)_iGG=-\sum_{i=1}^3(rGG)_i=\frac{1}{2}\sum_{i=1}^3k_{G,i}\,,
\end{align}
where $k_{G,i}$ is the flavor central charge of the $i$th $\CD_2(G)$ theory. For $\mathcal{D}_2(SU(2n+1))$, this is known to be
\begin{align}
    k_G = 2n+1 \,.
\end{align}
Thus, under this axial symmetry transformation with the angle $\theta$, the partition function acquires an anomalous phase
\begin{align}
    A:Z\mapsto \exp\left[2\times \frac{3}{2}(2n+1)i\theta\right]Z\,,
\end{align}
which breaks the axial symmetry to the $\IZ_{3(2n+1)}$ subgroup generated by the angles which are of the form:
\begin{equation}
    \theta=\frac{2\pi i k}{3(2n+1)} \,.
\end{equation} 
Then, it is obvious that the relation between $A$ and $A'$ is just $A'=A$. Thus, the discrete anomalies involving the $\IZ_{3(2n+1)}$ are
\begin{equation}
\begin{aligned}
    -3(6n+4)(6n+5)(4n^2+4n) \mod 18(2n+1) \,\,&=\,\, -24n(n+1) \mod\, 18(2n+1)\,,\\
    3(4n^2+4n)\mod\,3(2n+1) \,\,&=\,\, -3\mod\,3(2n+1) \,,
\end{aligned}
\end{equation}
where we have simplified the expressions for convenience. It is easy to see that these quantities are non-zero for all $n$, and thus the theory always has nontrivial discrete anomalies. We can also study the anomalies for the $\mathbb{Z}_3$ subgroup of $\mathbb{Z}_{3(2n+1)}$; these are
\begin{equation}\label{eqn:anom1}
    -24n(n+1)\equiv-6n(n+1) \mod\,18 \,, \qquad\qquad 0 \mod\,3 \,.
\end{equation}
The second anomaly is always trivial; the first is nontrivial when 
\begin{equation}
    n = 3k + 1 \,,
\end{equation}
and is trivial otherwise. 

We now show how to match these anomalies involving the discrete zero-form symmetry across the proposed duality to $\mathcal{N}=4$ SYM. From the $\mathcal{N}=1$ perspective, $\mathcal{N}=4$ super-Yang--Mills consists of an $SU(2n+1)$ vector multiplet together with three adjoint-valued chiral multiplets, $\Phi_1$, $\Phi_2$, and $\Phi_3$, and the superpotential
\begin{equation}
    W = [\Phi_1, [\Phi_2, \Phi_3]] \,.
\end{equation}
The naive $U(3)$ global symmetry rotating the chiral multiplets is broken to $SU(3)$ by the superpotential, and this $SU(3)$ contains the following subgroup
\begin{equation}\label{eqn:subsubsub}
    SU(3) \supset U(1)^2 \supset \mathbb{Z}_3 \,.
\end{equation}
The full $SU(3)$ symmetry survives in the quantum theory, however, we will attempt to match the anomalies of the subgroup in equation \eqref{eqn:subsubsub} with the proposed dual.\footnote{In fact, this $SU(3)$ combines with the $U(1)_R$ to form the usual $SU(4)$ R-symmetry of the $\mathcal{N}=4$ superconformal algebra, but this enhancement is not important for our discussion.} The $\mathbb{Z}_3$ subgroup in equation \eqref{eqn:subsubsub} has no mixed anomaly with the $\mathbb{Z}_{2n+1}$ one-form symmetry and has discrete anomalies
\begin{equation}\label{eqn:anom2}
    \begin{aligned}
    4\times5\times3\times(4n^2+4n)= 6n(n+1)\,\,&\mod\,18\,,\\
    2\times3\times(4n^2+4n)= 0\,\,&\mod\,3\,.
    \end{aligned}
\end{equation}
The anomaly coefficients in equation \eqref{eqn:anom2} matches with an extra minus sign, which reflects the fact that the $\IZ_3$ symmetry act oppositely across the duality. For example, the $\IZ_3$ rotates the adjoint chiral 
\begin{align}
    \Phi_i\mapsto\omega\Phi_i \,,
\end{align} 
where $\omega$ is the third root of unity. On the other hand, the moment map has $\CN=2$ R-charges which are 
\begin{align}
    \mu:(r,I_3)=(0,1)\,,
\end{align}
and consequently $\mu_i$ has charge 2 under the axial $\IZ_3$ symmetry generated by:
\begin{align}
    \IZ_3:\mu_i\mapsto \omega^2\mu_i=\omega^{-1}\mu_i\,.
\end{align} 
Then, the pairs of operators $\tr \Phi_i\Phi_j$ and $\tr\mu_i\mu_j$, which are matched across the duality, acquire the opposite phase under $\IZ_3$. With this extra minus sign, the anomaly coefficients in equation \eqref{eqn:anom2} perfectly matches with the discrete anomalies of the $\mathbb{Z}_3$ subgroup of the flavor group in equation \eqref{eqn:globenc} for the gauged Argyres--Douglas construction. This is not surprising since, as we have already noted, this particular $\mathbb{Z}_3$ subgroup is actually contained inside of the $U(1)^2$ part of the flavor group via the global quotient; therefore the matching of the continuous anomalies that was done in \cite{Kang:2023dsa} implies the matching of these discrete anomalies. More interestingly, we notice that the infrared of the gauged Argyres--Douglas theory is an $\mathcal{N}=1$ SCFT that appears to sit at a point of enhanced global symmetry on the ($\mathcal{N}=1$)-preserving conformal manifold of $\mathcal{N}=4$ super-Yang--Mills with $G = SU(2n+1)$, where the $\mathbb{Z}_3$ that exists at the generic point is enhanced to the $\mathbb{Z}_{3(2n+1)}$; though, we note that this may also be the infinite-distance IR-free point on the conformal manifold.\footnote{The study of the particular nature of this enhanced symmetry point, and how the S-duality of $\mathcal{N}=4$ which swaps the $SU(2n+1)$ and $SU(2n+1)/\mathbb{Z}_{2n+1}$ gauge groups interplays with the symmetry enhancement, we leave for future work.}

\subsection{The non-Lagrangian dual of two-adjoint SQCD}\label{sec:hound}

In this subsection, we present a non-Lagrangian dual theory to $SU(N)$ gauge theory coupled to two adjoint chiral multiplets, in the sense that they flow in the IR to the same conformal manifold \cite{LANDSCAPE}.\footnote{This duality is already hinted in \cite{Kang:2021ccs, Kang:2022vab, Maruyoshi:2023mnv}.}

\paragraph{Duality}
Let us consider gauging two copies of $\CD_3(SU(N))$ theory with $\gcd(3, N)=1$ via an $\CN=1$ gauge multiplet. This theory flows to an interacting SCFT with identical central charges $a=c$ \cite{Kang:2021ccs, Kang:2022vab}. We claim that this theory is dual to $SU(N)$ gauge theory coupled to two chiral multiplets in the adjoint representation. 

Upon gauging via an $\CN=1$ vector multiplet, the $\CN=2$ R-symmetry $SU(2)_R \times U(1)_r$ gets broken to $U(1)_{R_0} \times U(1)_F$. Since there are two copies of the Argyres--Douglas theory, we have $U(1)_F^2$ symmetry, which gets broken to a single $U(1)_\CF$, where the generator is given by $\CF = F_1 - F_2$, via the ABJ anomaly. 
The axial part of the global symmetry $U(1)_A$, with generator given by $A=F_1+F_2$, gets partially broken. The anomaly coefficient for the $U(1)_A$ upon gauging the diagonal $SU(N)$ flavor symmetry of the two copies of $\CD_3(SU(N))$ is
\begin{equation}
    \tr A GG = 2 \times \frac{2}{3} N \,,
\end{equation}
so that the axial $U(1)_A$ symmetry is broken to $\IZ_{8N}$ (by normalizing $A \to 3A$) for $\gcd(3, N)=1$. 
The faithful zero-form discrete symmetry is $\IZ_{4N}$, instead of $\IZ_{8N}$ since all the operators turn out to have even charges. We will discuss this momentarily when we match the operator spectrum under the duality. Therefore the (manifest) symmetry of the gauged Argyres--Douglas theory is $U(1)_R \times U(1)_\CF \times \IZ_{4N}$. 

The gauged Argyres--Douglas theory has an $\IZ_N$ one-form center symmetry. Furthermore, there is a mixed 't-Hooft anomaly between the $\IZ_{4N}$ zero-form symmetry and the $\IZ_N$ one-form symmetry where the fractional instanton breaks $\IZ_{4N}$ to $\IZ_4$. This symmetry and the mixed anomaly is identical to that of the $SU(N)$ gauge theory with two adjoint chiral multiplets, when both sides have vanishing superpotential.  

The theory flows to a superconformal theory whose R-symmetry is fixed by requiring the absence of the anomaly $\tr RGG$. Let us write the superconformal $R$-charge as
\begin{align}
    R = R_0 + \epsilon_1 F_1 + \epsilon_2 F_2 \ , 
\end{align}
where $R_0$ and $F = F_1 + F_2$, with $F_{1, 2}$ representing the $U(1)_F$ symmetry that acts on each of the $\CD_3(SU(N))$ theory as given in equation \eqref{eq:R0F}. 
One can fix the mixing parameter by evaluating
\begin{align}\label{eqn:peccarino}
    0 = \tr RGG = N + \sum_{i=1}^2 \left( \left(\frac{1}{3} - \epsilon_i \right) \tr_i r GG + \left(\frac{4}{3} + 2 \epsilon_i \right)\tr_i I_3 GG \right) \,, 
\end{align}
where the trace is over only the $i$th $\CD_3(SU(N))$ block. Imposing $\epsilon_1=\epsilon_2$ by symmetry,\footnote{The ABJ anomaly restricts $F_1+F_2=0$ for the continuous part.} we obtain $\epsilon = - 5/12$ \cite{Kang:2021ccs}. The R-symmetry acting on each individual $\CD_p(SU(N))$ theory reads 
\begin{align}
    R = R_0 - \frac{5}{12} F = \frac{3}{4} r + \frac{1}{2} I_3 \,,
\end{align}
where $r$ and $I_3$ denotes the Cartans of $\CN=2$ R-symmetry before gauging. From this, we can compute the central charges for the IR theory using 
\begin{align}
 a = \frac{3}{32} (3 \tr R^3 - \tr R) \ , \quad c = \frac{1}{32}(9\tr R^3 - 5 \tr R) \ , 
\end{align}
to obtain
\begin{align}
    a = c = \frac{27}{128} (N^2 - 1) \ .
\end{align}
Indeed their values are identical to the central charges of the $SU(N)$ gauge theory coupled to two chiral multiplets $(X, Y)$ in the adjoint representation. The anomaly-free condition for the R-symmetry determines the $R$-charges to be 
\begin{align}
    R(X)=R(Y)=\frac{1}{2}\,.
\end{align}
We can also check that the rest of the 't Hooft anomalies match across the duality, which we do shortly.

Let us now compare the spectrum of gauge-invariant operators across the duality. Each $\CD_3 (SU(N))$ theory contains 
\begin{enumerate}[(1)]
    \item a moment map operator $\mu$ with $(I_3, r)=(1, 0)$ in the adjoint representation of $SU(N)$,
    \item Coulomb branch operators of dimensions $4/3, 5/3, \cdots $ (with $\Delta = r/2$).
\end{enumerate}

\begin{table}[H]
\centering
    \begin{threeparttable}
    \setlength{\arraycolsep}{8pt}
    \renewcommand{\arraystretch}{1.6}
    $\begin{array}{cccc}
    \toprule
        R & \mathcal{F} & \text{Gauged } 2\times\mathcal{D}_3(SU(N)) & \text{Two adjoint chirals} \\\midrule
        1 & \pm 4/3 & Q^2 u_{1,1},\, Q^2 u_{1,2} &  \tr X^2, \, \tr Y^2 \\
        1 & 0 & \tr \mu_1 \mu_2  & \tr X Y \\
        3/2 & \pm2 & \tr \mu_1^2 \mu_2,\, \tr \mu_1 \mu_2^2 &  \tr X^3,\, \tr Y^3 \\
        3/2 & \pm 2/3 & Q^2u_{2,1},\,Q^2u_{2,2} &  \tr X^2Y,\,\tr XY^2 \\
        2 & \pm 8/3 & u_{1,1}, \, u_{1,2}, \, (Q^2u_{1,1})^2, \, (Q^2u_{1,2})^2 &  \tr X^4, \, \tr Y^4, (\tr X^2)^2, (\tr Y^2)^2 \,\,\\
        2 & 0 & \begin{array}{@{}c@{}}
        \tr\mu_1^2\mu_2^2,\,\tr \mu_1\mu_2\mu_1\mu_2,\\[-6pt](Q^2u_{1,1})(Q^2u_{1,2}), \, (\tr\mu_1\mu_2)^2 \end{array} & \begin{array}{@{}c@{}}
        \tr X^2Y^2,\,\tr XYXY, \\[-6pt]
        \tr X^2\,\tr Y^2, \, (\tr XY)^2 \end{array} \\
        2 & \pm 4/3 & (\tr \mu_1\mu_2)(Q^2u_{1,1}), \, (\tr \mu_1\mu_2)(Q^2u_{1,2}) & \tr X^3Y, \, \tr XY^3 \\\bottomrule
    \end{array}$
    \end{threeparttable}
    \caption{We enumerate the relevant and marginal operators of the SCFT obtained in the infrared of the gauging of two copies of $\mathcal{D}_3(SU(N))$ with $\gcd(N,3)=1$, together with their charges $R$, $\mathcal{F}$ under $U(1)_R$ and $U(1)_\mathcal{F}$. In the two-adjoint column, we have not written operators that become marginally irrelevant when breaking $SU(2)$ to $U(1)_B$. This matches with the spectrum of relevant and marginal operators of the $SU(N)$ gauge theory with two adjoint chiral multiplets upon identifying the $U(1)_B \subset SU(2)$ charge of the latter as $B = 3\mathcal{F}/2$.}
    \label{tb:N1dualOps}
\end{table}

The conformal dimensions of each Coulomb branch operator $u$ of the $\mathcal{D}_3(SU(N))$ building blocks is modified as follows along the flow into the infrared via
\begin{equation}
    \Delta_\text{IR}(u) = \left( 1 + 3 \varepsilon \right)\Delta_\text{UV}(u) = \frac{9}{4}\Delta_\text{UV}(u) \,.
\end{equation}
We should also consider the $\mathcal{N}=1$ multiplets whose primary, $Q^2u$, is a $\CN=2$ descendent of the Coulomb branch operators. Their scaling dimensions are modified as
\begin{equation}
    \Delta_\text{IR}(Q^2u) = \frac{9}{4}\Delta_\text{UV}(u) - \frac{2}{3} \,.
\end{equation}
For $\mathcal{D}_3(SU(N > 2))$, the only Coulomb branch operators that contribute to the relevant and marginal sector after gauging are those with 
\begin{equation}
    \Delta(u) = \frac{4}{3} , \, \frac{5}{3} \,.
\end{equation}
We refer to these operators as 
\begin{equation}
    u_{1,\alpha} \quad \text{ and } \quad u_{2,\alpha} \,,
\end{equation}
respectively, where $\alpha=1,2$ runs over the $\mathcal{D}_{3}(SU(N))$ building blocks. Under the $U(1)_\mathcal{F}$ after gauging, the R-charges of these operators and their $\mathcal{N}=2$ descendants are
\begin{equation}
    \mathcal{F}[u_{1,\alpha}] = \pm \frac{8}{3} \,, \quad \mathcal{F}[Q^2u_{1,\alpha}] = \pm \frac{4}{3} \,, \quad \mathcal{F}[Q^2u_{2,\alpha}] = \pm \frac{2}{3} \,,
\end{equation}
where the plus/minus signs respectively corresponds to when $\alpha=1,2$. Another class of half-BPS operators in the gauged theory come from the moment maps $\mu_1$, $\mu_2$ of the two $\mathcal{D}_3(SU(N))$s. The gauge-invariant non-vanishing operators are traces subject to the conditions that
\begin{equation}
    \tr \cdots \mu_i^{k \geq 3} \cdots \, = \, 0 \qquad \text{ and } \qquad \tr \mu_i^{k} \, = \, 0 \,.
\end{equation}
The charges of these operators under $U(1)_R$ and $U(1)_\mathcal{F}$ can be worked out directly from
\begin{equation}
    R[\mu_\alpha] = \frac{1}{2} \qquad \text{ and } \qquad \mathcal{F}[\mu_\alpha] = \pm 2 \,.
\end{equation}

Now, we can enumerate the relevant and marginal operators (with scalar primaries) across the duality. We begin with the $N = 2$ case, which the Coulomb branch operators with conformal dimension $5/3$ before gauging are absent. For the gauging of two copies of $\mathcal{D}_3(SU(2))$, the operators with R-charge one are
\begin{equation}\label{eqn:Rcharge1}
    Q^2u_{1,1} \,, \quad Q^2u_{1,2} \,, \quad \tr \mu_1\mu_2 \,.
\end{equation}
The next scalar operators are the marginal operators at R-charge two. We list all possible candidates of R-charge two as
\begin{equation}
  \begin{gathered}
    u_{1,1} \,, \quad u_{1,2} \,, \quad (Q^2u_{1,1})^2 \,, \quad (Q^2u_{1,2})^2 \,, \quad (Q^2u_{1,1})(Q^2u_{1,2}) \,, \\[0.6em] (\tr \mu_1\mu_2)(Q^2u_{1,1}) \,, \quad (\tr \mu_1\mu_2)(Q^2u_{1,2}) \,, \quad (\tr \mu_1\mu_2)^2 \,.
  \end{gathered}
\end{equation}
As discussed in \cite{Kang:2022vab}, not all of these operators are independent due to nontrivial relations coming from the OPE of the R-charge one multiplets. These chiral ring relations remove the following operators:
\begin{equation}\label{eqn:CRR}
    (Q^2u_{1,1})^2 \,, \quad (Q^2u_{1,2})^2 \,, \quad (\tr \mu_1\mu_2)(Q^2u_{1,1}) \,, \quad (\tr \mu_1\mu_2)(Q^2u_{1,2}) \,.
\end{equation}
Taking into account the current for the $U(1)$ flavor symmetry, we remove one additional operator; hence, there are three exactly marginal operators, spanning a three-dimensional conformal manifold. This analyses matches that of the superconformal conformal index in \cite{Kang:2022vab}, and which is reproduced below in equation \eqref{eqn:GADN2index}. We now compare this operator spectrum to that of the $SU(2)$ gauge theory with two-adjoint chiral multiplets $X,Y$ and with $W = 0$. We utilize their R-charges and their charges under $U(1)_B \subset SU(2)$:
\begin{equation}
    R[X] = R[Y] = \frac{1}{2}\,, \quad B[X] = 1\,, \quad B[Y] = -1 \,.
\end{equation}
The R-charge one operators are
\begin{equation}
    \tr X^2 \,, \quad \tr XY \,, \quad \tr Y^2 \,.
\end{equation}
At R-charge two we can see that there are six marginal operators:
\begin{equation}\label{eqn:2adjMARG}
  \begin{gathered}
    (\tr X^2)^2 \,, \quad (\tr XY)^2 \,, \quad (\tr Y^2)^2 \,, \\
    \tr X^2\,\tr XY  \,, \quad
    \tr X^2\,\tr Y^2 \,, \quad 
    \tr XY\,\tr Y^2 \,.
  \end{gathered}
\end{equation}
At the $W = 0$ point there is an $SU(2)$ flavor symmetry. Subtracting off the currents of this flavor symmetry we see that there are three exactly marginal operators, again spanning a three-dimensional conformal manifold. When moving on the conformal manifold to the locus where the $SU(2)$ is broken to $U(1)$, the marginal operators
\begin{equation}
    (\tr X^2)^2 \,, \quad (\tr XY)^2 \,, \quad (\tr Y^2)^2 \,,\quad \tr X^2\,\tr Y^2\,,
\end{equation}
remain marginal, and the rest become marginally irrelevant. Therefore, by comparing the $U(1)_\mathcal{F}$ and $U(1)_B$ charges, we find the following mapping between relevant operators,
\begin{equation}
    \begin{aligned}
      Q^2u_{1,1} \quad &\leftrightarrow \quad \tr X^2 \,,\\
      Q^2u_{1,2} \quad &\leftrightarrow \quad \tr Y^2 \,, \\
      \tr \mu_1\mu_2 \quad &\leftrightarrow \quad \tr XY  \,,
    \end{aligned}
\end{equation}
and marginal operators,
\begin{equation}\label{eqn:dualMARG}
    \begin{aligned}
      u_{1,1} \quad &\leftrightarrow \quad \tr X^4 \sim (\tr X^2)^2 \,, \\
      u_{1,2} \quad &\leftrightarrow \quad \tr Y^4 \sim (\tr Y^2)^2 \,, \\
      (Q^2u_{1,1})(Q^2u_{1,2}) \quad &\leftrightarrow \quad \tr X^2\,\tr Y^2 \,, \\
      (\tr\mu_1\mu_2)^2 \quad &\leftrightarrow \quad (\tr XY)^2 \,,
    \end{aligned}
\end{equation}
across the duality. Notice that $X, Y$ are $SU(2)$-valued so that there is a trace relation in the first two lines. 

It is straightforward, to perform the same analysis for $N > 2$. At R-charge one, we have exactly the operators as in equation \eqref{eqn:Rcharge1}. However, there are now operators that appear at R-charge $3/2$; we have
\begin{equation}
    Q^2u_{2,1} \,, \quad Q^2u_{2,2} \,, \quad \tr \mu_1^2 \mu_2 \,, \quad \tr \mu_1 \mu_2^2 \,. 
\end{equation}
For the two-adjoint theory, there are also R-charge $3/2$ operators that exist only when $N > 2$; these are
\begin{equation}
    \tr X^2 Y \,, \quad \tr XY^2 \,, \quad \tr X^3 \,, \quad \tr Y^3 \,.
\end{equation}
By matching the $U(1)_B$ and $U(1)_\mathcal{F}$ charges, we can expand the duality dictionary as follows:
\begin{equation}
    \begin{aligned}
      Q^2u_{2,1} \quad &\leftrightarrow \quad \tr X^2Y \,, \\
      Q^2u_{2,2} \quad &\leftrightarrow \quad \tr XY^2 \,, \\
      \tr \mu_1^2 \mu_2 \quad &\leftrightarrow \quad \tr X^3 \,, \\
      \tr \mu_1 \mu_2^2 \quad &\leftrightarrow \quad \tr Y^3 \,.
    \end{aligned}
\end{equation}
Finally, we discuss how the marginal operators for arbitrary $N$ map across the duality. On the gauged Argyres--Douglas side of the duality, we have two new operators for $N > 2$:
\begin{equation}
    \tr \mu_1^2\mu_2^2 \,, \qquad \tr \mu_1\mu_2\mu_1\mu_2 \,.
\end{equation}
Notice that certain chiral ring relations that rule out the operators in equation \eqref{eqn:CRR} for $N=2$ do not exist for $N > 2$. On the two-adjoint theory side of the duality, due to the lack of trace relations, the R-charge two operators are (in addition to those in equation \eqref{eqn:2adjMARG}):
\begin{equation} \label{eq:opmapAD}
    \begin{gathered}
        (\tr X^2)^2 \,, \quad
        \tr X^3 Y \,, \quad
        \tr XY \tr X^2 \,, \quad
        \tr X^2 Y^2 \,, \quad \\
        \tr XYXY \,, \quad
        \tr X Y^3 \,, \quad
        \tr XY \tr Y^2 \,, \quad
        (\tr Y^2)^2 \,.
    \end{gathered}
\end{equation}
The mapping of the additional marginal operators is then proposed to be
\begin{equation}
    \begin{aligned}
      (Q^2u_{1,1})^2 \quad &\leftrightarrow \quad (\tr X^2)^2 \,, \\
      (Q^2u_{1,2})^2 \quad &\leftrightarrow \quad (\tr Y^2)^2 \,, \\
      \tr \mu_1\mu_2\mu_1\mu_2 \quad &\leftrightarrow \quad \tr XYXY \,, \\
      \tr\mu_1^2\mu_2^2 \quad &\leftrightarrow \quad \tr X^2Y^2 \,, \\
      (\tr \mu_1\mu_2) (Q^2u_{1,1}) \quad &\leftrightarrow \quad \tr XY \tr X^2 \,, \\
      (\tr \mu_1\mu_2) (Q^2u_{1,2}) \quad &\leftrightarrow \quad \tr XY \tr Y^2 \,, \\
    \end{aligned}
\end{equation}
augmenting those in equation \eqref{eqn:dualMARG}. 
More accurately, we claim that $u_{1,1}$ and $(Q^2u_{1,1})^2$ map to independent linear combinations of $\tr X^4$ and $(\tr X^2)^2$, and so on for the other operators with the same charges under $U(1)_R$ and $U(1)_B$/$U(1)_\mathcal{F}$. 
Similarly, $\tr XY \tr X^2$ and $\tr X^3 Y$ have the same charges so that they can be mixed under the duality map. A linear combination of these operators are removed from the chiral ring on a generic point in the conformal manifold since it breaks $SU(2)$ flavor symmetry. 
We have summarized this matching in Table \ref{tb:N1dualOps}. We can see that the theories on each side of the duality both have nine-dimensional conformal manifolds.

\paragraph{Matching of 't Hooft anomalies for continuous symmetry} 
Identifying the symmetry structure is the first thing to check for the duality. Indeed, continuous global symmetries on each side of duality are both $U(1)_R\times U(1)$.\footnote{For the two-adjoint theory, the $U(1)$ enhances to $SU(2)$ at the $W=0$ point on the conformal manifold.} If these symmetries are indeed identical, their 't Hooft anomalies should match as well. For the $SU(N)$ gauge theory with two adjoint chiral multiplets, the anomaly coefficients of the continuous global symmetries are
\begin{align}\label{eqn:peanuts}
    \tr R^3=\frac{3}{4}(N^2-1)\,,\quad\tr R=0\,,\quad\tr RB^2=-(N^2-1)\,,
\end{align}
while the other anomaly coefficients are all trivial.
Here, $B$ is the generator of the baryonic $U(1)_B$ symmetry, which is identified with the Cartan of the enhanced $SU(2)$ flavor symmetry at the $W=0$ point on the conformal manifold. The vanishing of $\tr R$ is consistent with the fact that this theory has equal central charges. 

For the gauged $\CD_3(SU(N))$ theory, the anomaly coefficients  are
\begin{align}
\begin{split}
    \tr R^3&=(N^2-1)+2\times\left[\tr \left(\frac{3}{4}r+\frac{1}{2}I_3\right)^3\right]_{\CD_3(SU(N))}=\frac{3}{4}(N^2-1)\,,\\
    \tr R&\,=(N^2-1)+2\times\left[\tr \left(\frac{3}{4}r+\frac{1}{2}I_3\right)\right]_{\CD_3(SU(N))}=0\,,\\
    \tr R\mathcal{F}^2&\,=2\times\left[\tr \left(\frac{3}{4}r+\frac{1}{2}I_3\right)(-r+2I_3)^2\right]_{\CD_3(SU(N))}=-\frac{4}{9}(N^2-1)\,.
    \end{split}
\end{align}
We can see that this matches the anomalies in equation \eqref{eqn:peanuts}, where the generator $\mathcal{F}$ of the $U(1)_\mathcal{F}$ is identified with the generator $B$ of $U(1)_B$ as follows:
\begin{equation}\label{eqn:facebook}
    \mathcal{F} = \frac{2}{3} B \,.
\end{equation}

\paragraph{Matching of superconformal indices}
We indeed find that superconformal indices for the gauged $\CD_3(SU(N))^{\otimes 2}$ theory and the two adjoint $SU(N)$ theory agrees \cite{Kang:2022vab}, which is a strong check of the duality. 
For an arbitrary 4d $\mathcal{N}=1$ theory, the superconformal index is defined via \cite{Kinney:2005ej,Romelsberger:2005eg}
\begin{align}\label{eqn:SCI}
I=\tr\,(-1)^F\,t^{3(R+2j_2)}y^{2j_1}\prod_{i}v_i^{f_i}\,,
\end{align}
where the trace runs over states satisfying the following condition on the conformal dimension:
\begin{align}
    \Delta=\frac{3}{2}R+2j_2\,,
\end{align}
where, $j_1$ and $j_2$ are the Lorentz spins, $R$ is the $U(1)_R$ R-charge, and the flavor charges are denoted by schematically by $f_i$. 

We can first check that the full superconformal index matches when $N = 2$, as the full index of $\mathcal{D}_3(SU(2))$ is known from \cite{Agarwal:2016pjo,Maruyoshi:2016aim}. In fact, the superconformal index for the gauging of two copies of $\mathcal{D}_3(SU(2))$ was already determined in \cite{OPSPEC}. The reduced superconformal index is defined as
\begin{align}\label{eqn:reducedindex}
    \widehat{I}=\left(1-t^3y\right)\left(1-\frac{t^3}{y}\right)\left(I-1\right) \,,
\end{align}
and which does not include contributions from the conformal descendants. The reduced index of the gauged $\CD_3(SU(N))^{\otimes 2}$ theory is
\begin{align}
\label{eqn:GADN2index}
\begin{split}
\widehat{I}_{(3,3)}=&\;t^3\left(1+v^{\frac{4}{3}}+v^{-\frac{4}{3}}\right)-t^{\frac{9}{2}}\chi_{\bm{2}}(y)\left(v^{\frac{2}{3}}+v^{-\frac{2}{3}}\right)+t^6\left(1+v^{\frac{8}{3}}+v^{-\frac{8}{3}}\right)\\
&-t^{\frac{15}{2}}\chi_{\bm{2}}(y)\left(v^2+v^{-2}\right)+t^9\left(1+v^4+v^{-4}+\chi_{\bm{2}}(y)\left(v^{\frac{4}{3}}+v^{-\frac{4}{3}}\right)\right)+O\left(t^\frac{21}{2}\right) \,,
\end{split}
\end{align}
where $v$ is the fugacity for the surviving $U(1)$ flavor symmetry.
For the two-adjoint theory, the index can be written, for general $N$, as
\begin{equation}\label{eqn:2ADJschemeindex}
    I^{\text{two-adjoint}}(p, q) = \int [dz] I_{\text{vec}}(z) I_{\text{chi}}(z)^2 \,,
\end{equation}
where we have reparametrized $p = t^3y$ and $q = t^3/y$ from equation \eqref{eqn:SCI} and  
\begin{equation}
    I_\text{chi}(z) = \text{PE} \left[ \frac{(pq)^{1/4} - (pq)^{3/4}}{(1-p)(1-q)} \,\chi_{\textbf{adj}}^{SU(N)}(z) \right] \,.
\end{equation}
For $N = 2$, it is possible to perform the integral and expand order-by-order and recover the reduced index written in equation \eqref{eqn:GADN2index}. To match the flavor fugacities, we find that 
\begin{equation}
    v' = v^{\frac{2}{3}} \,,
\end{equation}
where $v'$ is the fugacity for the Cartan of the $SU(2)$ flavor symmetry of the two-adjoint theory. As we can see, this is exactly the same rescaling as in equation \eqref{eqn:facebook}. 

Beyond this case, however, we cannot compute the full superconformal index, as the superconformal index of the individual $\mathcal{D}_3(SU(N>2))$ is unknown. However, we still can study a particular limit of the full index for arbitrary $N$. We will consider the Schur limit where the index is still computable and verify that they match. This check was similarly carried out for the duality between the gauging of three copies of $\mathcal{D}_2(SU(2n+1))$ and the $\mathcal{N}=4$ super-Yang--Mills theory in \cite{Kang:2023dsa}.

The superconformal index of a theory with $\mathcal{N}=2$ supersymmetry can be formally written as
\begin{equation}
    I_{\CN=2}(p, q, \mathfrak{t}) = \tr (-1)^F p^{j_1 + j_2 + r} q^{j_2 - j_1 + r} \mathfrak{t}^{I_3 - \frac{1}{2} r} \,,
\end{equation}
where $(j_1, j_2)$, $r$, and $I_3$ are the charges under the Lorentz group, the $U(1)_R$, and the Cartan of the $SU(2)_R$, respectively; the trace runs over all states such that 
\begin{align}
    \Delta - 2j_2 - 2I_3 - r/2=0\,. 
\end{align}
The Schur limit involves taking $q = \mathfrak{t}$ after which the dependence on $p$ is removed \cite{Gadde:2011uv}. The Schur limit of the $\mathcal{D}_3(SU(N))$ index, for $\gcd(N, 3) = 1$ is  \cite{Song:2015wta,Xie:2016evu,Song:2017oew} 
\begin{equation}\label{eqn:schurlimit}
    \mathcal{I}_S^{\mathcal{D}_3(SU(N))}(q; z) = \operatorname{PE}\left[\frac{q-q^3}{(1-q)(1-q^3)} \,\chi_{\textbf{adj}}^{SU(N)}(z) \right]  \,,
\end{equation}
where $\chi_{\textbf{adj}}^{SU(N)}(z)$ is the character of the adjoint representation of $SU(N)$, $z$ is the collective fugacities for $SU(N)$, and $\operatorname{PE}$ denotes the plethystic exponential. For the gauged Argyres--Douglas theory, the full superconformal index can be schematically written as
\begin{equation}\label{eqn:GADschemeindex}
    I(p, q) = \int [dz] I_{\text{vec}}(z) \prod_{i=1}^2 I^{\CD_3} (z)\Big|_{\mathfrak{t} \to (pq)^{2/3+\varepsilon_i}} \,,
\end{equation}
where $I^{\CD_3} (z)$ is the (unknown) index for the $\mathcal{D}_3(SU(N))$ SCFT, and $\varepsilon_i$ are the mixing parameters determined from equation \eqref{eqn:peccarino}. We would like to compare the index in equation \eqref{eqn:GADschemeindex} with the index of the two-adjoint theory. 
To compare the two indices, equations \eqref{eqn:GADschemeindex} and \eqref{eqn:2ADJschemeindex}, we take the Schur limit of the index on both sides. The $I^{\CD_3} (z)$ in equation \eqref{eqn:GADschemeindex} becomes $\mathcal{I}_S^{\mathcal{D}_3(SU(N))}(q; z)$ in equation \eqref{eqn:schurlimit}. When we take the limit $q = \mathfrak{t} = (pq)^{\frac{1}{4}}$, i.e.~$p \rightarrow q^3$, in equation \eqref{eqn:2ADJschemeindex}, we find that the $I_\text{chi}(z)$ becomes
\begin{equation}
    \operatorname{PE}\left[\frac{q(1+q)}{(1-q^3)} \,\chi_{\textbf{adj}}^{SU(N)}(z) \right] \,,
\end{equation}
which matches equation \eqref{eqn:schurlimit}. Therefore, the Schur limits of the superconformal indices match across the duality, for arbitrary $N$ such that $\gcd(N, 3) = 1$.

\paragraph{Discrete anomalies and conformal manifold}
What we have so far seems to be a strong enough condition to support our duality between gauged AD theory and two-adjoint gauge theory, we find that there is a very interesting subtlety. 
This phenomenon is visible if we look into the discrete symmetry and its anomalies, as we now elaborate.

One natural check of the duality is to match global symmetries. It is discussed in \cite{Maruyoshi:2023mnv} and above that continuous global symmetries on both sides of duality match. Here, we consider the discrete $\IZ_{4N}$ symmetries and their remnants after the superpotential deformation. On the Lagrangian theory side, the $\IZ_{4N}$ symmetry acts as
\begin{align}
    \IZ_{4N}:(X,Y)\mapsto \omega (X,Y) \,,
\end{align}
where $\omega$ is the $4N$th root of unity satisfying $\omega^{4N}=1$. On the other hand, on the gauged $\CD_3(SU(N))^{\otimes 2}$ side, the $\IZ_{4N}$ is generated by $\frac{3}{2}A=\frac{3}{2}((-r+2I_3)_1+(-r+2I_3)_2)$. Then the local operators in $i$th $\CD_3(SU(N))$ theory whose $R$-charges are
\begin{align}
    \mu_i:\,(r,I_3)_i=(0,1)\,,\quad u_i:\,\left(\frac{8}{3},0\right)\,,
\end{align}
are rotated by the $\IZ_{4N}$ symmetry as
\begin{align}
    \IZ_{4N}:(\mu_i,u_i)\mapsto(\omega^3\mu_i,\,\omega^{-4}u_i)\,.
\end{align}
One can immediately see that the two $\IZ_{4N}$ do not match. For example, the $\IZ_{4N}$ rotates $\tr X^nY^{4-n}$ equally by $\omega^4$ on the Lagrangian theory side. On the other hand, it rotates $u_i\mapsto\omega^{-4}u_i$ while it rotates $\tr\mu_i^n\mu_j^{4-n}\mapsto\omega^{12}\left(\tr\mu_i^n\mu_j^{4-n}\right)$ on the dual gauged $\CD_3(SU(N))^{\otimes 2}$ side. 

It turns out they do not match across the proposed duality. The discrete anomaly in equation \eqref{eq:disanom} of the axial $\IZ_{4N}$ symmetry in the $SU(N)$ gauge theory with two adjoint chirals is characterized by two integers
\begin{equation}
    \begin{aligned}
        (4N+1)(4N+2)\times 2\times(N^2-1)\,\,&\mod\,24N\,,\\
        2\times2\times(N^2-1)\,\,&\mod\,4N\,.
    \end{aligned}
\end{equation}
On the other hand, the 't Hooft anomaly of the axial $\IZ_{4N}$ symmetry of the gauged $\CD_3(SU(N))$ theories can be computed by
\begin{equation}
\begin{aligned}
    (4N+1)(4N+2)\tr {A'}^3\,\,&\mod\,24N\,,\\
    2\times\tr A'\,\,&\mod\,4N\,,
\end{aligned}
\end{equation}
where the axial symmetry $A'$ should be properly normalized as $A'=\frac{3}{2}A$ (the factor 3 follows from above, and the divisor 2 comes from the natural embedding of the $\IZ_{4N}\in \IZ_{8N}$). The 't Hooft anomaly coefficients 
\begin{align}
    \tr {A'}^3&=\left(\frac{3}{2}\right)^3\times\sum_{i=1,2}\left[\tr(-r+2I_3)^3\right]_i=\frac{27}{8}\times2\times\left[-\tr r^3-12\tr rI_3^2\right]_{\CD_3(SU(N))}\,,\\
    \tr A'&=\frac{3}{2}\times 2\times\left[-\tr r\right]_{\CD_3(SU(N))}\,,
\end{align}
can be computed by using the relations
\begin{align}
\begin{split}
    \tr r=\tr r^3=48(a-c)\,,&\quad\tr r I_3^2=4a-2c\,,\\
    a=\frac{(4p-1)(p-1)}{48p}\operatorname{dim}(G)\,,&\quad c=\frac{p-1}{12}\operatorname{dim}(G)\,,
\end{split}
\end{align}
for $\CD_p(G)$ theories with $\gcd(p,h^\vee_G)=1$.
At last, the 't Hooft anomaly of the axial $\IZ_{4N}$ symmetry turns out to be
\begin{equation}
\begin{aligned}
-(4N+1)(4N+2)\times18(N^2-1)\,\,&\mod\,24N\,,\\
2\times2(N^2-1)\,\,&\mod 4N\,.
\end{aligned}
\end{equation}
The second part of the anomaly matches, while the first part does not. One can check that the difference of the first part is trivial only when $N=5$. On the other hand, there are some chances that the generators of those two symmetries are identified with opposite sign. Then the second parts of the anomalies match only when $N=2$, where the first parts of the anomalies also match.

The mismatch between the 't Hooft anomalies of two $\IZ_{4N}$ symmetries vanishes if we move around the conformal manifold by marginal deformations, which breaks the discrete symmetry from $\IZ_{4N}$ to $\IZ_4$. Then, the 't Hooft anomalies of $\IZ_4$ are respectively
\begin{align}
\begin{aligned}
    5\times 6\times 2(N^2-1)&\mod\,24\,,\\
    2\times2(N^2-1)&\mod 4\,,
\end{aligned}
\end{align}
and
\begin{align}
\begin{aligned}
    -\,5\times 6\times 18(N^2-1)&\mod\,24\,,\\ 
    2\times 2(N^2-1)&\mod 4\,.
\end{aligned}
\end{align}
One can easily find they always match with an extra minus sign upon identifying the generators. And this minus sign agrees with the fact that $\IZ_4$ acts on $\phi_i\mapsto e^{\frac{2\pi ik}{4}}\phi_i$ while it acts on $\mu_i\mapsto e^{-\frac{2\pi ik}{4}}\mu_i$ on the other side of duality.  
Away from the $W=0$ point, the 't Hooft anomalies match. Therefore, we find that $W=0$ fixed points of the gauged AD theory do not agree with the $W=0$ fixed point of the Lagrangian gauge theory with two adjoints. This does not necessarily mean that the duality fails at the $W=0$ fixed points. It may mean that $W=0$ RG fixed point of the AD theory in the conformal manifold arises from the fixed RG flow of the two adjoint theory with a nontrivial superpotential.

\section{Non-supersymmetric generalizations}\label{sec:nonsusy} 

In Questions \ref{ques:first} and \ref{ques:second}, we have considered 4d $\mathcal{N}=1$ QFTs that have a one-form symmetry, a nontrivial mixed axial-electric anomaly, and which flow to an interacting SCFT in the infrared. In fact, only the third condition, the flow to an interacting SCFT in the IR, is challenging to generalize to non-supersymmetric setups. Instead, we could ask:
\begin{ques}{ques:4}
\renewcommand{\thempfootnote}{\arabic{mpfootnote}}
What are the 4d $\mathcal{N}=0$ theories with gauge group $G$, a simple and simply-connected group, and massless chiral fermions in a (possibly reducible) representation $\bm{R}$ of $G$ that have a nontrivial one-form symmetry, a one-form and zero-form mixed anomaly, and a non-positive one-loop $\beta$-function? What can be said about their infrared fate?
\end{ques} 

The presence of a one-form symmetry follows again from equation \eqref{eqn:CONDITION_1FS}, the mixed anomaly can be determined via consultation with Table \ref{tbl:mixedanom}, and the non-positivity of the $\beta$-function requires that $\bm{R}$ satisfies
\begin{equation}\label{eqn:onelooponeloop}
    \sum_{i} n_i \, T(\bm{R}_i) \leq \frac{11}{2} h_G^\vee \,,
\end{equation}
where we have decomposed $\bm{R}$ as in equation \eqref{eqn:Rgeneral}. It is straightforward to enumerate such $\bm{R}$ that satisfy the conditions and produce $\mathcal{N}=0$ analogues of Tables \ref{tbl:genericN}, \ref{tbl:N1SU}, \ref{tbl:N1SpinUSp}. For the convenience of the reader, we briefly discuss all solutions to these conditions here. 

For a simple and simply-connected group $G$, we find the generic solution to equation \eqref{eqn:onelooponeloop} with a nontrivial one-form symmetry, which is $Z(G)$, to be
\begin{equation}
    \bm{R} = n_a \, \textbf{adj} \,, \qquad \text{ where } \qquad n_a = 0, \cdots, 5 \,.
\end{equation}
When $n_a = 0$, we have pure Yang--Mills. This is the only solution to the conditions in Question \ref{ques:4} that exists for all simple gauge groups.

We now turn to the solutions where there is matter in other representations than the adjoint. We begin by considering cases where there is an $SU(2N)$ gauge group and the matter preserves a $\mathbb{Z}_2$ subgroup of the $\mathbb{Z}_{2N}$ center as the one-form symmetry. For generic $N$, there are only five representations consistent with preserving a $\mathbb{Z}_2$ subgroup of the center: $\textbf{adj}$, $\bm{S}^2$, $\overline{\bm{S}}^2$, $\bm{A}^2$, and $\overline{\bm{A}}^2$. When combined with cancellation of the gauge anomaly, the only consistent combination takes the form
\begin{equation}
    \bm{R} = n_a \, \textbf{adj} \oplus n_S \, (\bm{S}^2 \oplus \overline{\bm{S}}^2) \oplus n_A \, (\bm{A}^2 \oplus \overline{\bm{A}}^2) \,,
\end{equation}
where the non-positivity of the one-loop $\beta$-function imposes the following relation among the coefficients:
\begin{equation}
    n_a \, (2N) + n_S \, (2N + 2) + n_A \, (2N - 2) \leq 11 N \,.
    \label{eq:betanonpositivity}
\end{equation}
We assume that at least one of $n_S$ and $n_A$ are non-zero, otherwise the one-form symmetry enhances to $\mathbb{Z}_{2N}$. In addition to the combinations of $(\bm{S}^2 \oplus \overline{\bm{S}}^2)$ and $(\bm{A}^2 \oplus \overline{\bm{A}}^2)$, we are allowed to have another combination involving $\bm{S}^2$ and $\overline{\bm{A}}^2$ irreps,
\begin{equation}\label{eqn:mixedrep}
    \bm{r} = \frac{N-2}{\gcd(N + 2, N - 2)}\,\bm{S}^2 + \frac{N+2}{\gcd(N + 2, N - 2)}\overline{\bm{A}}^2 \,,
\end{equation}
and its conjugate, both of which have trivial triangle anomaly coefficient: 
\begin{align}
    A(\bm{r}) = A(\overline{\bm{r}}) = 0 . 
\end{align}
The Dynkin index of $\bm{r}$ grows fast with $N$. We can write the condition that a chiral multiplet in the representation $\bm{r}$ is compatible with a non-positive one-loop $\beta$-function as
\begin{equation}
    T(\bm{r}) = \frac{2(N^2 - 2)}{\gcd(N+2, N-2)}
    \leq 11 N \,.
    \label{eq:betarestriction}
\end{equation}
We note that this condition is only satisfied for particular small values of $N$. We list all such $\mathfrak{g}$ and $\bm{r}$ in Table \ref{tbl:reprr}. In these cases, by utilizing equations \eqref{eq:betanonpositivity} and \eqref{eq:betarestriction}, we should instead consider solutions to
\begin{equation}
    n_a \, (2N) + n_S \, (2N + 2) + n_A \, (2N - 2) + n_m \, \left( \frac{2(N^2 - 2)}{\gcd(N+2, N-2)} \right) \leq 11 N \,,
    \label{eq:solutionbound}
\end{equation}
specifying the representation 
\begin{equation}
    \bm{R} = n_a \, \textbf{adj} \oplus n_S \, (\bm{S}^2 \oplus \overline{\bm{S}}^2) \oplus n_A \, (\bm{A}^2 \oplus \overline{\bm{A}}^2) \oplus n_m \, \frac{\left((N-2)\,\bm{S}^2 + (N+2)\overline{\bm{A}}^2\right)}{\gcd(N + 2, N - 2)} \,.
    \label{eq:solutionR}
\end{equation}

\begin{table}[t]
    \centering
    \begin{threeparttable}
    \setlength{\arraycolsep}{8pt}
    \renewcommand{\arraystretch}{1.2}
    \begin{tabular}{cc}
        \toprule
         $\mathfrak{g}$ & $\bm{r}$ \\\midrule
         $\mathfrak{su}(6)$ & $\bm{21} \oplus 5\,\overline{\bm{15}}$ \\
         $\mathfrak{su}(8)$ & $\bm{36} \oplus 3\,\overline{\bm{28}}$ \\
         $\mathfrak{su}(10)$ & $3\,\bm{55} \oplus 7\,\overline{\bm{45}}$ \\
         $\mathfrak{su}(12)$ & $\bm{78} \oplus 2\,\overline{\bm{66}}$ \\
         $\mathfrak{su}(16)$ & $3\,\bm{136} \oplus 5\,\overline{\bm{120}}$ \\
         $\mathfrak{su}(20)$ & $2\,\bm{210} \oplus 3\,\overline{\bm{190}}$ \\
         $\mathfrak{su}(28)$ & $3\,\bm{406} \oplus 4\,\overline{\bm{378}}$ \\
         $\mathfrak{su}(36)$ & $4\,\bm{666} \oplus 5\,\overline{\bm{630}}$ \\
         $\mathfrak{su}(44)$ & $5\,\bm{990} \oplus 6\,\overline{\bm{946}}$ \\
         \midrule
    \end{tabular}
    \end{threeparttable}
    \caption{For certain $\mathfrak{g} = \mathfrak{su}(N)$, there are reducible representations $\bm{r}$ built out of $\bm{S}^2$ and $\overline{\bm{A}}^2$ such that the triangle anomaly coefficient $A(\bm{r})$ vanishes and $T(\bm{r}) \leq \frac{11}{2}N$. We list them exhaustively here.}
    \label{tbl:reprr}
\end{table}

While equation \eqref{eq:solutionbound} provides solutions for high values of $N$ with the representation in equation \eqref{eq:solutionR}, there are several special cases that occur for low values of $N$. These special exceptions can be studied case by case. 

When $G = SU(2)$, the $\bm{S}^4 = \bm{5}$ representation does not screen any of the $\mathbb{Z}_2$ center of $SU(2)$, and thus the reducible representation
\begin{equation}
    \bm{R} = n_a \bm{3} + n_s \bm{5}
\end{equation}
realizes a $\mathbb{Z}_2$ one-form symmetry whenever the coefficients solve
\begin{equation}
    2 n_a + 10 n_s \leq 11 \,. 
\end{equation}

For $G=SU(4)$, the $\bm{A}^2 = \bm{6}$ representation is, in fact, real instead of complex, and thus it does not have a non-zero triangle anomaly coefficient. Furthermore, there is the $\bm{20'}$ representation which can be incorporated. The generic $\mathbb{Z}_2$ one-form symmetry preserving matter spectrum is then given by the representation
\begin{equation}
    \bm{R} = n_a \, \bm{15} \oplus n_S \, (\bm{10} \oplus \overline{\bm{10}}) \oplus n_A \, \bm{6} \oplus n_{\bm{20'}} \, \bm{20'} \,,
\end{equation}
where the coefficients are non-negative integers satisfying
\begin{equation}
    4n_a + 6n_S + n_A + 8n_{\bm{20'}} \leq 22 \,.
\end{equation}

For $SU(8)$, the (real) $\bm{A}^4 = \bm{70}$ has a sufficiently small Dynkin index that it can be included in $\bm{R}$ without violating the non-positivity of the one-loop $\beta$-function. The matter spectrum then, up to conjugation, takes the form:
\begin{equation}
    \bm{R} = n_a \, \bm{63} \oplus n_m \, (\bm{36} \oplus 3\,\overline{\bm{28}}) \oplus n_S \, (\bm{36} \oplus \overline{\bm{36}}) \oplus n_A \, (\bm{28} \oplus \overline{\bm{28}}) \oplus n_{\bm{70}} \, \bm{70} \,.
\end{equation}
The coefficients must satisfy the constraint that
\begin{equation}
    8n_a + 14 n_m + 10n_S + 6n_A + 10 n_{\bm{70}} \leq 44 \,.
\end{equation}

The final such special case occurs when $G = SU(10)$. The $\bm{A}^4 = \bm{210}$ and $\overline{\bm{A}}^4 = \overline{\bm{210}}$ representations have triangle anomaly coefficients $A(R) = \pm 14$, respectively, which is precisely the same as the $\bm{S}^2$ and $\overline{\bm{S}}^2$. Therefore, the generic matter representation can be written, up to conjugation, as
\begin{equation}
    \bm{R} = n_a \, \bm{99} \oplus n_s \, (\bm{210} \oplus \overline{\bm{55}}) \oplus n_m \, (3\,\bm{55} \oplus 7\,\overline{\bm{45}}) \oplus n_S \, (\bm{55} \oplus \overline{\bm{55}}) \oplus n_A \, (\bm{45} \oplus \overline{\bm{45}}) \,,
\end{equation}
where the coefficients satisfy
\begin{equation}
    10n_a + 34 n_s + 46 n_m + 12n_S + 8n_A \leq 55 \,.
\end{equation}

When considering pairs $(SU(N), \bm{R})$ where the theory has a one-form symmetry $\mathbb{Z}_K$ for $K > 2$, there are only a small number of possible $SU(N)$ gauge groups which have such accommodating (non-adjoint) matter. We list all such cases in Table \ref{tbl:specialN01FS}.

For $G = USp(2N)$, a similar story holds. The generic representation which has a nontrivial one-form symmetry in the associated gauge theory is
\begin{equation}
    \bm{R} = n_a \, \textbf{adj} \oplus n_A \, \bm{A}^2 \,.
\end{equation}
Here, $\bm{A}^2$ denotes the non-singlet part of the second anti-symmetric power. The non-negative integer coefficients must satisfy
\begin{equation}
    (N+1) n_a + (N-1) n_A \leq \frac{11(N+1)}{2} \,,
\end{equation}
for the one-loop $\beta$-function to be non-positive. For certain small values of $N$, there are additional irreducible representations that can appear in $\bm{R}$; we list these configurations in Table \ref{tbl:specialN01FSUSp}.

\begin{table}[H]
    \centering
    \begin{threeparttable}
    \setlength{\arraycolsep}{8pt}
    \renewcommand{\arraystretch}{1.5}
    \begin{tabular}{cccc}
         \toprule
         $G$ & $\bm{R}$ & $\Gamma$ & $\beta_g \leq 0$ \\\midrule
         $SU(3)$ & $n_a \, \bm{8} \oplus n_s \, (\bm{10} \oplus \overline{\bm{10}})$ & $\mathbb{Z}_3$ & $3n_a + 15n_s \leq \dfrac{33}{2}$ \\ 
         $SU(4)$ & $n_a \, \bm{15} \oplus n_s \, \bm{20'}$ & $\mathbb{Z}_4$ & $3n_a + 8n_s \leq 22$ \\ 
         $SU(6)$ & $n_a \, \bm{35} \oplus n_s \, \bm{20}$ & $\mathbb{Z}_3$ & $6n_a + 3n_s \leq 33$ \\ 
         $SU(8)$ & $n_a \, \bm{63} \oplus n_s \, \bm{70}$ & $\mathbb{Z}_4$ & $8n_a + 10n_s \leq 44$ \\ 
         $SU(9)$ & $n_a \, \bm{80} \oplus n_s \, (\bm{84} \oplus \overline{\bm{84}})$ & $\mathbb{Z}_3$ & $9n_a + 21n_s \leq \dfrac{99}{2}$ \\ 
         $SU(10)$ & $n_a \, \bm{99} \oplus n_s \, \bm{252}$ & $\mathbb{Z}_5$ & $10n_a + 35n_s \leq 55$ \\ 
         $SU(12)$ & $n_a \, \bm{143} \oplus n_s \, (\bm{220} \oplus \overline{\bm{220}})$ & $\mathbb{Z}_3$ & $12n_a + 45n_s \leq 66$ \\ 
         $SU(15)$ & $n_a \, \bm{224} \oplus n_s \, (\bm{455} \oplus \overline{\bm{455}})$ & $\mathbb{Z}_3$ & $15n_a + 78n_s \leq \dfrac{165}{2}$ \\\midrule
         $SU(N)$ & $n_a \, \textbf{adj}$ & $\mathbb{Z}_N$ & $n_a \leq 5$ \\
         \bottomrule
    \end{tabular}
    \end{threeparttable}
    \caption{Pairs $(SU(N), \bm{R})$ that are associated with gauge theories with non-positive one-loop $\beta$-function and which have a $\mathbb{Z}_{K > 2}$ one-form symmetry.}
    \label{tbl:specialN01FS}
\end{table}

\begin{table}[H]
    \centering
    \begin{threeparttable}
    \setlength{\arraycolsep}{8pt}
    \renewcommand{\arraystretch}{1.5}
    \begin{tabular}{cccc}
         \toprule
         $G$ & $\bm{R}$ & $\Gamma$ & $\beta_g \leq 0$ \\\midrule
         $USp(4)$ & $n_a \, \bm{10} \oplus n_A \, \bm{5} \oplus n_s \, \bm{14}$ & $\mathbb{Z}_2$ & $3n_a + n_A + 7n_s \leq \dfrac{33}{2}$ \\
         $USp(6)$ & $n_a \, \bm{21} \oplus n_A \, \bm{14} \oplus n_s \, \bm{70}$ & $\mathbb{Z}_2$ & $4n_a + 2n_A + 20n_s \leq 22$ \\
         $USp(8)$ & $n_a \, \bm{36} \oplus n_A \, \bm{27} \oplus n_s \, \bm{42}$ & $\mathbb{Z}_2$ & $5n_a + 3n_A + 7n_s \leq \dfrac{55}{2}$ \\
         $USp(10)$ & $n_a \, \bm{55} \oplus n_A \, \bm{44} \oplus n_s \, \bm{165}$ & $\mathbb{Z}_2$ & $6n_a + 4n_A + 24n_s \leq 33$ \\
         \bottomrule
    \end{tabular}
    \end{threeparttable}
    \caption{Pairs $(USp(2N), \bm{R})$, where $\bm{R}$ contains irreps other than the adjoint and anti-symmetric representations, that are associated with gauge theories with non-positive one-loop $\beta$-function and nontrivial one-form symmetry.}
    \label{tbl:specialN01FSUSp}
\end{table}

Finally, we analyze $G = Spin(N)$. We first consider the cases where a $\mathbb{Z}_2$ one-form symmetry is realized in the resulting gauge theory. For generic values of $N$, the only possible representations that can appear in an $\bm{R}$ such that the one-loop $\beta$-function is non-positive are the adjoint, the $\bm{S}^2$, and the vector $\bm{V}$. That is, we consider
\begin{equation}
    \bm{R} = n_a \, \textbf{adj} \oplus n_S \, \bm{S}^2 \oplus n_V \, \bm{V} \,,
\end{equation}
such that
\begin{equation}\label{eqn:dora}
    (N-2)n_a + (N+2)n_S + n_V \leq \frac{11}{2}(N-2) \,.
\end{equation}
When $N$ is even, it is possible to preserve the full center as the one-form symmetry (either $\mathbb{Z}_4$ or $\mathbb{Z}_2 \oplus \mathbb{Z}_2$ depending on whether $N$ is a multiple of four). In such a case, the only representations that can appear, for generic even $N$, are the adjoint and the $\bm{S}^2$:
\begin{equation}
    \bm{R} = n_a \, \textbf{adj} \oplus n_S \, \bm{S}^2 \,,
\end{equation}
where the coefficients must satisfy equation \eqref{eqn:dora} with $n_V = 0$. For a $Spin(N)$ gauge theory with $N$ sufficiently small, there can be additional irreducible representations appearing in $\bm{R}$ which are consistent with both the realization of one-form symmetry and $\beta_g \leq 0$. We have listed these low rank examples exhaustively in Table \ref{tbl:specialN01FSSO}.

\afterpage{
\begin{landscape}
\pagestyle{empty}
\begin{table}[p]
    \centering    
    \resizebox{!}{.26\paperheight}{%
    \begin{threeparttable} 
    \renewcommand{\arraystretch}{1.2}
    \begin{tabular}{cccc}
         \toprule
         $G$ & $\bm{R}$ & $\Gamma$ & $\beta_g \leq 0$ \\\midrule
         $Spin(7)$ & $n_a \, \bm{21} \oplus n_S \, \bm{27} \oplus n_V\,\bm{7} \oplus n_s \, \bm{35}$ & $\mathbb{Z}_2$ & $5n_a + 9n_S + n_V + 10n_s \leq \dfrac{55}{2}$ \\\midrule
         \multirow{4}{*}{$Spin(8)$} & $n_a \, \bm{28} \oplus n_{Sv} \, \bm{35_v} \oplus n_{Ss} \, \bm{35_s} \oplus n_{Sc} \, \bm{35_c}$ & $\mathbb{Z}_2^L \oplus \mathbb{Z}_2^R$ & $6n_a + 10(n_{Sv} + n_{Ss} + n_{Sc}) \leq 33$ \\
         & $n_a \, \bm{28} \oplus n_{Sv} \, \bm{35_v} \oplus n_{Ss} \, \bm{35_s} \oplus n_{Sc} \, \bm{35_c} \oplus n_v \,\bm{8_v} \oplus n_{z} \, \bm{56_v}$ & $\mathbb{Z}_2^D$ & $6n_a + 10(n_{Sv} + n_{Ss} + n_{Sc}) + n_v + 15n_z \leq 33$ \\
         & $n_a \, \bm{28} \oplus n_{Sv} \, \bm{35_v} \oplus n_{Ss} \, \bm{35_s} \oplus n_{Sc} \, \bm{35_c} \oplus n_s\,\bm{8_s} \oplus n_z \, \bm{56_s}$ & $\mathbb{Z}_2^L$ & $6n_a + 10(n_{Sv} + n_{Ss} + n_{Sc}) + n_s + 15n_z \leq 33$ \\
         & $n_a \, \bm{28} \oplus n_{Sv} \, \bm{35_v} \oplus n_{Ss} \, \bm{35_s} \oplus n_{Sc} \, \bm{35_c} \oplus n_c\,\bm{8_c} \oplus n_z \, \bm{56_c}$ & $\mathbb{Z}_2^R$ & $6n_a + 10(n_{Sv} + n_{Ss} + n_{Sc}) + n_c + 15n_z \leq 33$ \\\midrule
         $Spin(9)$ & $n_a \, \bm{36} \oplus n_S \, \bm{44} \oplus n_V\,\bm{9} \oplus n_s \, \bm{84} \oplus n_s'\,\bm{126}$ & $\mathbb{Z}_2$ & $7n_a + 11n_S + n_V + 21n_s + 35n_s' \leq \dfrac{77}{2}$ \\\midrule
         $Spin(10)$ & $n_a \, \bm{45} \oplus n_S \, \bm{54} \oplus n_V\,\bm{10} \oplus n_s \, \bm{120} \oplus n_s'\,\bm{126} \oplus n_s''\,\overline{\bm{126}}$ & $\mathbb{Z}_2$ & $8n_a + 12n_S + n_V + 28n_s + 35n_s' + 35n_s'' \leq 44$ \\\midrule
         $Spin(11)$ & $n_a \, \bm{55} \oplus n_S \, \bm{65} \oplus n_V\,\bm{11} \oplus n_s \, \bm{165}$ & $\mathbb{Z}_2$ & $9n_a + 13n_S + n_V + 36n_s \leq \dfrac{99}{2}$ \\\midrule
         \multirow{3}{*}{$Spin(12)$} & $n_a \, \bm{66} \oplus n_S \, \bm{77} \oplus n_V \, \bm{12} \oplus n_s \, \bm{220}$ & $\mathbb{Z}_2^D$ & $10n_a + 14n_S + n_V + 45n_s \leq 55$ \\
         & $n_a \, \bm{66} \oplus n_S \, \bm{77} \oplus n_s \, \bm{32}$ & $\mathbb{Z}_2^L$ & $10n_a + 14n_S + 4n_s \leq 55$ \\
         & $n_a \, \bm{66} \oplus n_S \, \bm{77} \oplus n_s \, \bm{32'}$ & $\mathbb{Z}_2^R$ & $10n_a + 14n_S + 4n_s \leq 55$ \\\midrule
         $Spin(13)$ & $n_a \, \bm{78} \oplus n_S \, \bm{90} \oplus n_V\,\bm{13} \oplus n_s \, \bm{286}$ & $\mathbb{Z}_2$ & $11n_a + 15n_S + n_V + 55n_s \leq \dfrac{121}{2}$ \\\midrule
         $Spin(14)$ & $n_a \, \bm{91} \oplus n_S \, \bm{104} \oplus n_V\,\bm{14} \oplus n_s \, \bm{364}$ & $\mathbb{Z}_2$ & $12n_a + 16n_S + n_V + 66n_s \leq 66$ \\\midrule
         \multirow{2}{*}{$Spin(16)$} & $n_a \, \bm{120} \oplus n_S \, \bm{135} \oplus n_s \, \bm{128}$ & $\mathbb{Z}_2^L$ & \multirow{2}{*}{$14n_a + 18n_S + 16n_s \leq 77$} \\
         & $n_a \, \bm{120} \oplus n_S \, \bm{135} \oplus n_s \, \bm{128'}$ & $\mathbb{Z}_2^R$ & \\\midrule
         \multirow{2}{*}{$Spin(20)$} & $n_a \, \bm{190} \oplus n_S \, \bm{209} \oplus n_s \, \bm{512}$ & \multirow{2}{*}{$\mathbb{Z}_2^L$} & \multirow{2}{*}{$18n_a + 22n_S + 64n_s \leq 99$} \\
         & $n_a \, \bm{190} \oplus n_S \, \bm{209} \oplus n_s \, \bm{512'}$ & \\
         \bottomrule
    \end{tabular}
    \end{threeparttable}}
    \caption{Pairs $(Spin(N), \bm{R})$, where $\bm{R}$ contains irreps other than the adjoint and symmetric representations, that are associated with gauge theories with non-positive one-loop $\beta$-function and nontrivial one-form symmetry.}
    \label{tbl:specialN01FSSO}
\end{table}
\end{landscape}
}

The following question remains: given one of these just-derived asymptotically-free chiral gauge theories with a one-form symmetry and nontrivial mixed axial-electric anomaly, what is the behavior in the deep infrared?\footnote{When discussing supersymmetric gauge theories in Section \ref{sec:2}, we made a brief digression on theories where the global form of the gauge group was not simply-connected, corresponding to different choices of polarization of the intermediate defect group. Analogous considerations also arise in the non-supersymmetric case, and when the gauge group is non-simply-connected there can also be interesting infrared behavior.} As emphasized in Section \ref{sec:intro}, the infrared may be a unique gapless symmetry-preserving vacuum, a unique gapped symmetry-preserving vacuum, or else there must be spontaneous symmetry breaking. Much recent work has focused on studying the infrared behavior of 4d gauge theories from the generalized symmetry perspective, see, for example, \cite{Csaki:2021aqv,Csaki:2021xhi,Sulejmanpasic:2020zfs,Smith:2021vbf,Bolognesi:2021hmg,Bolognesi:2021yni,Bolognesi:2020mpe,Bolognesi:2019fej,Bolognesi:2019wfq,Bolognesi:2017pek,Anber:2021lzb,Anber:2020qzb,Anber:2020xfk,Anber:2019nze,Nakajima:2022jxg,Bolognesi:2022beq,Anber:2023mlc,Anber:2023yuh,Anber:2021iip,Anber:2023urd,Anber:2023pny,Bolognesi:2023sxe,Konishi:2024rjz,Apruzzi:2021phx,Antinucci:2024ltv} and references therein. In this paper, we do not aim to give a detailed analysis; we make only surface level observations, taking into account on a small subset of the data of the gauge theory, that indicate which IR behavior is forbidden or favored. Understanding the IR fate requires a holistic perspective, weaving together myriad threads of information. 

The most constrained possibility is the existence of a unique gapped symmetry-preserving vacuum; in this case the anomaly must be saturated by a topological quantum field theory. In \cite{Cordova:2019bsd}, the authors determined an obstruction which, given a particular anomaly, must vanish for a TQFT to exist that realizes that anomaly. Furthermore, in \cite{Cordova:2019jqi}, this obstruction was applied to theories with discrete zero-form symmetries with nontrivial anomalies: this is precisely the setup we are studying here. In \cite{Cordova:2019jqi}, the focus is on 4d theories with a $\mathbb{Z}_M$ zero-form symmetry and a mixed anomaly between the $\mathbb{Z}_M$ and the Poincare zero-form symmetry -- this is often called the mixed-gravitational anomaly. There is also a discrete cubic anomaly (as was discussed in Section \ref{sec:global0}), which provides further constraints on the IR behavior. In addition, we have the mixed axial-electric anomaly; in a similar way, we can use the obstruction of \cite{Cordova:2019bsd} to determine whether there exists any consistent TQFT that realizes this mixed zero-form/one-form anomaly. We briefly summarize this obstruction here. For a five-dimensional anomaly TFT, given by Lagrangian $\omega(A, B)$, then it is necessary that
\begin{equation}
    \exp\left(2\pi i \int_{S^1 \times S^2 \times S^2} \omega(A, B) \right) = 1 \,,
\end{equation}
for $\omega(A, B)$ to define a TFT that satisfies the necessary axioms. Using \cite{Cordova:2019bsd}, we can see that
\begin{equation}
    2\pi i \int_{S^1 \times S^2 \times S^2} \omega(A, B) \,\,=\,\, 2 \pi i \, \frac{N(N-1)}{p^2} \,,
\end{equation}
using the expression for the anomaly TFT given in equation \eqref{eqn:anomTFT}. Therefore, whenever 
\begin{equation}\label{eqn:dora2}
    \frac{N(N-1)}{p^2} \not\in \mathbb{Z} \,,
\end{equation}
there does not exist a TFT that saturates the anomaly. However, the only case where there can exist a nontrivial mixed anomaly and equation \eqref{eqn:dora2} is satisfied occurs when considering spacetime backgrounds that do not admit spin structure and $N(N-1)/p^2$ is an odd integer; c.f., equation \eqref{eqn:nscond}.\footnote{Perhaps the simplest example is $SU(4)$ gauge theory with fermions in the $\bm{10} \oplus \overline{\bm{10}}$ representation.} Therefore, this is the only situation in which the anomaly can be saturated by topological degrees of freedom.

A hint towards the presence of a conformal field theory in the infrared occurs if there exists a Banks--Zaks fixed point \cite{Banks:1981nn}. That is, if there exists a zero of the perturbative $\beta$-function at small values of the gauge coupling, then a CFT fixed point can appear while the perturbative calculation remains within its regime of validity. The existence of a weakly-coupled conformal field theory itself opens doors to a perturbative analysis of the infrared physics. Examples include the extraction of an effective potential, or more generally, the calculation of the anomalous dimensions of operators. It would be interesting to determine what we can learn about the symmetry structure of these CFT fixed points through such perturbative calculations, though it is beyond the scope of this paper to do so.

We consider a 4d $\mathcal{N}=0$ gauge theory with simple gauge algebra $\mathfrak{g}$, and massless chiral fermions in a representation 
\begin{equation}
    \bm{R} = \bigoplus_i n_i \bm{R}_i \,,
\end{equation}
of $\mathfrak{g}$. As usual, we let $\bm{R}_i$ denote irreducible representations. The $\beta$-function of the gauge coupling associated with $\mathfrak{g}$ has been determined up to three loops in \cite{Tarasov:1980au,Larin:1993tp} (see also \cite{Zoller:2016sgq}). We have 
\begin{equation}\label{eqn:betafn}
    \beta(g) = - \left(\beta_0 \frac{g^3}{16\pi^2} + \beta_1 \frac{g^5}{(16\pi^2)^2} + \beta_2 \frac{g^7}{(16\pi^2)^3} + \cdots \right) \,,
\end{equation}
where the coefficients are
\begin{equation}
    \begin{aligned}
        \beta_0 &= \frac{11}{3} h_G^\vee - \frac{2}{3} \sum_i n_i T(\bm{R}_i) \,, \\
        \beta_1 &= \frac{34}{3} (h_G^\vee)^2 - \frac{10}{3} \sum_i n_i h_G^\vee T(\bm{R}_i) - 2 \sum_i n_i T(\bm{R}_i) C_2(\bm{R}_i) \,, \\
        \beta_2 &= \frac{2857}{54} (h_G^\vee)^3 - \frac{1415}{54} (h_G^\vee)^2 \sum_i n_i T(\bm{R}_i) + \frac{79}{54} h_G^\vee \sum_{i,j} n_i n_j T(\bm{R}_i) T(\bm{R}_j) \\ &\qquad\qquad\qquad - \frac{205}{18} h_G^\vee \sum_i n_i T(\bm{R}_i) C_2(\bm{R}_i) + \frac{11}{9} \sum_{i,j} n_i n_j T(\bm{R}_i) T(\bm{R}_j) C_2(\bm{R}_i) \\ &\qquad\qquad\qquad + \sum_i n_i T(\bm{R}_i) C_2(\bm{R}_i)^2 \,.
    \end{aligned}
\end{equation}
Here, we have introduced the quadratic Casimir invariant of the irreducible representation $\bm{R}_i$, which is related to its Dynkin index via
\begin{equation}
    C_2(\bm{R}_i) = \frac{\operatorname{dim}(G)}{\operatorname{dim}(\bm{R}_i)} \, T(\bm{R}_i) \,.
\end{equation}

Now that we have the closed-form expressions for the three-loop $\beta$-function to hand, it is straightforward to determine if there exists a $g \ll 1$ such that $\beta(g) = 0$. Many of the theories answering Question \ref{ques:4} have such a Banks--Zaks fixed point, especially those where the inequality in equation \eqref{eqn:onelooponeloop} is close to being saturated. For example, in \cite{Anber:2023yuh}, the authors study Banks--Zaks type fixed points for $SU(N)$ gauge theories where the massless chiral fermions are in the representation $\bm{r}$ as given in Table \ref{tbl:reprr}. We provide a few additional interesting examples to conclude this section. 

First, we consider a family of examples that admit a large $N$ limit. For simplicity, we take
\begin{equation}
    G = SU(2N) \qquad \text{ and } \qquad \bm{R} = 5 \, \left( \bm{S}^2 \oplus \overline{\bm{S}}^2 \right) \,,
\end{equation}
where we have assumed that $N > 10$ so that the gauge coupling is asymptotically-free. Plugging the appropriate values for $\beta_\ell$ into equation \eqref{eqn:betafn}, and solving for $g_*$ such that $\beta(g_*) = 0$, it is easy to see that
\begin{equation}
    \frac{g_*^2}{(4\pi)^2} \,\, \xrightarrow{\,\, N \rightarrow \infty \,\,} \,\, 0 \,,
\end{equation}
indicating the existence of a Banks--Zaks fixed point under control in this large $N$ limit.

We also study some examples where there does not exist a large $N$ limit, but the numerical value of the gauge coupling when the $\beta$-function vanishes is sufficiently small that we may be comfortable to trust the perturbative results. Consider $G = SU(10)$ with a single chiral fermion in the $\bm{252}$ representation. This theory has a $\mathbb{Z}_5$ one-form symmetry which has a mixed anomaly with the $\mathbb{Z}_{35}$ axial zero-form symmetry.\footnote{Following the discussion around equation \eqref{eqn:careful}, we find that the naive $\mathbb{Z}_{70}$ axial symmetry is reduced. The transformation of the fermion by the $\ell = 35$ element of $\mathbb{Z}_{70}$ is the same as the transformation of the fermion by any odd $n$ in the center of the gauge group: $\mathbb{Z}_{10}$. Therefore, the axial symmetry is really $\mathbb{Z}_{35}$. We note that this $\mathbb{Z}_2$ subgroup of $\mathbb{Z}_{70}$ which is gauged is actually the fermion number; hence there are no gauge-invariant composite fermions, and thus the anomalies cannot be matched in the infrared by massless composite fermions. This rules out one possible infrared scenario. See, for example, \cite{Anber:2021iip} for an analysis of the infrared of certain $SU(N)$ gauge theories where all operators are bosonic.} We find that the zero of the three-loop $\beta$-function is at
\begin{equation}
    \frac{g_*^2}{(4\pi)^2} \sim 0.0092 \,.
\end{equation}
Another example where a single chiral fermion almost saturates the inequality in equation \eqref{eqn:onelooponeloop} is when we take
\begin{equation}
    G = SU(2) \qquad \text{ and } \qquad \bm{R} = \bm{5} \,.
\end{equation}
In this case, the zero of the three-loop $\beta$-function occurs at
\begin{equation}
    \frac{g_*^2}{(4\pi)^2} \sim 0.0046 \,.
\end{equation}

We leave an in-depth analysis of the infrared fate of the non-supersymmetric simple gauge theories (including those where the global form of the gauge group is not simply-connected) with nontrivial one-form symmetries enumerated in this section for future work.

\subsection*{Acknowledgements}

We thank Minseok Cho, Jacques Distler, Jonathan Heckman, Sungwoo Hong, Sungkyung Kang, Heeyeon Kim, Ho Tat Lam, Ling Lin, Hyejung Moon, Xingyang Yu, and Hao Zhang for discussions. M.J.K.~and C.L.~thank KAIST for hospitality during the initiatory phase of this work; M.J.K.~is grateful to the Aspen Center for Physics where the work is performed in part, which is supported by a grant from the Simons Foundation (1161654, Troyer). M.J.K., C.L., and J.S.~thank the Simons Summer Workshop 2024 for hospitality, especially Martin Roček for 35 Beach Road, 
during the final stage of this work. 
M.J.K.~is supported by the U.S.~Department of Energy, Office of Science, Office of High Energy Physics, under Award Numbers DE-SC0013528 and QuantISED Award DE-SC0020360.
C.L.~acknowledges support from DESY (Hamburg, Germany), a member of the Helmholtz Association HGF; C.L.~also acknowledges the Deutsche Forschungsgemeinschaft under Germany's Excellence Strategy - EXC 2121 ``Quantum Universe'' - 390833306 and the Collaborative Research Center - SFB 1624 ``Higher Structures, Moduli Spaces, and Integrability'' - 506632645. K.H.L.~and J.S.~are supported by the National Research Foundation of Korea (NRF) grant RS-2023-00208602. J.S. is also supported by the National Research Foundation of Korea (NRF) grant RS-2024-00405629, and POSCO Science Fellowship of POSCO TJ Park Foundation.

\bibliographystyle{sortedbutpretty}
\bibliography{references}
	
\end{document}